\documentclass[twocolumn]{aastex62}

\usepackage{graphicx,amssymb,verbatim,times,rotating}

\received{}
\revised{}
\accepted{}
\submitjournal{ApJ}

\newcommand\lya{Ly$\alpha$}
\newcommand\ha{H$\alpha$}
\newcommand\hb{H$\beta$}
\newcommand\hg{H$\gamma$}
\newcommand{\oiiiweak}{[O\,{\sc iii}]$\lambda$4363}
\newcommand{\niiweak}{[N\,{\sc ii}]$\lambda$5755}
\newcommand{\oiiweak}{[O\,{\sc ii}]$\lambda\lambda$7319,7330}

\newcommand{\oii}{[O\,{\sc ii}]$\lambda\lambda$3726,3729}
\newcommand{\sii}{[S\,{\sc ii}]$\lambda\lambda$6717,6731}

\newcommand{\hii}{H\,{\sc ii}}
\newcommand{\oiiisingle}{[O\,{\sc iii}]$\lambda$5007}
\newcommand{\niisingle}{[N\,{\sc ii}]$\lambda$6584}
\newcommand{\oiisingle}{[O\,{\sc ii}]$\lambda$3727}

\begin{document}

\title{VLT/XShooter spectroscopy of Lyman Break Analogs: Direct method O/H abundances and nitrogen enhancements}

\correspondingauthor{Roderik A. Overzier}
\email{roderikoverzier@gmail.com}

\author{Maryory Loaiza-Agudelo}
\affiliation{Observat\'{o}rio Nacional/MCTIC, Rua General Jos\'{e} Cristino, 77, S\~{a}o Crist\'{o}v\~{a}o, Rio de Janeiro, RJ 20921-400, Brazil}

\author{Roderik A. Overzier}
\affiliation{Observat\'{o}rio Nacional/MCTIC, Rua General Jos\'{e} Cristino, 77, S\~{a}o Crist\'{o}v\~{a}o, Rio de Janeiro, RJ 20921-400, Brazil}
\affiliation{Institute of Astronomy, Geophysics and Atmospheric Sciences, University of S\~{a}o Paulo, Rua do Mat\~{a}o, 1226, S\~{a}o Paulo, SP 05508-090, Brazil}

\author{Timothy M. Heckman}
\affiliation{Department of Physics \& Astronomy, John Hopkins University, Bloomberg Centre, 3400 N. Charles Street, Baltimore, MD 21218, USA}

\shorttitle{Oxygen and nitrogen abundances of Lyman Break Analogs}
\shortauthors{Agudelo, Overzier \& Heckman}

\begin{abstract}
We used VLT/XShooter to target a sample of nearby analogs of Lyman Break Galaxies (LBGs). These Lyman Break Analogs (LBAs) are similar to the LBGs in many of their physical properties. We determine electron temperatures using the weak \oiiiweak\ emission line, and determine the oxygen abundance (O/H) using the direct and strong line methods. We show that the direct and strong line abundances are consistent with established relations within $\sim$0.2 dex. The analogs have nitrogen-to-oxygen ratios (N/O) and ionization parameters ($q$) that are, on average, offset with respect to typical local galaxies but similar to galaxies at $z\sim2$ and other analogs. The N/O and $q$ excesses correlate with the offsets observed in the strong line ratios, again similar to $z\sim2$. The star formation rate surface densities are consistent with the high electron density and ionization, indicating that the interstellar medium (ISM) pressure is set by feedback from the starbursts. For a given O/H, the apparent N/O excess arises due to the offset in O/H with respect to the local mass-metallicity relation. This can be explained by recent inflow of relatively metal-poor gas which lowers O/H while leaving N/O unchanged. The difficulties in determining even basic ISM parameters in these nearby analogs illustrates some of the challenges we face at much higher redshifts, where similar rest-frame optical diagnostics for large samples of galaxies can be accessed with JWST. 
\end{abstract}

%

\section{Introduction}

Many of today's most actively pursued open questions in galaxy evolution focus on the period between cosmic `dawn' and cosmic `noon'. A complete picture of how galaxy evolution proceeded within this period must include answers as to how galaxies received and recycled their gas, formed their stellar populations, and how the ionizing radiation from the first galaxies reionized the universe. While new observatories have begun to address these questions by directly targeting the most distant galaxies, a complementary path is to study similar processes occurring in the more nearby universe using so-called `analogs' in the hope of gaining insight into the processes occurring at less accessible wavelengths, resolutions or redshifts. One such class of objects are the Lyman Break Analogs (LBAs) first studied by \citet{heckman05} and \citet{hoopes07}. The LBA project was designed to find and study relatively nearby starburst galaxies that share typical characteristics of Lyman Break Galaxies (LBGs) at high redshift. A crossmatch of the Sloan Digital Sky Survey (SDSS) with the GALEX UV imaging survey \citep{martin05} was used to select luminous ($L_{FUV}>10^{10.3}$ L$_\odot$) and compact ($I_{FUV} > 10^9$ L$_\odot$ kpc$^{-2}$) star-forming galaxies at $z<0.3$ having similar rest-frame far-UV properties as typical (i.e. $L_{FUV}\simeq L^*_{z=3}$) LBGs.
These simple criteria select galaxies with relatively high SFRs that are furthermore relatively compact and experiencing modest extinction by dust similar to the high redshift LBGs. The sample of LBAs from SDSS-GALEX was significant because previous surveys lacked the depth and coverage in the UV to find significant numbers of such galaxies, making comparisons between local and distant galaxies less direct \citep[e.g.][]{meurer99}. 

Several useful samples of local analogs have been constructed in recent years. Examples include galaxies having high equivalent width \lya\ selected to be good local analogs of \lya\ emitters (LAEs) at high redshift \citep[e.g., the \lya\ Reference Sample (LARS);][]{ostlin14}, high equivalent width \ha\ sources selected to be local analogs of \ha\ emitters (HAEs) at $z\sim4$ \citep{shim13}, the Green Pea (GP) galaxies selected on the basis of strong optical emission lines that also span the properties of LAEs, HAEs and LBGs at high redshift \citep{cardamone09,lofthouse17}, high \oiiisingle/\oiisingle\ ratio sources that are good candidate Lyman continuum leakers \citep[e.g.][]{nakajima14,schaerer16}, and galaxies selected based on their offsets in the BPT diagram designed to match higher redshift samples \citep[e.g.][]{bian16,cowie16}. It is important to mention that there is typically significant overlap between all these samples and the UV-selected LBAs studied here \citep[e.g.][]{cardamone09,amorin12,ostlin14}. 

The sample of LBAs share numerous other physical characteristics of star-forming galaxies at high redshift. The sizes, morphologies and gas kinematics of LBAs have been compared in detail with those of star-forming galaxies at $z>2$, finding general good agreement in the main parameter distributions \citep{overzier08,overzier10,goncalves10}. Although the triggers of star formation are likely to be different for the low and high redshift starbursts, at least the distributions of star formation, dust and gas appear to be comparable. The star formation in LBAs is dominated by luminous clumpy emission \citep{overzier08,overzier09,overzier10} that resembles that seen in the clumpy galaxies at intermediate redshifts \citep[e.g.][]{elmegreen13,garland15}. Some LBAs have luminous unresolved super-star clusters that are structurally similar to those seen in (lensed) sources at $z\gtrsim6$ \citep[e.g.][]{bradley12,bouwens17}. Although a small subset of LBAs also appear to host low-luminosity, obscured AGN at their SNe-feedback dominated centers \citep{jia11,alexandroff12}, the radio and X-ray luminosities associated with these AGN are well below the current detection threshold in high redshift star-forming galaxies \citep[e.g.][]{habouzit17,latif18}. These structural and energetic similarities further indicate that LBAs may be good local laboratories for probing the extreme physical conditions expected to govern the interstellar medium of young galaxies at high redshifts. 

LBAs were important for testing the so-called $\beta$--IRX relation for dust-correcting UV-based star formation rate measurements of star-forming galaxies \citep{meurer99,salim18}. Because LBAs are more similar to LBGs compared to typical local starburst galaxies that are often either highly obscured or have SFRs orders of magnitudes lower compared to their high redshift counterparts, dust-correction methods that are based on this sample should be less biased than those based on other types of local populations \citep{overzier11,bouwens12}. For similar reasons, LBAs can be used to probe the conditions responsible for the emission of the far-infrared fine structure lines, molecular gas and dust continuum of ordinary galaxies on the main sequence at high redshifts \citep[e.g.][]{overzier11,goncalves14,contursi17,wu19}. 

\begin{figure}
\includegraphics[width=\columnwidth]{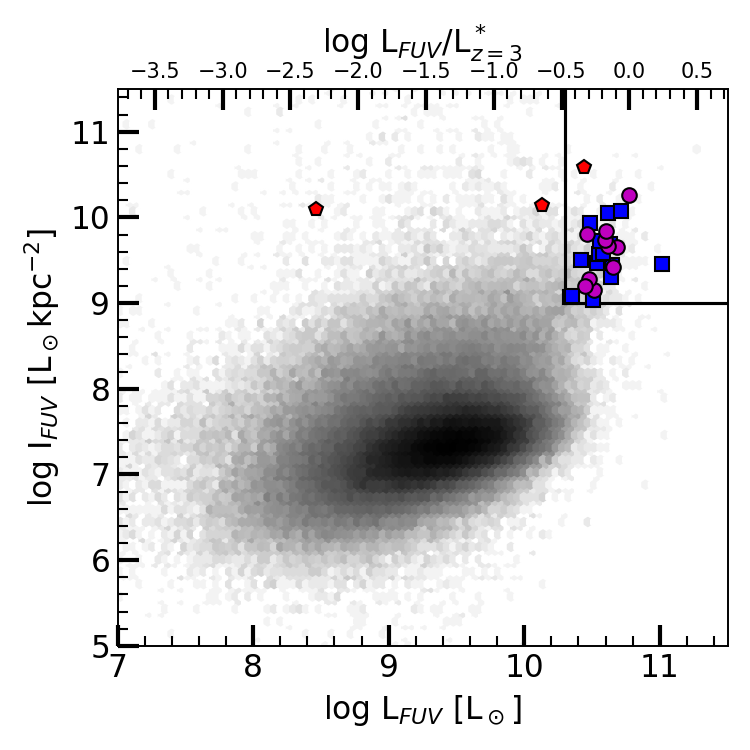}
\caption{\label{fig:sample}Sample properties. Far-UV luminosity versus surface brightness for all sources in the SDSS-GALEX crossmatched catalog (grey shaded region). Large symbols indicate objects for which VLT/XShooter spectroscopy was obtained. Blue squares: LBAs from \citet{overzier09}. Magenta circles: new LBAs selected on the basis of a large offset in the BPT-diagram (9 objects) or a high HST/COS far-UV flux (1 object). Red pentagons: Objects selected on the basis of their relatively low [S\,{\sc ii}]/H$\alpha$-ratios. See text and Table \ref{tab:sample} for details. 
}
\end{figure}

\begin{figure}
\includegraphics[width=\columnwidth]{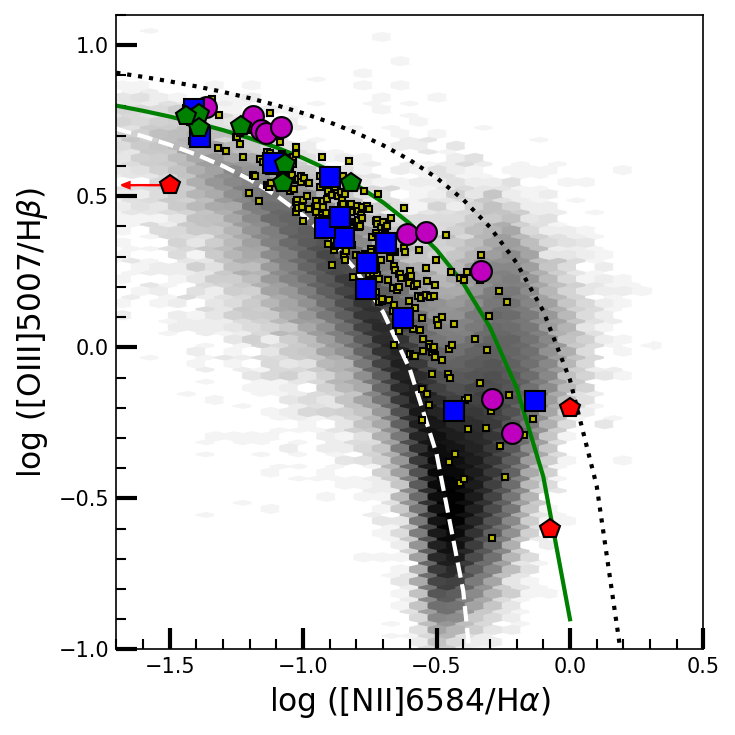}
\caption{\label{fig:bpt}BPT diagram of all sources in the SDSS-GALEX crossmatched catalog (grey shaded region) compared to LBAs. Small yellow squares indicate LBAs from the extended LBA catalog. Large symbols indicate objects with deep optical spectra discussed in this paper. Large blue squares are LBAs from \citet{overzier09}. Measurements for 7 objects with spectroscopic data from the literature are indicated with green pentagons (B14: \citet{brown14}; A12: \citet{amorin12}). Large magenta circles are new LBAs from the extended sample and selected for VLT/XShooter spectroscopy on the basis of a large offset in the BPT-diagram (9 objects) or a high HST/COS far-UV flux (1 object). The large red pentagons are objects selected on the basis of their relatively low [S\,{\sc ii}]/H$\alpha$-ratios (3 objects). See text and Table \ref{tab:sample} for details.}
\end{figure}

Far-UV spectroscopic observations with the Cosmic Origins Spectograph (COS) on the Hubble Space Telescope (HST) revealed that LBAs typically have strong outflows with some extreme cases up to $\sim1000-2000$ km s$^{-1}$ \citep{heckman11,heckman15,heckman16,borthakur14}. These outflows correlate strongly with the SFR per unit area \citep{heckman15}, as expected if the outflows are mainly driven by the momentum flux in compact starbursts. The LBA program also delivered a strong case of detection of Lyman continuum (LyC) photon escape \citep{borthakur14}, showing for the first time a strong connection between the SNe-driven outflows in compact starbursts and LyC escape. \citet{overzier09} showed that the extinction-corrected \ha\ was relatively weak compared to their far-UV and far-IR emission in some of the most compact LBAs, suggesting that ionizing radiation may be escaping (among several other explanations). \citet{heckman11} noted that besides the strong outflows and relative weakness of \ha, some LBAs have significant residual intensities in the cores of the saturated interstellar low-ionization absorption lines tracing the neutral gas, and that these sources also tend to show significant blue-shifted \lya\ in emission. All these properties can be explained by a simple model in which compact starbursts drive powerful winds that remove the neutral gas along certain lines of sight, allowing LyC and \lya\ photons to escape. In a follow-up study, \citet{alexandroff15} further found that objects either confirmed \citep{borthakur14} or suspected \citep{heckman11} of being LyC leakers based on the above-mentioned indicators, also have relatively weak \sii\ emission line doublets, indicating the presence of matter-bounded \hii\ regions, while other diagnostics proposed to be indicators for LyC escape, such as relatively weak dust-corrected \ha\ or the \oiiisingle/\oiisingle\ line flux ratio \citep[e.g.][]{nakajima14}, did not correlate with the other tracers. The indirect indicators of LyC escape based on the study of LBAs have thus offered a number of empirical probes that may be used as proxies for the escape fractions of galaxies during the EoR for which LyC emission cannot be measured directly \citep{borthakur14,alexandroff15}. 

Analysis of optical spectra from SDSS has shown that LBAs lie, on average, below the local stellar mass-metallicity relation \citep{hoopes07,overzier10}, and above the star-forming sequence in the `BPT’ diagram \citep[][and this paper]{overzier09,bian16,kojima17,patricio18}, again two important features that are analogous to those of typical star-forming galaxies at high redshift \citep[e.g.][]{erb06,steidel14}. Given that both LBAs and LBGs are presumed to be galaxies undergoing a phase of rapid build-up of their stellar populations from recent influx of relatively metal-poor gas, LBAs could perhaps also aid in answering a number of open questions related to the chemical enrichment history of early galaxies. However, a closely related problem that has played a central role in recent years is that star-forming galaxies at high redshift occupy different locations in the main optical diagnostic emission line diagrams \citep[e.g.][]{liu08,overzier09,steidel14,bian16,bian17,bian18}. Specifically, \citet{steidel14} showed that UV-selected galaxies at $z\sim2$ lie along a locus that is offset with respect to that of local star-forming galaxies in the \oiiisingle/\hb\ versus \niisingle/\ha\ ``BPT'' diagram \citep{baldwin81}. Given that the conditions in the interstellar medium in typical star-forming galaxies at local and high redshifts are likely very different, this is perhaps not so surprising. Possible explanations include contributions from AGN photoionization, shocks, different N/O ratios, higher ionization parameters and harder radiation fields. 

Understanding the BPT offsets and possible evolution is important, because they involve the same optical emission line ratios that are being employed to determine the nebular gas abundances of star-forming galaxies as well as to identify AGN \citep[e.g.][]{kewley13,hirschmann17}. This is especially important at high redshifts where the strong optical emission lines are often the only viable way of determining these ISM abundances. Proper local analogs can therefore aid in these studies as well. Strong line metallicities have been compared against direct method values in gravitationally-lensed galaxies with oxygen auroral lines detected out to $z\sim2.5$, showing that locally calibrated methods are reliable at high redshift within a dispersion of $\sim$0.2 dex \citep[e.g.][]{patricio18,gburek19}. \citet{bian16} selected a new set of local analogs from SDSS solely on the basis of their proximity to the locus of $z\sim2$ galaxies from \citet{steidel14} in the BPT diagram, showing that these objects have relatively high ionization parameters and electron densities that can only partly be explained by their increased (specific) star formation rates compared to typical star-forming galaxies. \citet{bian17} show that at low stellar masses these analogs lie $\sim0.2$ dex below the local $M_*-Z$ relation (MZR) when using metallicities based on the N2 or O3N2 diagnostics \citep{pettini04}, similar to samples at $z\sim2$ \citep[e.g.][]{steidel14,sanders15}. \citet{bian18} stacked the SDSS spectra of their analogs sample to measure direct oxygen abundances, finding that local strong line calibrations underestimate the abundances by $\lesssim0.1$ dex, and providing updated calibrations. Various authors have found that the offsets are often related to an increased N/O abundance at high redshift compared to local galaxies at a fixed oxygen abundance \citep[e.g.][]{masters14,shapley15,sanders16a,kojima17}. This could be an effect of enhanced nitrogen production by WR stars, or, more likely, the consequence of rapid accretion of low metallicity gas which reduces O/H while leaving N/O unchanged \citep{koppen05,amorin10,amorin12,masters16}. The N/O--O/H relation appears constant with redshift, and harder ionizing spectra may be required at high redshift to fully explain the BPT offsets \citep{steidel16,strom17,strom18,shivaei18}. These harder spectra could be a result of the differences in star formation histories, and thus the chemical enrichment, between typical local and high redshift galaxies. 

\citet{andrews13} used large stacks of star-forming galaxies from the SDSS along the star-forming main sequence to compare the direct-method oxygen abundances with several calibrations derived from strong optical lines. They show that the MZR has a clear SFR-dependence, and that N/O correlates with O/H, star formation history and stellar mass. \citet{brown16} improved local strong-line methods for use at high redshift by quantifying the dependence on specific star formation history. \citet{brown14} tested the direct method using deep follow-up spectra obtained for four LBAs from the sample of \citet{overzier09}, finding that the strong line method abundances are in agreement with those from the temperature sensitive method. \citet{amorin12} confirmed the low oxygen abundances but remarkably high nitrogen to oxygen ratios in three LBAs. \citet{kojima17} show that local analogs and $z\sim2$ galaxies with BPT offsets can be explained by excesses in either N/O or ionization parameter, or combinations thereof, while their data did not allow them to test for changes in hardness of the radiation. 

In this Paper, we analyze the oxygen and nitrogen abundances of LBAs determined through the direct and strong line methods, and investigate the cause of offsets in ionization, O/H and N/O in the context of typical star-forming galaxies at $z\sim2$. We use magnitudes in the AB system, a Chabrier initial mass function and the solar metallicity scale from \citet{asplund09} where 12 + log(O/H)$_\odot=8.69$ and log(N/O)$_\odot=-0.86$. The cosmological parameters are set to $\Omega_m = 0.286$, $\Omega_\Lambda = 0.714$ and $H_0=69.6$ km s$^{-1}$ Mpc$^{-1}$.

\section{Sample and observations}

\subsection{Sample selection}

The targets studied as part of this paper were selected from a number of sources related to the LBA project, and by no means represent a complete sample. 17 targets (marked `LBA' in Table \ref{tab:sample}) were selected from the original sample of \citet{heckman05} and studied in detail by \citet{overzier09}. Ten new LBAs (marked `LBA2' in Table \ref{tab:sample}) were taken from an extended sample of several hundred LBAs in the crossmatch between SDSS data release 7 \citep[DR7;][]{abazajian09} and GALEX GR6. Of these ten targets, nine targets (marked `BPT' in Table \ref{tab:sample}) were further selected based on having large BPT offsets and observability, while one target (marked `COS' in Table \ref{tab:sample}) was selected because it has been observed with the HST/COS as part of a program to target LBAs having high far-UV fluxes within the COS aperture. Finally, three more targets (marked `S2-Deficit' in Table \ref{tab:sample}) were selected on the basis of their relative weakness of the [SII]6717,6731 optical emission-lines. This was motivated by previous work on LBAs that showed that objects with abnormally low [SII] 6717,6731/H$\alpha$ flux ratios compared to typical star-forming galaxies may have matter-bounded conditions in the ISM which can result in the escape of ionizing photons \citep[e.g.][]{pellegrini12}. This has inspired a new category of local analogs that have potentially large escape fractions \citep{alexandroff15,wang19}. All three `S2-Deficit' objects satisfy the UV surface brightness criterion of LBAs, but one is of substantially lower FUV luminosity than LBAs, as defined by \citet{heckman05}. This object is a known low metallicity blue compact dwarf \citep[e.g., it was part of Subsample 1 from][]{izotov07}.

Originally part of the LBA sample, two objects (001054 and 005439) were found to be Type 1 AGN based on broad Mg II lines detected in the UVB-arm of VLT/XShooter. These two objects are considered to be contaminants of the LBA sample as the emission from the unobscured nucleus likely dominates the far-UV flux detected by GALEX. These two sources were removed from the sample. Object 210358 had problems with the observations, as the on-slit dither positions were chosen to be too close to eachother leading to problems with the background subtraction. This object is not studied further as part of this paper.

Several objects from the \citet{overzier09} sample have been observed with deep spectroscopic data before. \citet{amorin12} observed three of the sources with the GTC-OSIRIS spectrograph (004054, 113303, 232539), while \citet{brown14} observed four sources with the LBT-MODS1 spectrograph (004054, 005527, 020356, 092600), one of which (004054) is in common with \citet{amorin12}. These six unique sources include two LBAs not covered by our spectroscopy (092600 and 113303), and four that overlap with our data. Although these authors have already derived many of the parameters similar to the ones that we will derive here (i.e. stellar populations, densities, temperatures and abundances), we will use the stellar absorption and dust-corrected line fluxes as reported by these authors in order to compare their results with our own data. 
Fig. \ref{fig:sample} shows the rest-frame UV properties of the VLT/XShooter sample compared to the local galaxy population. The location of the objects in the BPT diagram as measured from the VLT/XShooter spectra or given by the literature data is shown in Fig. \ref{fig:bpt}.

\begin{figure*}
\begin{center}
\includegraphics[width=0.33\textwidth]{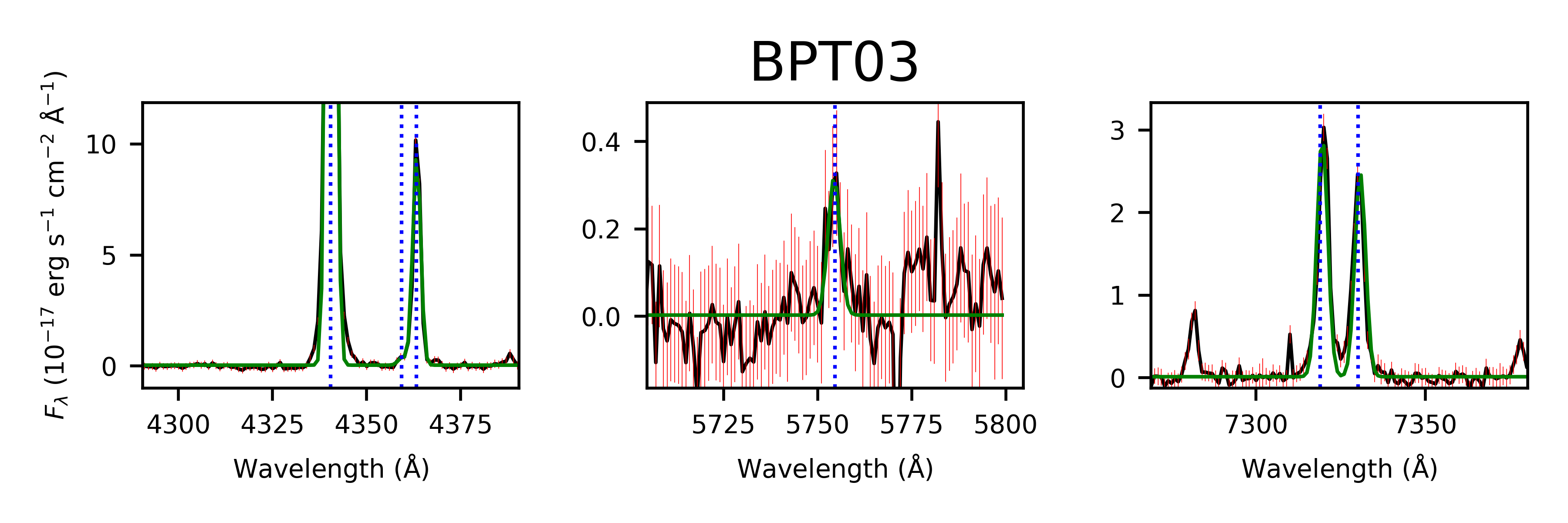}
\includegraphics[width=0.33\textwidth]{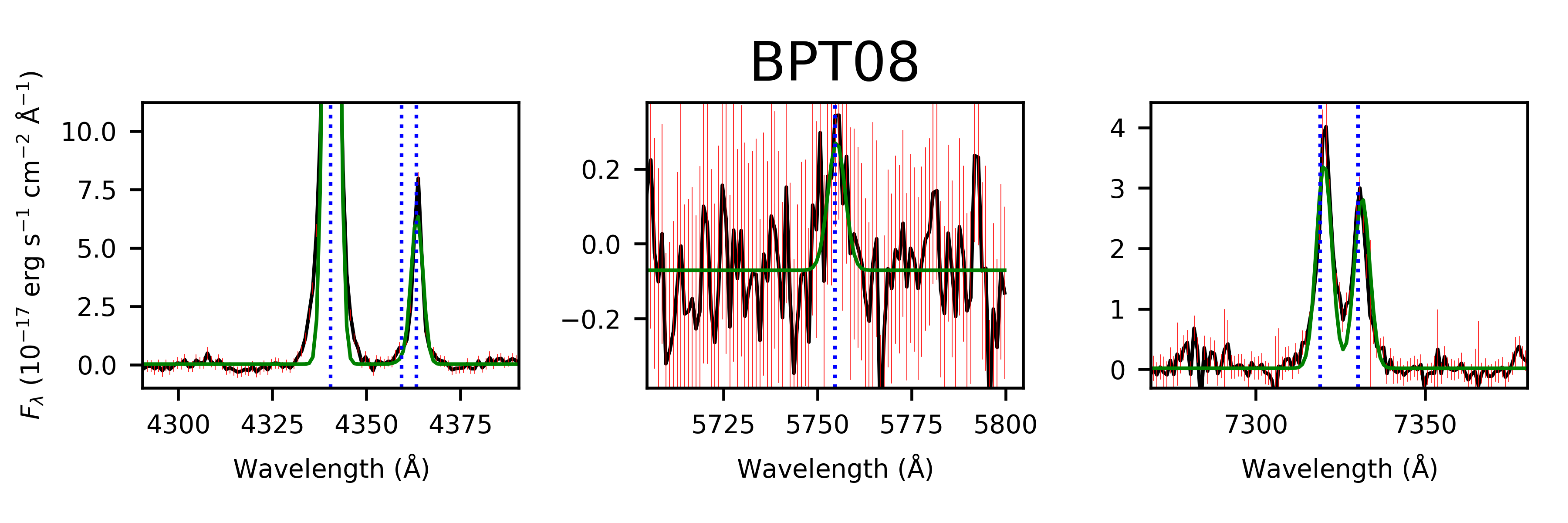}
\includegraphics[width=0.33\textwidth]{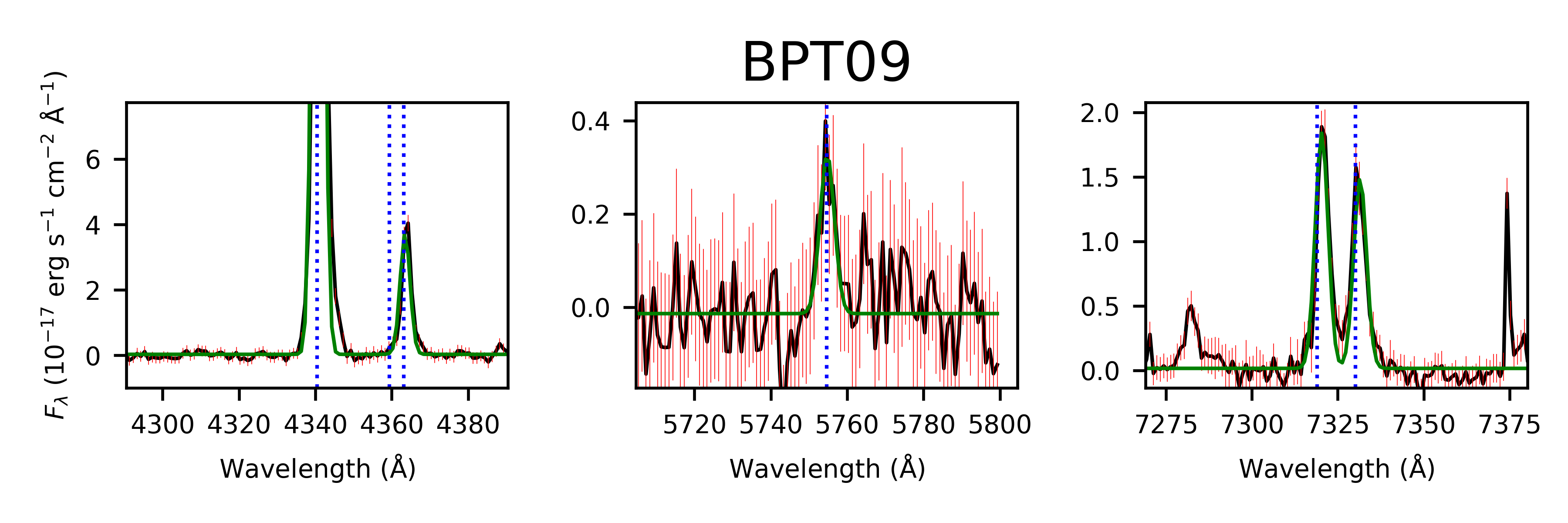}\\
\includegraphics[width=0.33\textwidth]{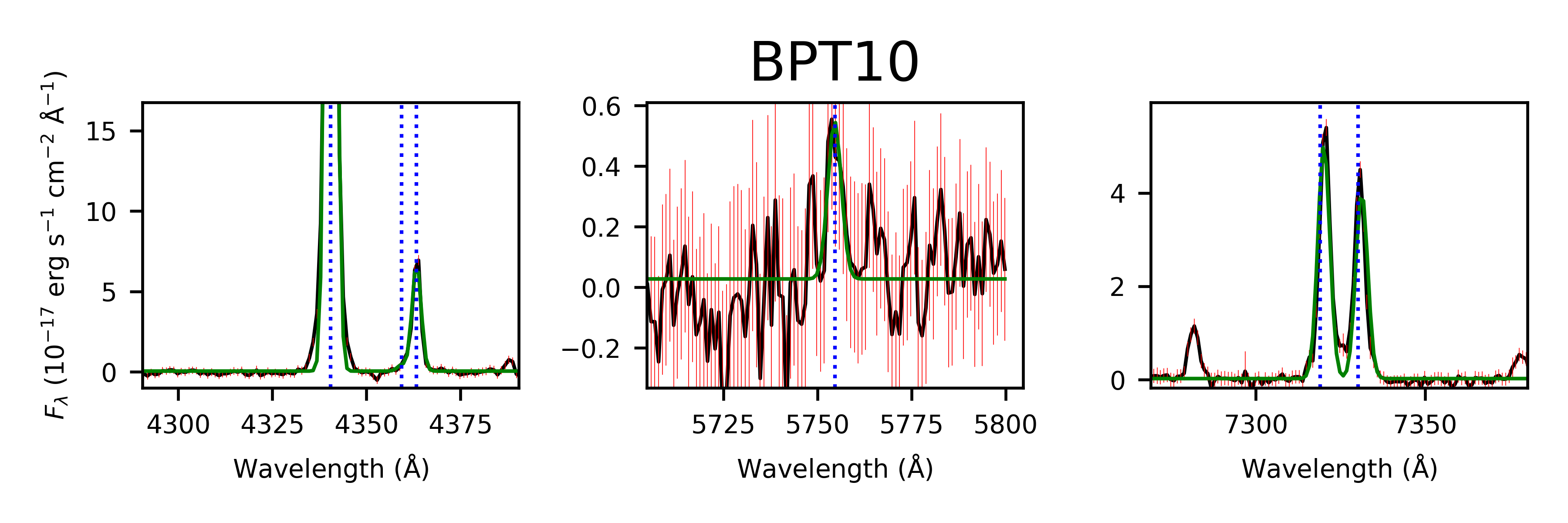}
\includegraphics[width=0.33\textwidth]{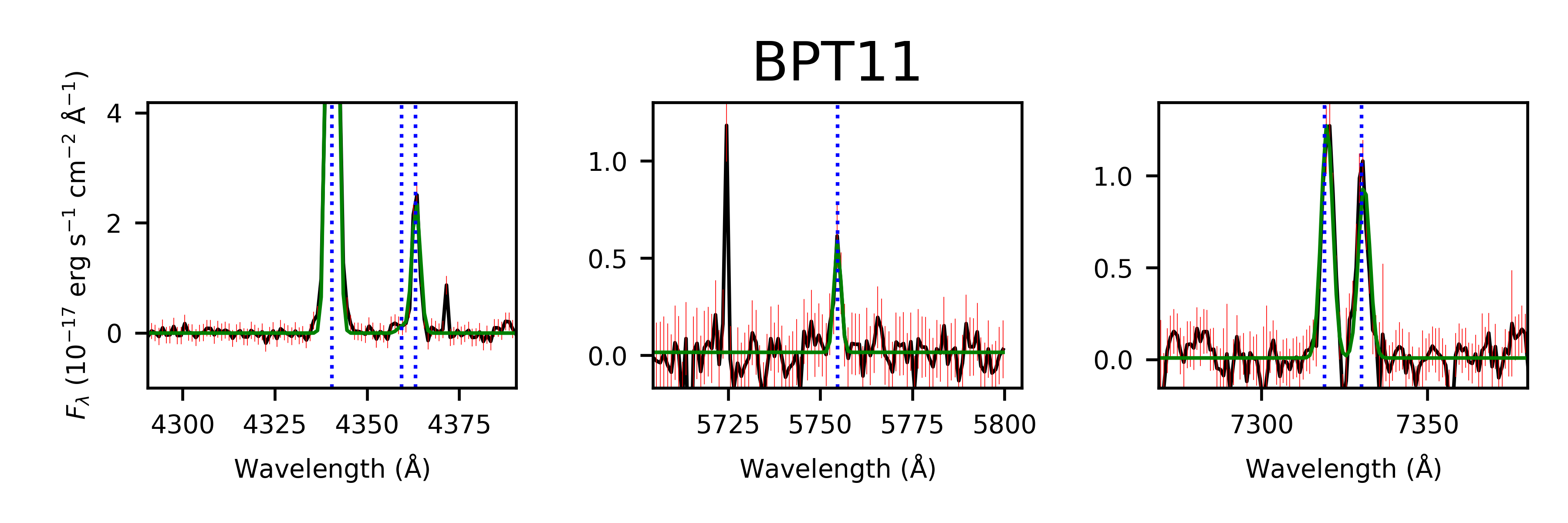}
\includegraphics[width=0.33\textwidth]{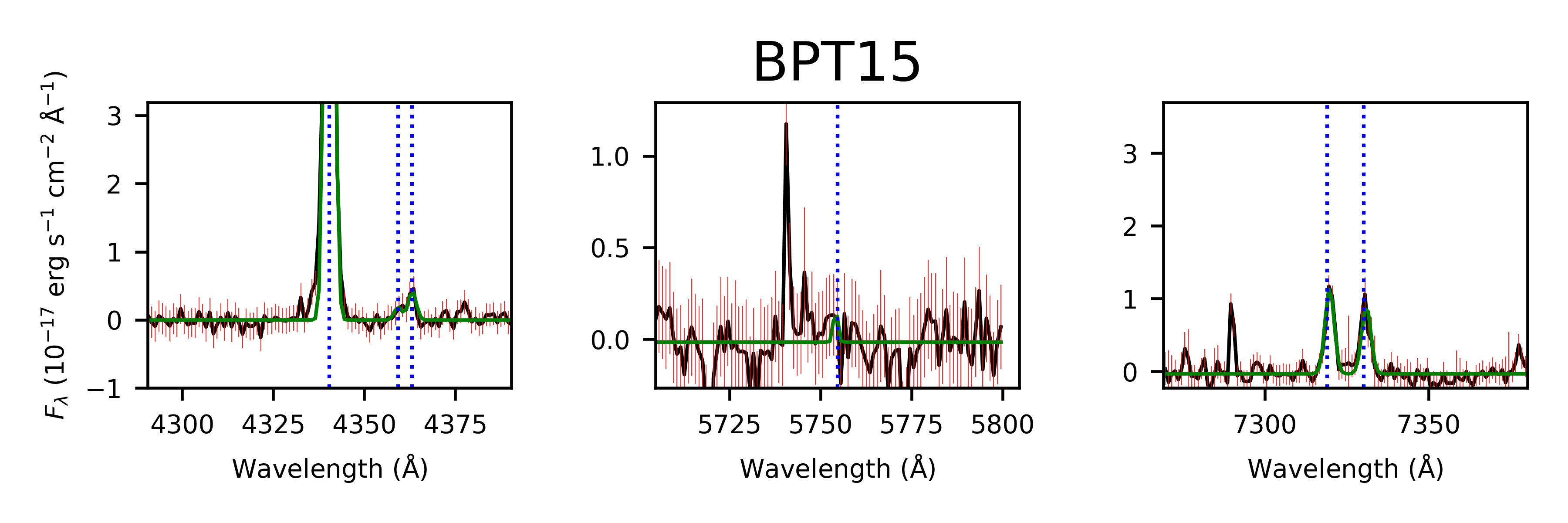}\\
\includegraphics[width=0.33\textwidth]{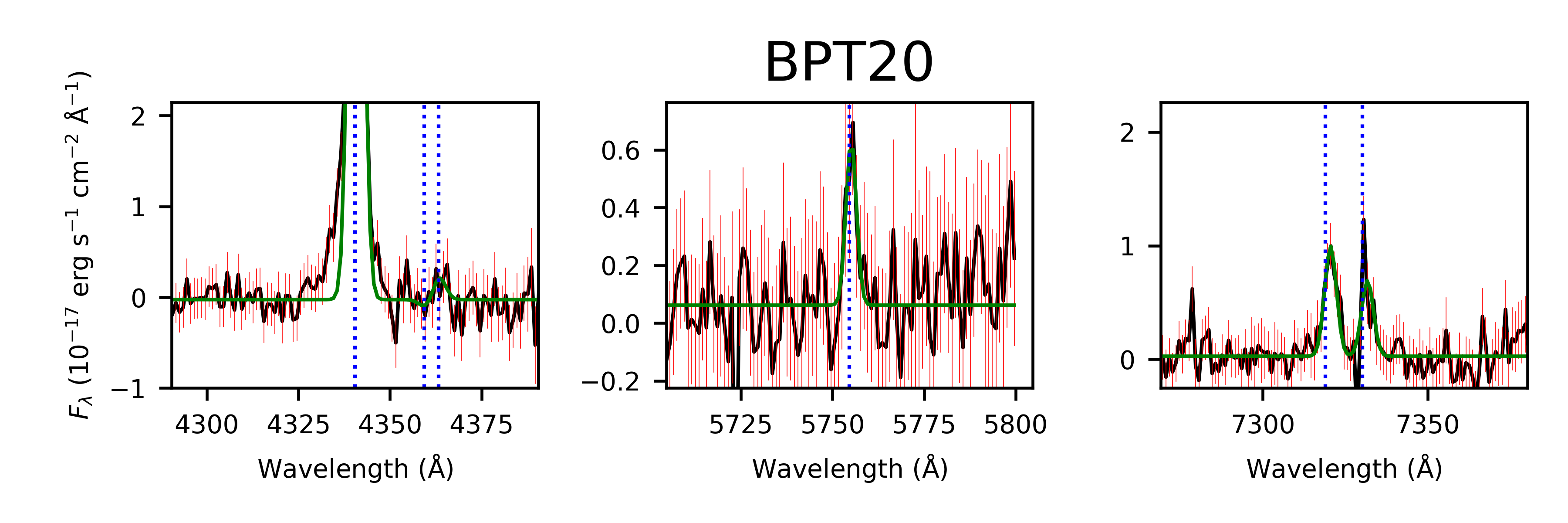}
\includegraphics[width=0.33\textwidth]{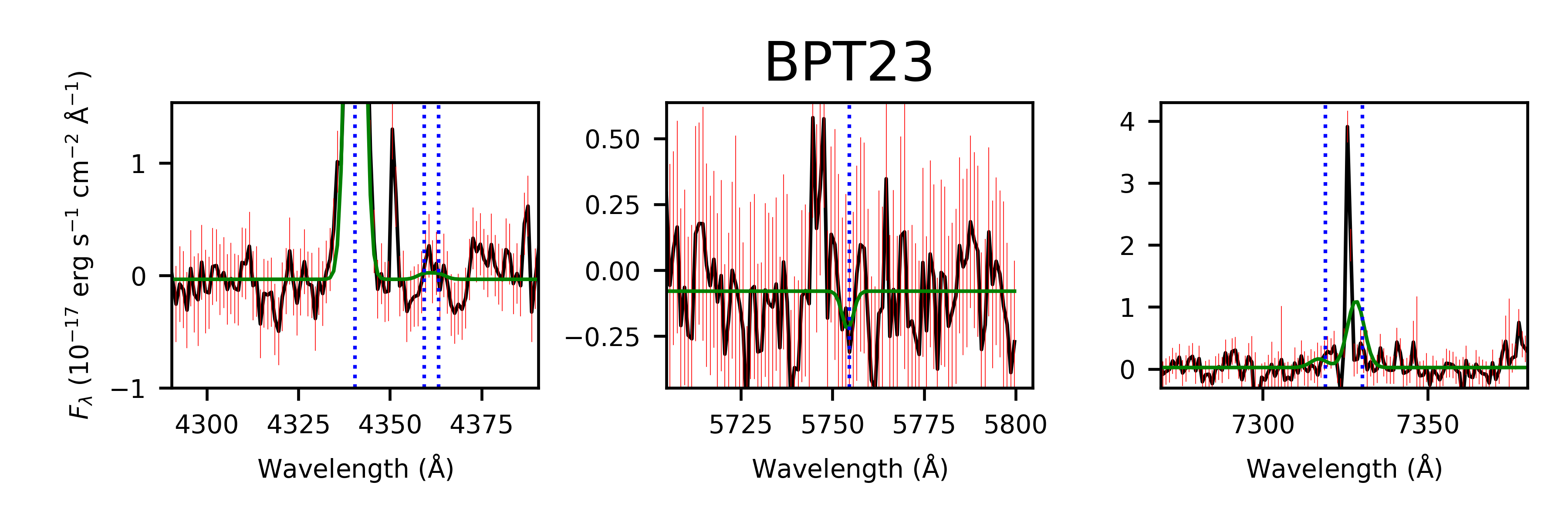}
\includegraphics[width=0.33\textwidth]{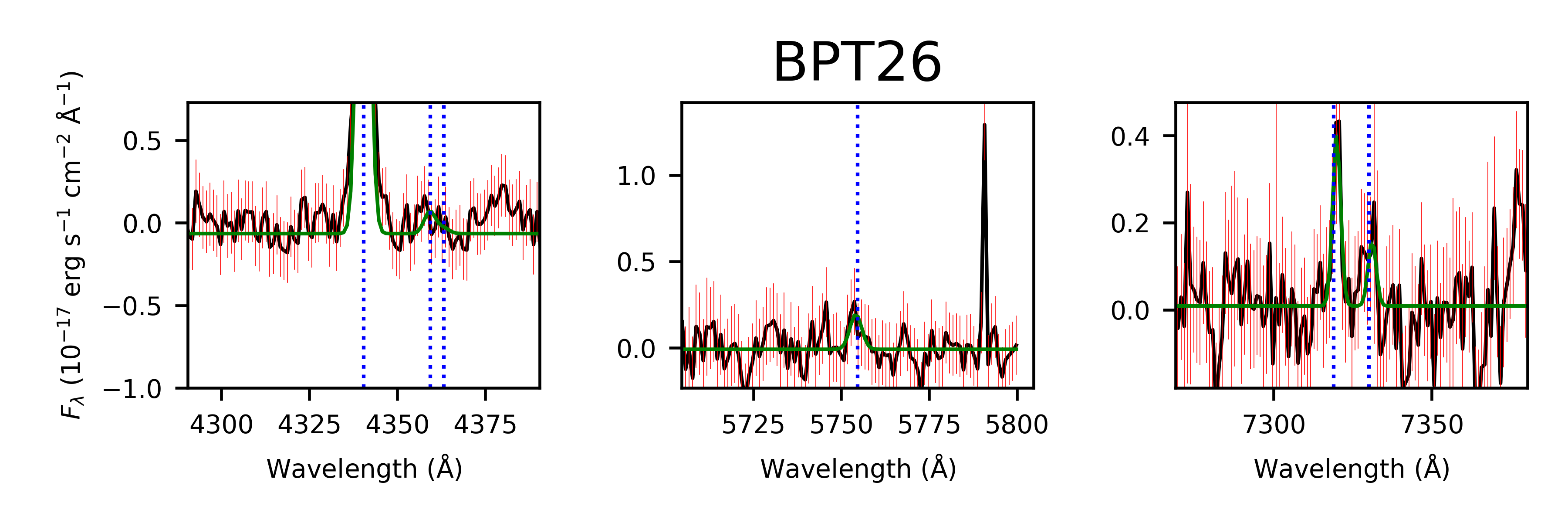}\\
\includegraphics[width=0.33\textwidth]{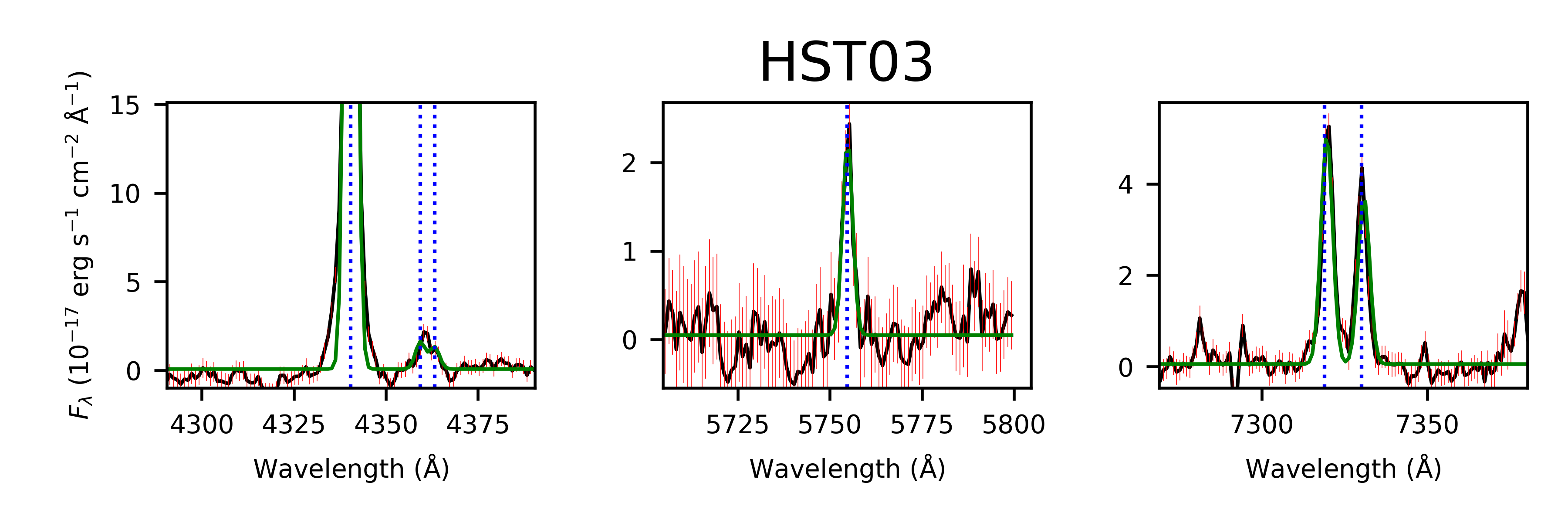}
\includegraphics[width=0.33\textwidth]{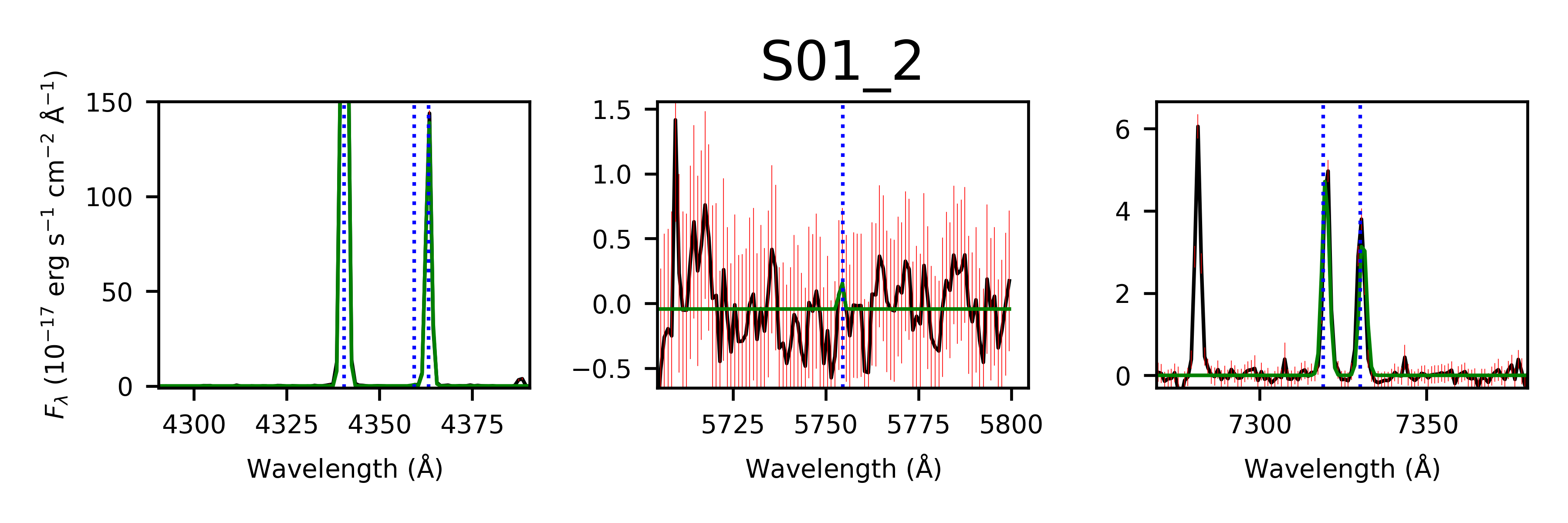}
\includegraphics[width=0.33\textwidth]{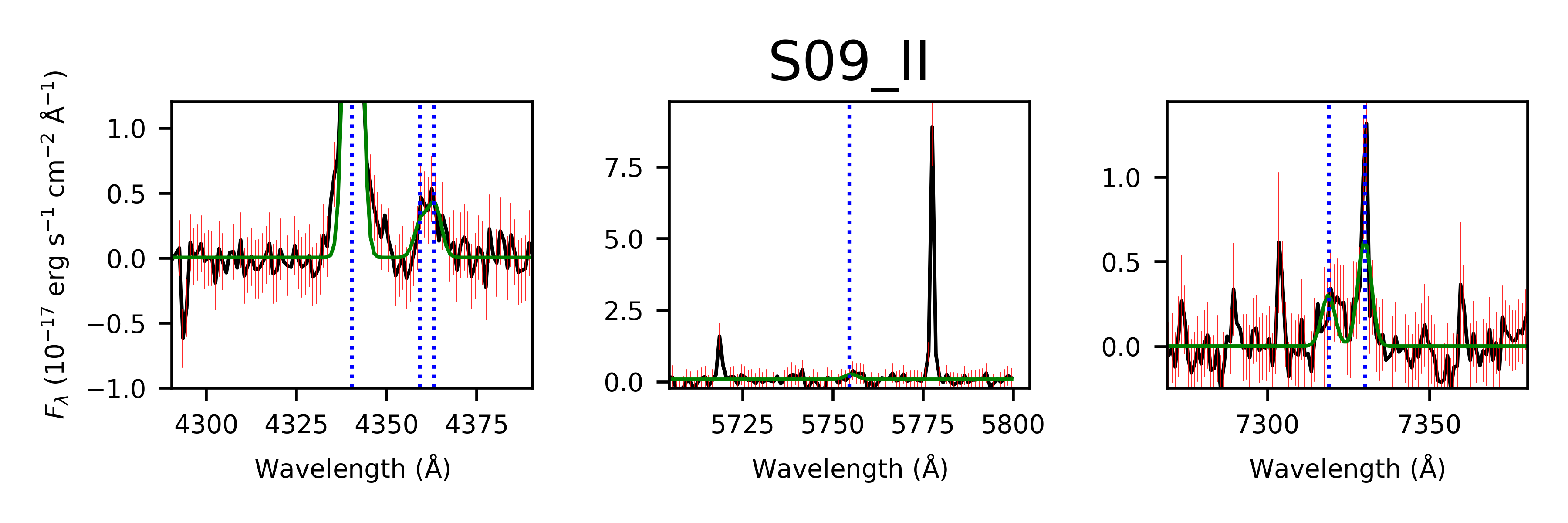}\\
\includegraphics[width=0.33\textwidth]{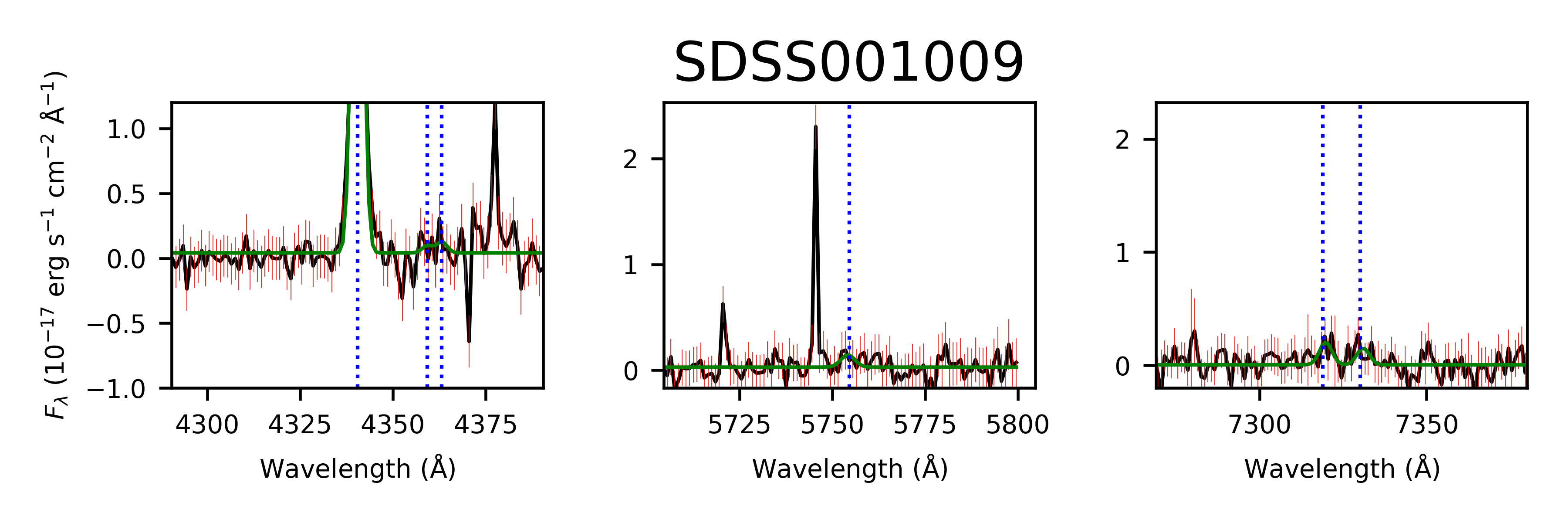}
\includegraphics[width=0.33\textwidth]{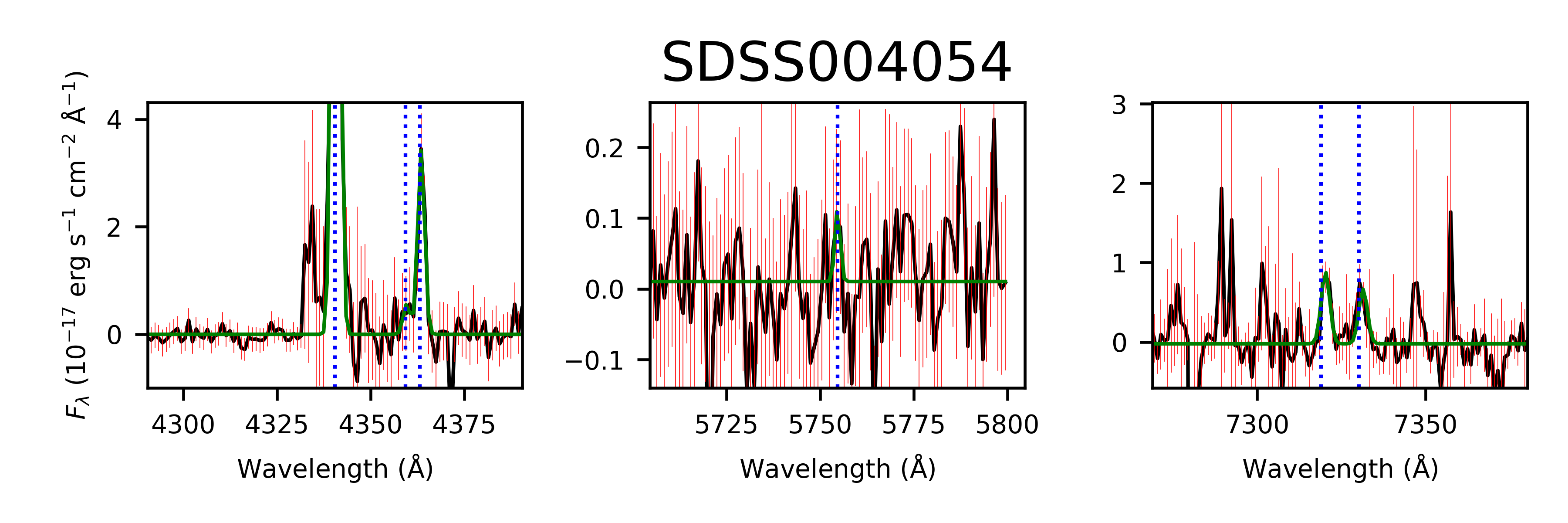}
\includegraphics[width=0.33\textwidth]{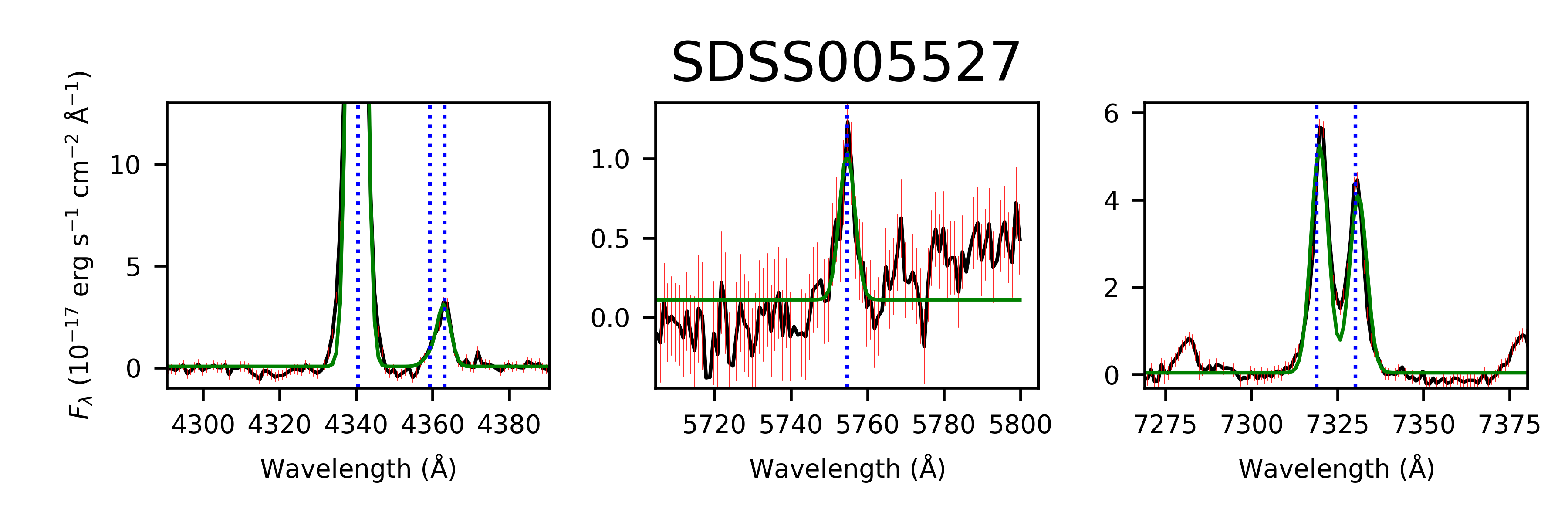}\\
\includegraphics[width=0.33\textwidth]{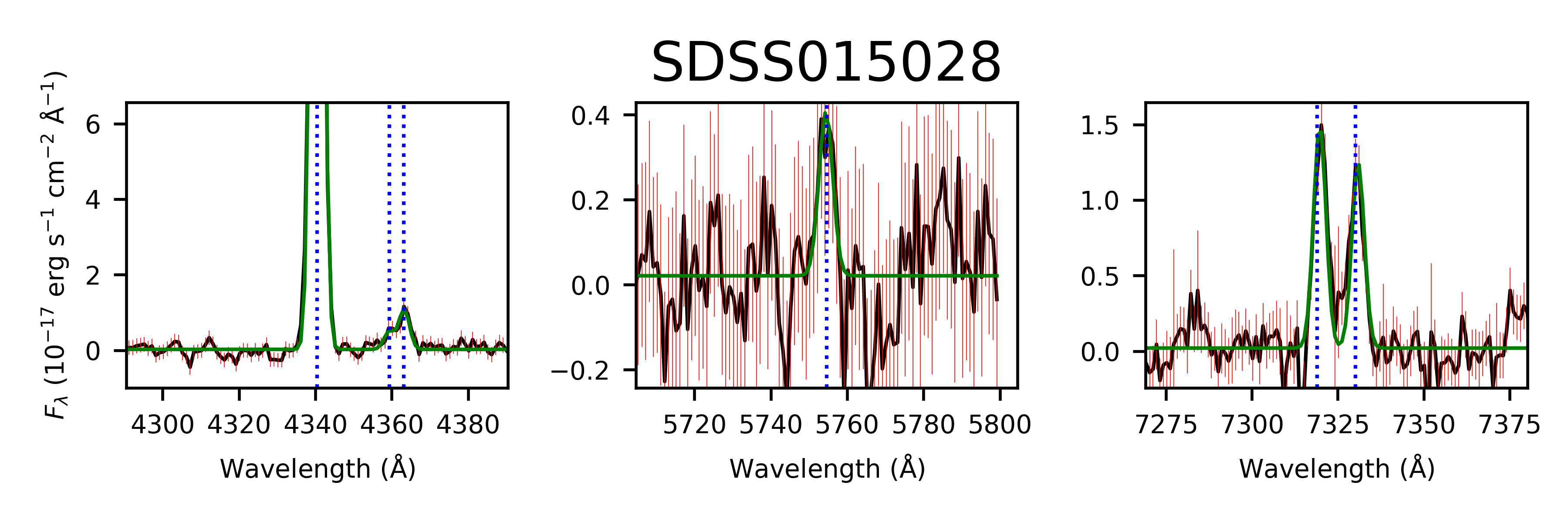}
\includegraphics[width=0.33\textwidth]{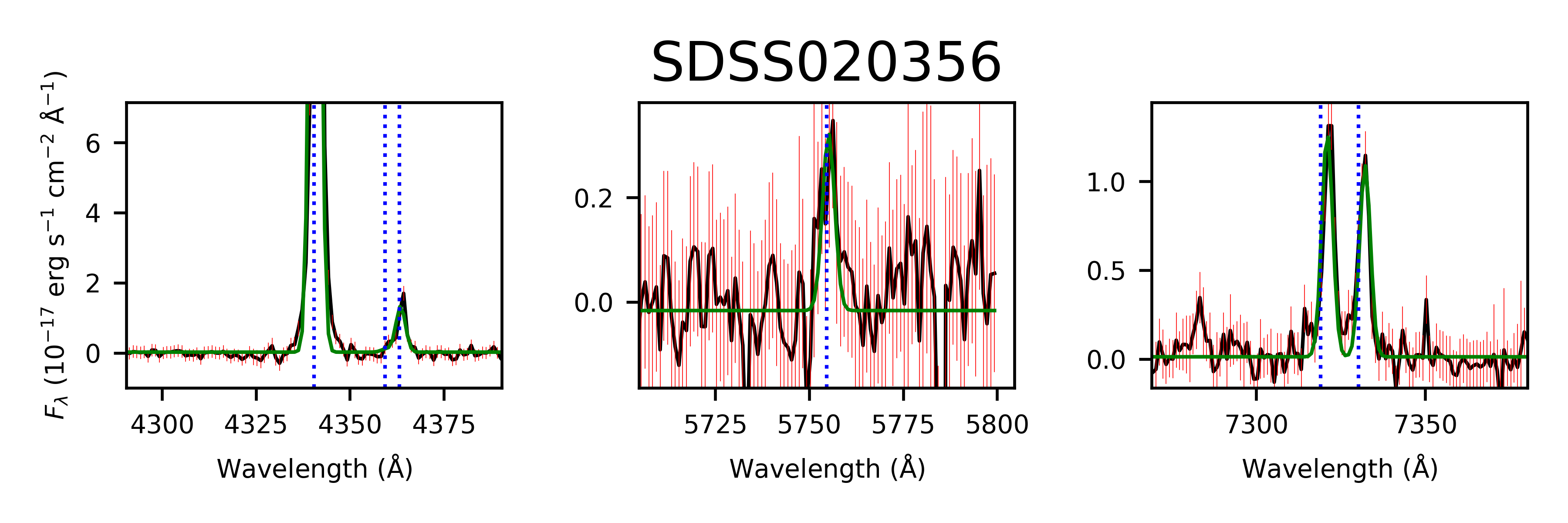}
\includegraphics[width=0.33\textwidth]{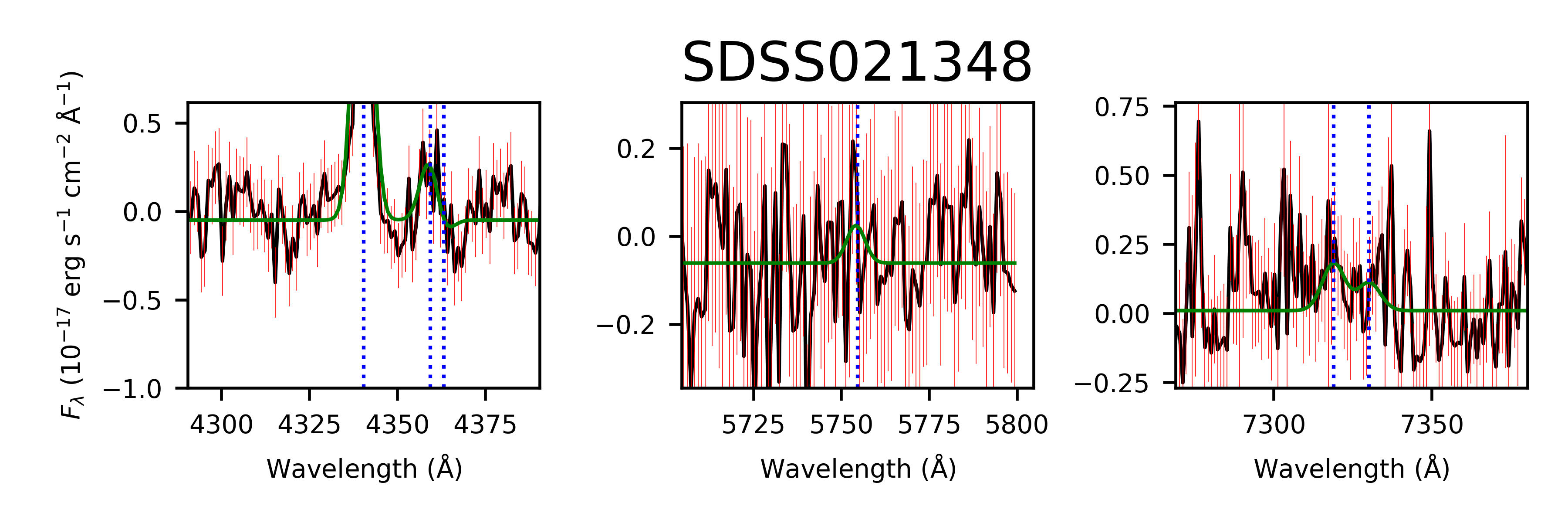}\\
\includegraphics[width=0.33\textwidth]{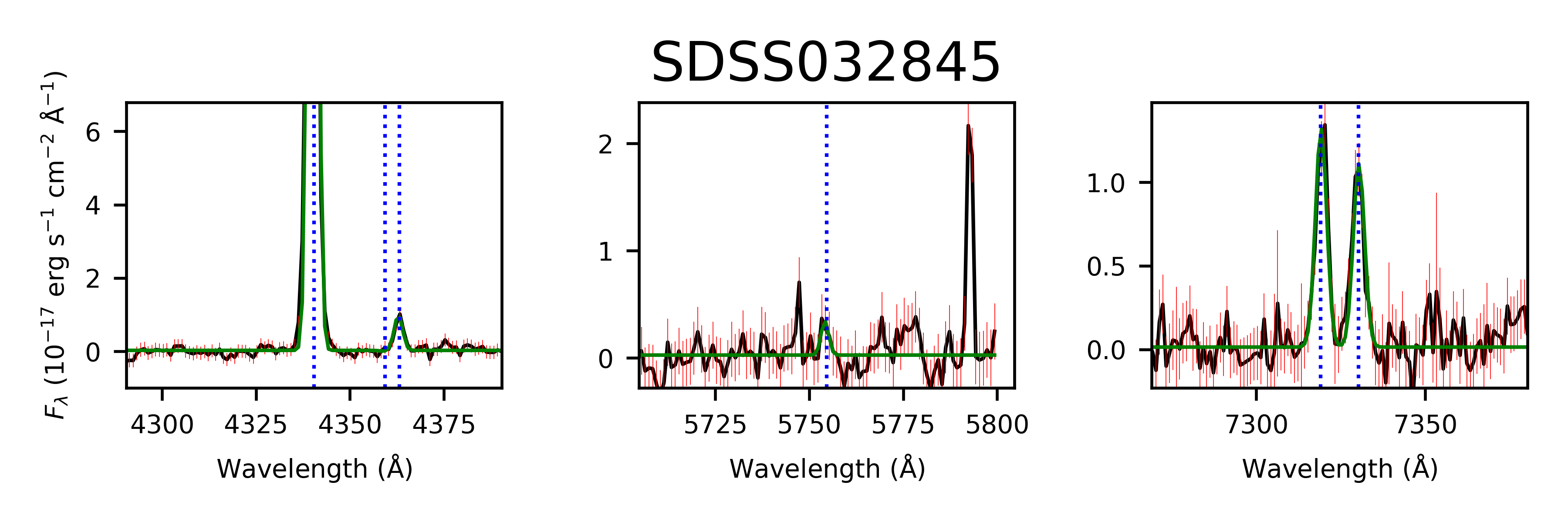}
\includegraphics[width=0.33\textwidth]{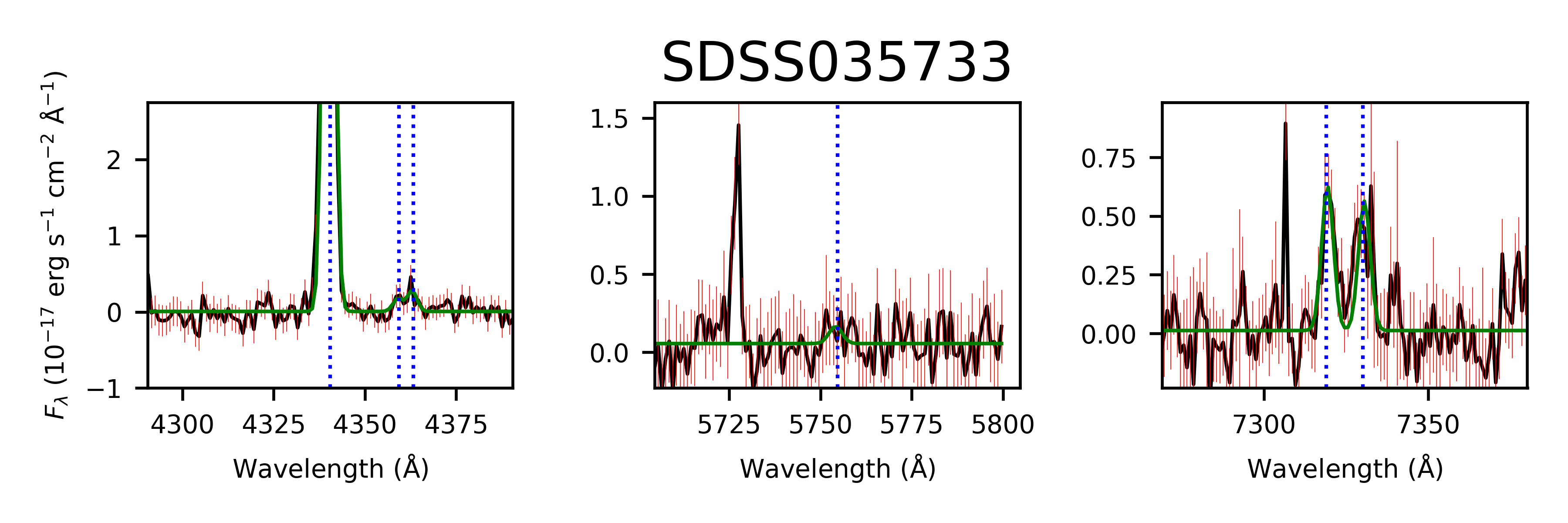}
\includegraphics[width=0.33\textwidth]{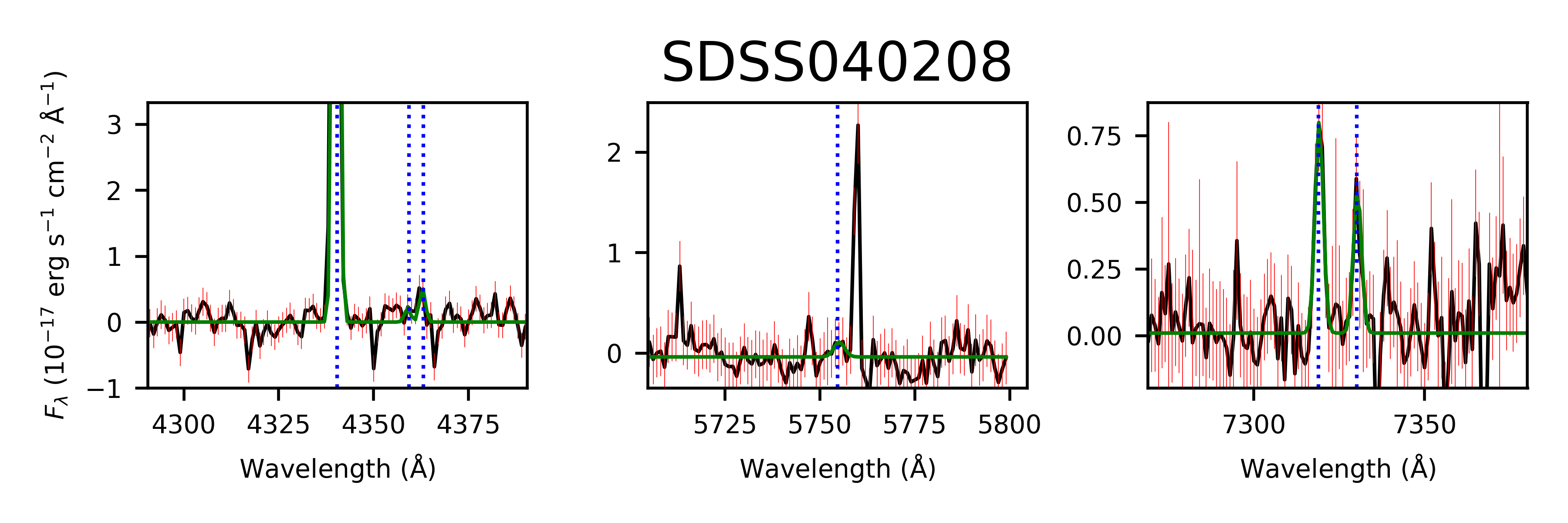}\\
\includegraphics[width=0.33\textwidth]{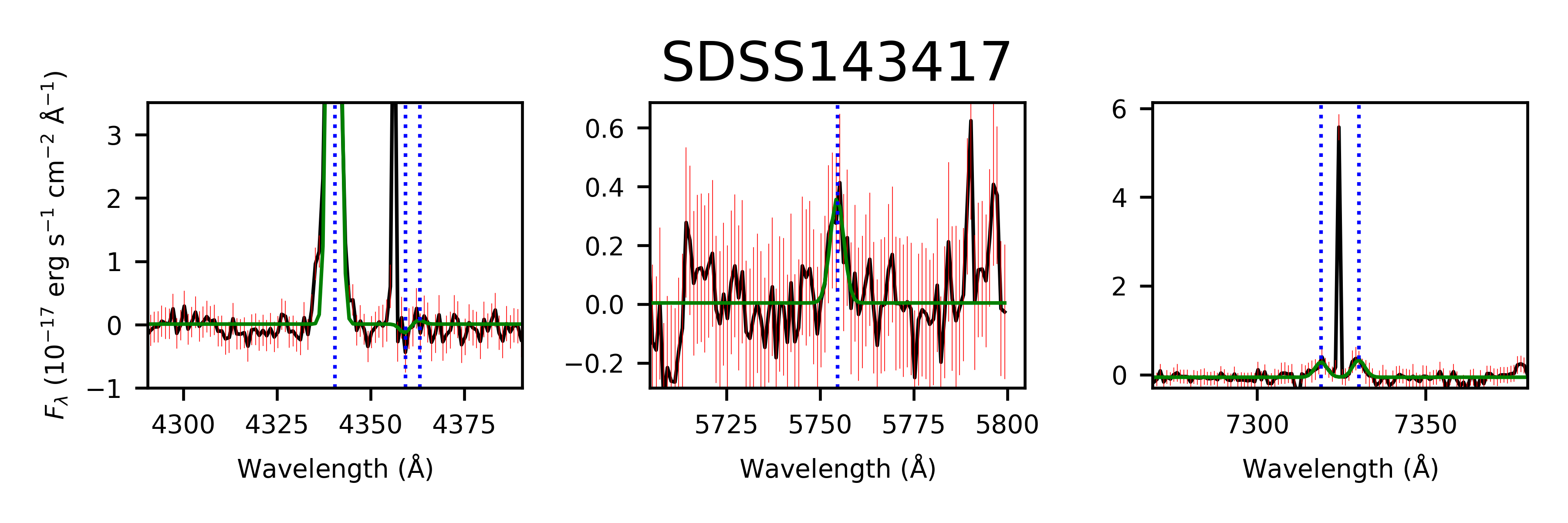}
\includegraphics[width=0.33\textwidth]{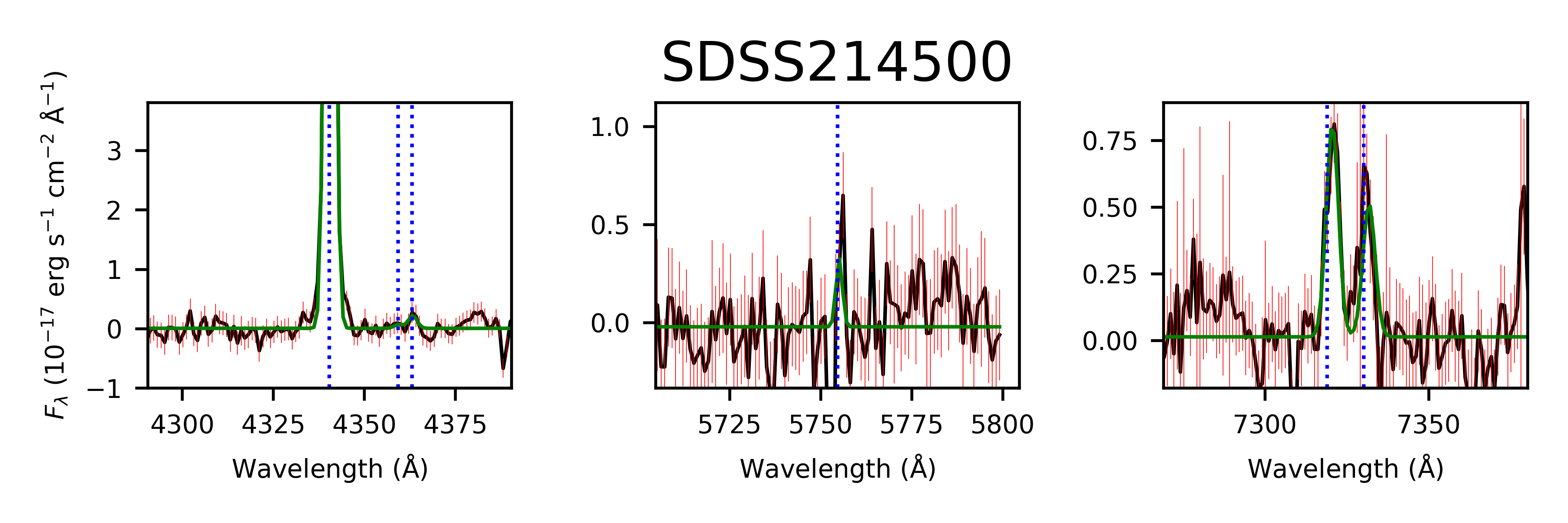}
\includegraphics[width=0.33\textwidth]{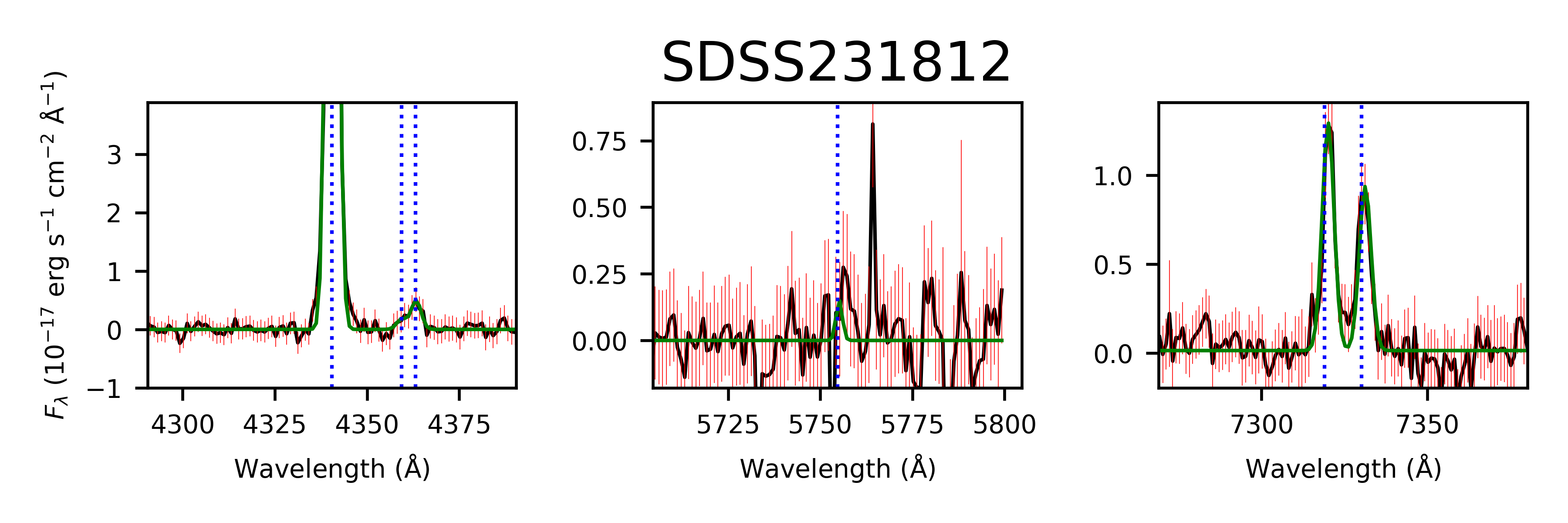}\\
\includegraphics[width=0.33\textwidth]{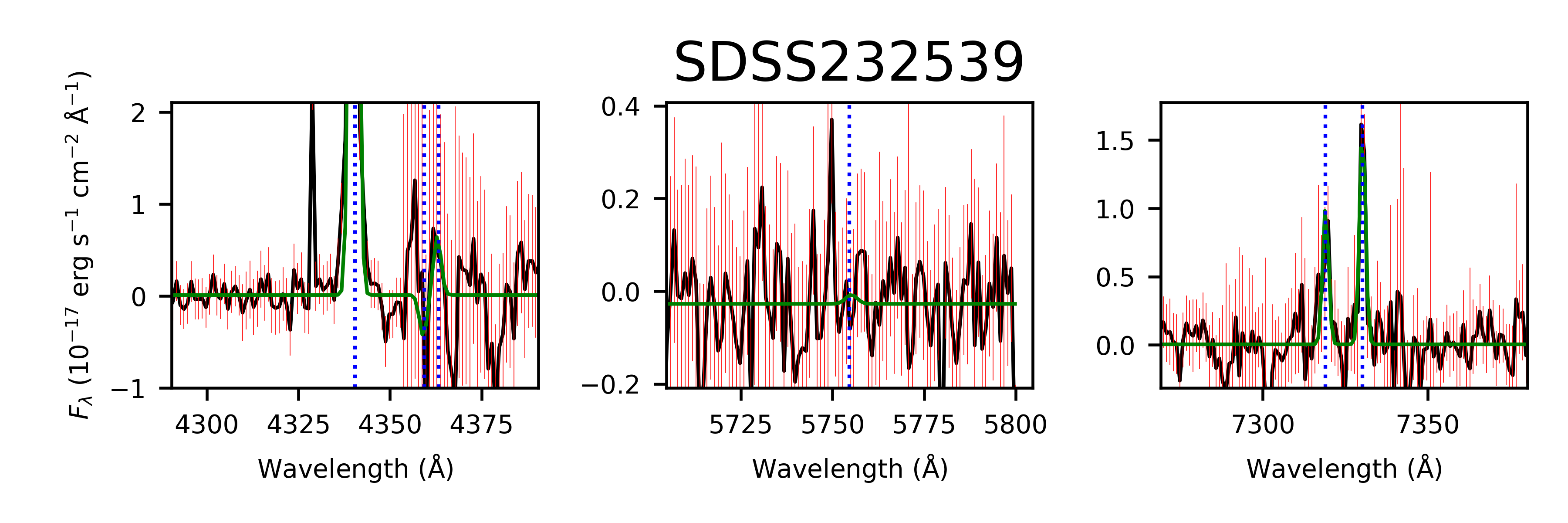}
\includegraphics[width=0.33\textwidth]{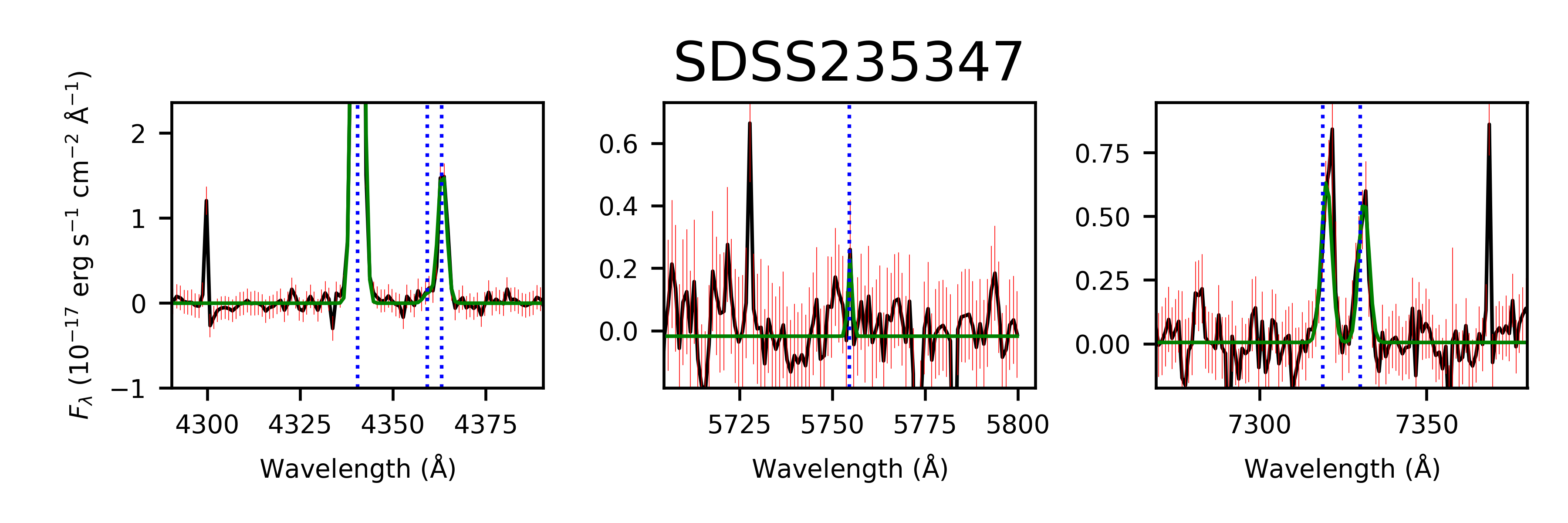}\\
\end{center}
\caption{\label{fig:4363}XShooter spectra around the weak \oiiiweak, \niiweak, and \oiiweak\ lines. The best-fit results of a multi-Gaussian function centered at the wavelengths of \hg, [Fe\,{\sc ii}]$\lambda$4352.78, [Fe\,{\sc ii}]$\lambda$4359.34 and [O\,{\sc iii}]$\lambda$4363.21 (vertical blue dashed lines) and including a flat continuum are indicated by the green solid line. The error spectrum is indicated in red. Absorption by the stellar continuum was subtracted prior to the fitting. Only lines detected at $\ge5\sigma$ were used in this paper.}
\end{figure*}

\subsection{Observations and data reduction}

Spectra were taken using the X-Shooter spectrograph installed at UT2 (Kueyen) of the ESO Very Large Telescope (VLT) in Chile during two separate observing campaings. The first sample of 17 objects was observed in service mode during 2010 (Program ID 085.B-0784(A)), and a second sample of 13 objects was observed in visitor mode during March 2016 (Program ID 096.B-0192(A)). The target list and log of observations is given in Table \ref{tab:sample}. We used X-Shooter in slit-mode (11\arcsec\ slit) to obtain simultaneous spectra from the $U$- to $K$-band using the UVB ($1\farcs0$ wide slit, $R\approx5100$), VIS ($0\farcs9$ wide slit, $R\approx8800$) and NIR ($0\farcs9$, $R\approx5100$) arms. The exposure times for each object were $4\times690$ s, $8\times320$ s and $12\times240$ s in nodding-on-slit mode, except for very extended objects that were observed in offset mode and received half the exposure time. The slits were centered on the brightest starburst regions but we note that the effective radii of most objects are comparable to the slit width. The seeing was estimated from the telluric standard star spectra taken each night at the same airmass as the science observations. 

We used the ESO X-Shooter pipeline \citep{modigliani10} and the EsoRex command-line recipes to reduce the data. The pipeline removes non-linear pixels, substracts the bias in the UVB and VIS arms and dark frames in the NIR arm. It then predicts the positions of arc-lines and order-edges on a format-check frame, determines the order positions and the two-dimensional wavelength  solution to resample the orders, divides the raw frames by a master flat field, measures the instrument resolving power, computes the instrument flexures, the instrument response and the total system efficiency. The spectra are then flux calibrated using standard stars that were observed during each night. The pipeline combines the science frames and subtracts the sky background. One-dimentional spectra were extracted using the X-Shooter pipeline at the location of the continuum along the slit. The size of the apertures was set to the larger value of the seeing (FWHM) measured from the telluric standard in the UVB arm (averaged over all wavelengths) and the spatial extent of \ha\ in the VIS arm. The spectra extracted from each of the arms were joined, and small flux offsets were applied to the UVB and NIR spectra in order to correct for small flux normalization mismatches with overlapping wavelength ranges at either end of the VIS spectra. Although telluric standards were observed and reduced in the same way, ultimately the telluric absorption corrections were done using the ESO \emph{Molecfit} tool \citep{kausch15,smette15}. Molecfit allows one to fit a model of the tropo- and strato-spheric telluric features directly to the science spectra and subtract this model from the data in order to remove the telluric features. Molecfit also updates the input error spectra that are produced by the ESO pipeline. 

\begin{figure}
\includegraphics[width=\columnwidth]{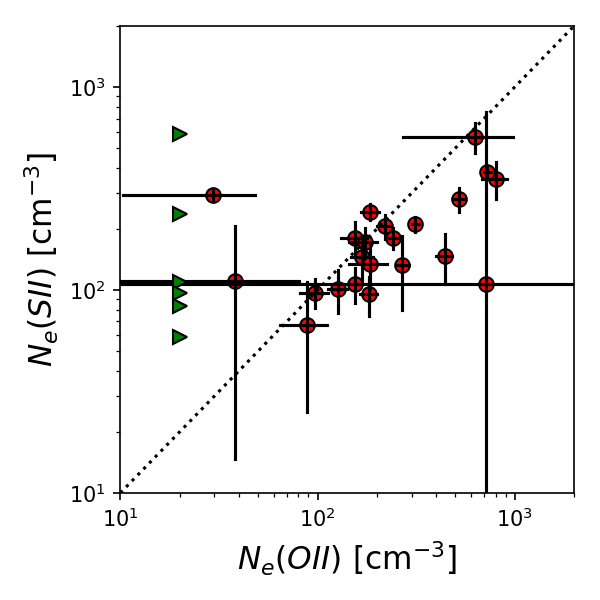}
\caption{\label{fig:Ne}Electron densities estimated from \oii\ and \sii\ using the model fit results obtained by \citet{sanders16a} for a 5-level atom aproximation and assuming an electron temperature of 10,000 K. Red circles are LBAs observed with XShooter, while the green pentagons are the literature LBAs for which only the \sii-doublet ratio is known. The black dotted line shows the one-to-one relation.}
\end{figure}

\subsection{Measurement methods}

The reduced, calibrated spectra are first corrected for foreground extinction assuming the \citet{cardelli89} extinction curve and reddening from \citet{schlegel98}. Emission line fluxes are determined by fitting single Gaussians in the case of isolated lines or series of Gaussians in the case of groups of lines. In order to correct the main lines for underlying stellar continuum absorption we use the STARLIGHT code version 04 \citep{cid-fernandes05,cid-fernandes11,mateus06,asari07}. We mask all emission lines, as well as the locations of possible WR bumps (4600--4700 \AA\ and 5700--5900 \AA). We apply an additional mask around the locations where the UVB and VIS arm spectra were joined due to increased noise, and possible residual offsets that exist in this wavelength range. The spectra are interpolated on a regular grid of 1 \AA\ bins. We run the code with 115 single stellar populations (SSPs) with a \citet{chabrier03} initial mass function for 5 different metallicities up to solar metallicity and ages up to 13 Gyr with the Padova 1994 tracks from \citet{bruzual03}. For objects with strong nebular continuum, we include a nebular continuum template. 
We use the \citet{calzetti01} extinction curve with a single reddening parameter that is applied equally to all SSPs with $A(V)$ in the range 0--4 mag. We fit the stellar kinematics allowing for a systematic shift of $\pm500$ km s$^{-1}$ and a stellar velocity dispersion of up to 500 km s$^{-1}$. In order to obtain the best fits to the stellar continuum in the vicinity of the weakest, temperature-sensitive lines (\oiiiweak, \niiweak, and \oiiweak), we follow \citet{andrews13} and limit the spectral range of the STARLIGHT fit to 400\AA\ around these lines (200\AA\ in either direction), which produced a good match to the observed continuum in all cases.     
For higher metallicity sources, in which \oiiiweak\ is weak, there can be substantial contamination of \oiiiweak\ from [Fe\,{\sc ii}]$\lambda$4352.78 and [Fe\,{\sc ii}]$\lambda$4359.34 \citep[e.g.][]{andrews13,curti17}. We took this into account by including the possible presence of the [FeII] lines into our fits and fixing the widths of these weak lines to that of \hg. The line fluxes were measured after subtracting the best STARLIGHT model, and corrected for dust extinction based on the Balmer decrement and assuming a \citet{cardelli89} extinction curve. For \oiiweak\ we were able to estimate the fraction of recombination excitation that contributes to the auroral line using Equation 2 from \citet{liu00} that uses the temperature, the \hb\ line flux and the $\mathrm{O}^{2+}/\mathrm{H}^+$ ratio. In all cases in which \oiiiweak\ was detected (thus allowing a direct estimate of the temperature and $\mathrm{O}^{2+}/\mathrm{H}^+$), the recombination excitation fraction of the measured \oiiweak\ line flux was at the 2--4\% level. We did not correct for this small contribution. The errors on the emission line fluxes were obtained by performing a monte carlo simulation of the line fitting process using the error spectrum. For the analysis involving the weak \oiiiweak, \niiweak, and \oiiweak\ lines we required a detection of at least 5$\sigma$. The continuum-subtracted spectra and the best-fit results in the spectral region around \oiiiweak, \niiweak\ and \oiiweak\ are shown in Fig. \ref{fig:4363}. The measurements used and derived in this paper are summarized in Tables \ref{tab:fluxes}, \ref{tab:basic_pars}, and \ref{tab:Z_pars}.

\begin{figure}
\includegraphics[width=\columnwidth]{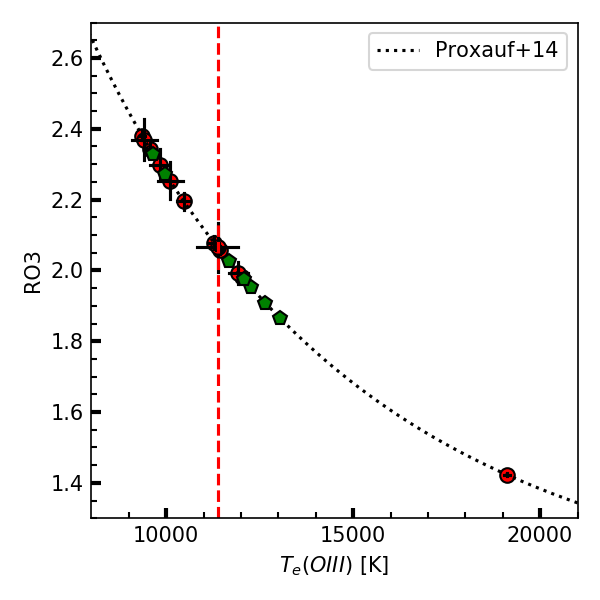}
\caption{\label{fig:R4363}Electron temperature $T_e$(O\,{\sc iii}) based on RO3 and the calibration of \citet{proxauf14}. Literature LBAs are marked with green pentagons. The median value of $T_e$(O\,{\sc iii})$\approx$11,000 K is indicated by the dashed line.}
\end{figure}

\begin{figure*}
\begin{center}
\includegraphics[width=0.33\textwidth,height=0.3\textwidth]{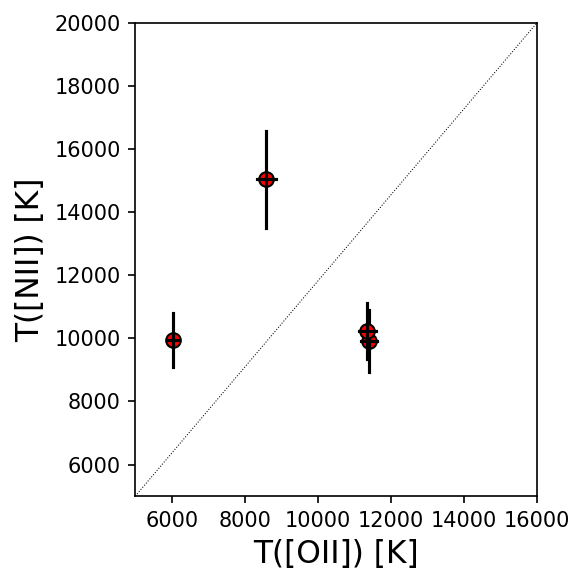}\includegraphics[width=0.33\textwidth,height=0.3\textwidth]{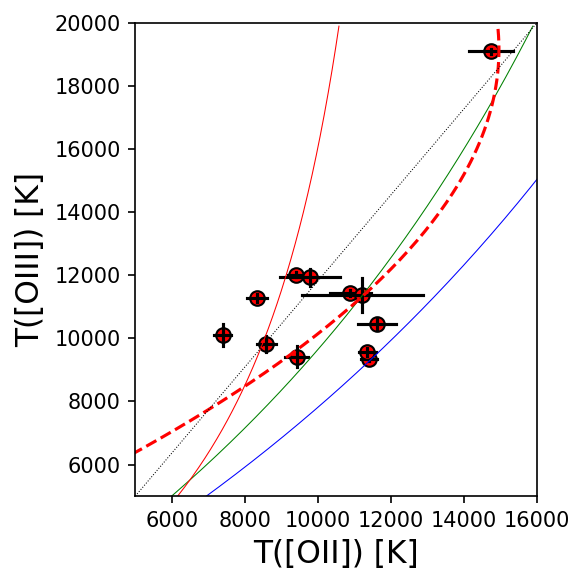}\includegraphics[width=0.33\textwidth,height=0.3\textwidth]{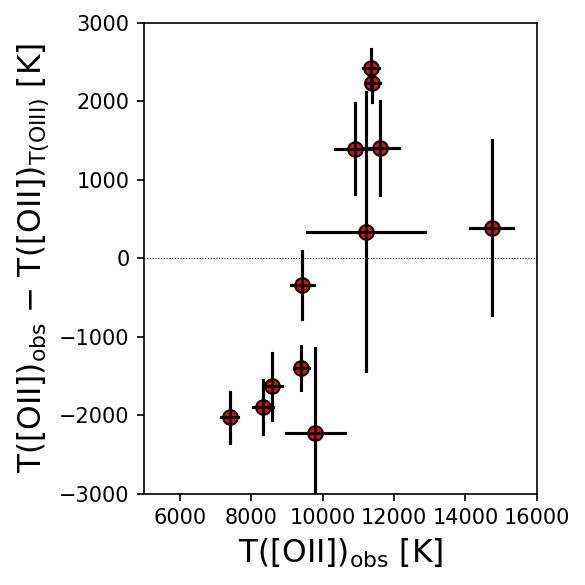}
\end{center}
\caption{\label{fig:temperatures}Relations between the temperatures measured for the different ionization states of O and N estimated using the temperature-sensitive RO3, RO2 and RN2 line ratios. Left panel: $T_e$(O\,{\sc ii}) versus $T_e$(N\,{\sc ii}). Middle panel: $T_e$(O\,{\sc ii}) versus $T_e$(O\,{\sc iii}). The red-dashed line is the relation from \citet{proxauf14}. Thin solid lines indicate the density-dependent calibrations from \citet{perez-montero17} for electron densities of 10 (blue line), 100 (green line) and 1000 cm$^{-3}$ (red line). Right panel: measured value of $T_e$(O\,{\sc ii}) versus the difference between that value and that inferred from $T_e$(O\,{\sc iii}) and the electron density following \citet{perez-montero17}. The black dotted lines indicate the one-to-one relation (left and middel panel) or the zero offset relation (right panel).}
\end{figure*}

\begin{figure}
\includegraphics[width=\columnwidth]{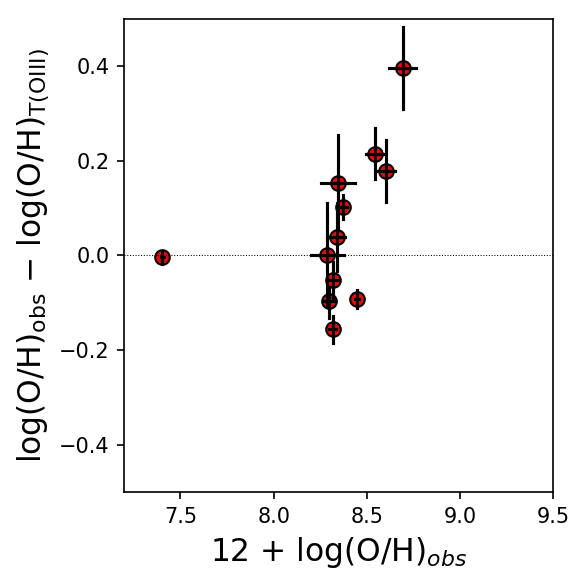}
\caption{\label{fig:ZOH_bothmethods}The difference in direct method oxygen abundances determined using our two temperatures $T_e$(O\,{\sc ii}) (based on RO2) and $T_e$(O\,{\sc iii}) (based on RO3), $12 + \mathrm{log(O/H)_{obs}}$, and that determined using $T_e$(O\,{\sc ii}) inferred from $T_e$(O\,{\sc iii}) and the electron density using Equation 14 from \citet{perez-montero17}, as function of the two-temperature value ($12 + \mathrm{log(O/H)_{obs}}$). The black dotted line indicates the zero offset relation.}
\end{figure}

\begin{figure}
\includegraphics[width=\columnwidth]{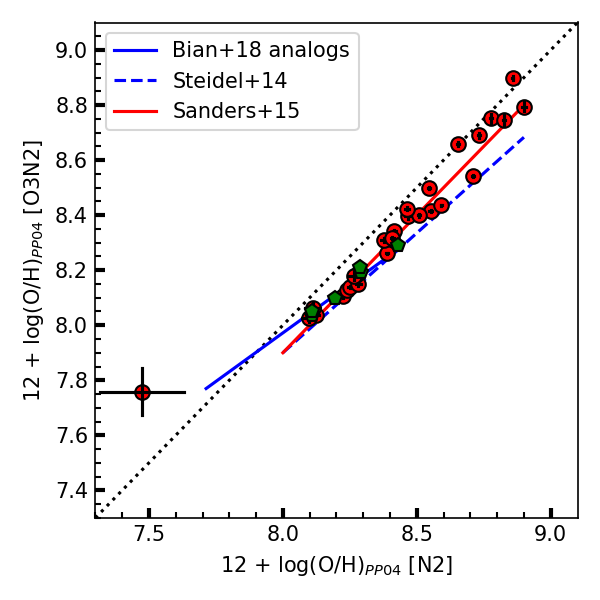}
\caption{\label{fig:ZN2_vs_ZO3N2}Strong line method oxygen abundances determined through the N2 and O3N2 methods of \citet{pettini04}. Several best-fit relations from the literature are indicated for the range in O/H for which they were determined \citep{steidel14,sanders15,bian18}. The black dotted line indicates the one-to-one relation. Literature LBAs are indicated by the green pentagons.}
\end{figure}

\section{Results}

\subsection{Electron density and temperature}
\label{sec:Ne}

The electron density ($n_e$) plays an important role in abundance measurements, since this parameter affects the fluxes of collisionally excited lines in star-forming regions. We determine $n_e$ based on the sulfur \sii\ and \oii\ line ratios following \citet{sanders16a}. The results are shown in Fig. \ref{fig:Ne} and show good agreement between the two values.  However, given the wider separation of the \sii\ doublet, in this paper we adopt the electron density based on \sii. For two sources (040208 and 232539), the \sii\ ratio produced an unphysical density and in the analysis below we used their \oii-based densities instead. For the LBAs with literature spectroscopic data, we were only able to calculate densities on the basis of \sii, showing that they are in the same range as the other objects (green triangles drawn as lower limits on $N_e$(O\,{\sc ii}) in Fig. \ref{fig:Ne}).

We measure electron temperatures, $T_e$(O\,{\sc iii}), through the $RO3\equiv(I(4959)+I(5007))/I(4363)$ ratio that probes conditions in the innermost high-excitation zone in the \hii\ region, which is relatively insensitive to the density. The relative proximity of the LBAs allow us to detect the faint \oiiiweak\ line and hence determine RO3 (see the spectra in Fig. \ref{fig:4363}). However, the electron temperature decreases with increasing metallicity, making the auroral lines too weak to be measured, especially for the more massive systems. In addition, these higher metallicity sources also often show an increased strength of [Fe\,{\sc ii}]$\lambda$4359.34, which needs to be deblended from \oiiiweak. We detected \oiiiweak\ in 12 of the cases (S/N of \oiiiweak\ larger than 5). The remaining sources in which \oiiiweak\ was not detected were excluded from the parts of the analysis that require RO3. To estimate $T_e$(O\,{\sc iii}) we use the fitting function from \citet{proxauf14} that relates the temperature to RO3. The result is shown in Fig. \ref{fig:R4363}, and indicates that the temperatures are in the range 9,000 to 13,000 K, with the exception of S01\_2 which has 19,000 K. In order to derive total oxygen and nitrogen abundances, we also require the temperatures of the lower ionization states of O and N, $T_e$(O\,{\sc ii}) and $T_e$(N\,{\sc ii}). These can be estimated from the density-dependent $RO2\equiv(I(3726)+I(3729))/(I(7319)+I(7330))$\ and $RN2\equiv(I(6548)+I(6584))/(I(5755)$\ line ratios, respectively \citep[see][]{perez-montero17}. Because these ratios depend on the usually weak auroral lines [O\,{\sc ii}]$\lambda\lambda$7319,7330 and [N\,{\sc ii}]$\lambda$5755, many studies rely on standard conversions between $T_e$(O\,{\sc iii}) and $T_e$(O\,{\sc ii}) \citep[e.g.][]{garnett92,proxauf14,perez-montero17}. Requiring at least a 5$\sigma$ detection in each of the temperature sensitive lines, we were able to measure $T_e$(O\,{\sc iii}) in 12, $T_e$(O\,{\sc ii}) in 22, and $T_e$(N\,{\sc ii}) in 4 sources. The results are compared in Fig. \ref{fig:temperatures}. The discrepancies between $T_e$(N\,{\sc ii}) and $T_e$(O\,{\sc ii}) are up to a factor of $\sim$2, i.e. much larger than the formal measurement errors, although there are only 4 sources with both determinations (left panel). The scatter between $T_e$(O\,{\sc iii}) and $T_e$(O\,{\sc ii}) is a few thousand K, and the points lie around the $T_e$(O\,{\sc iii})--$T_e$(O\,{\sc ii}) relations typically assumed for an \hii\ region with an electron density of $\sim$100 cm$^{-3}$ (middle panel). Last, we compare the observed $T_e$(O\,{\sc ii}) with that estimated from $T_e$(O\,{\sc iii}) and the electron density using Equation 14 from \citet{perez-montero17}. The right panel of Fig. \ref{fig:temperatures} shows the difference between the two values as funcion of the $T_e$(O\,{\sc ii}) measured from RO2, with a full range of $\approx\pm$2000 K. Because our abundances results below are quite sensitive to the correct value for $T_e$(O\,{\sc ii}), in this paper we will evaluate our results both using the RO2-based values and the values inferred from $T_e$(O\,{\sc iii}). For the literature LBAs, we always estimate $T_e$(O\,{\sc ii}) from $T_e$(O\,{\sc iii}) in the same way as done for our own spectra.  

\begin{figure*}
\begin{center}
\includegraphics[width=\columnwidth]{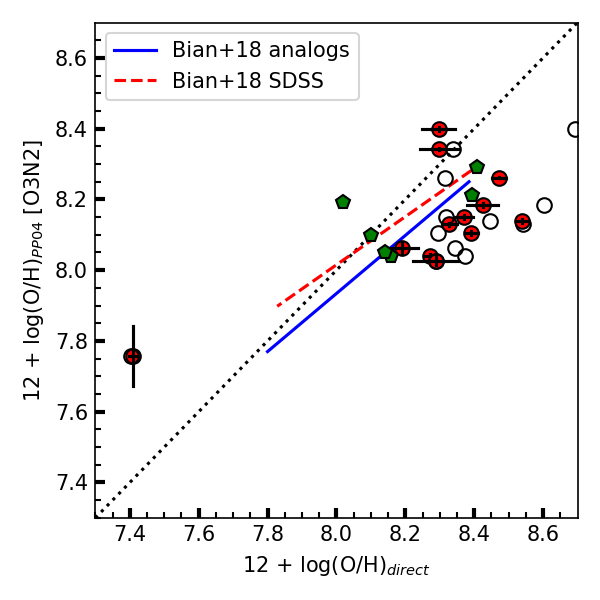}
\includegraphics[width=\columnwidth]{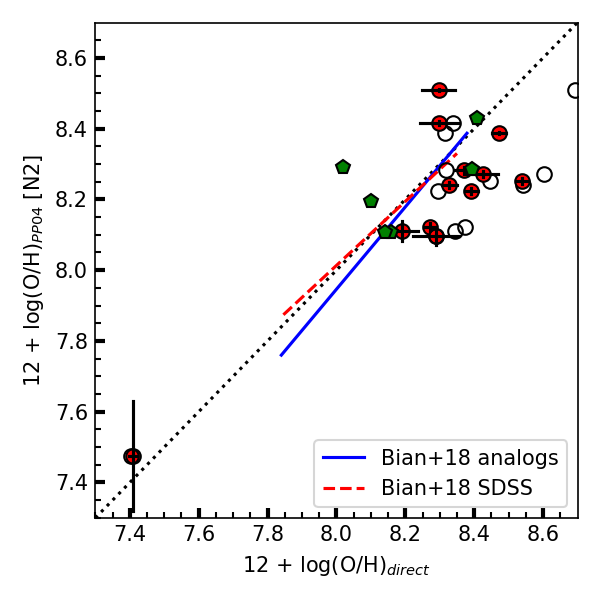}
\end{center}
\caption{\label{fig:Zdirect_vs_ZO3N2}Left: Direct method oxygen abundance determined in Sect. \ref{sec:direct} versus the O3N2 oxygen abundance from \citet{pettini04}. The relations found by \citet{bian18} for typical SDSS star-forming galaxies (red dashed line) and their sample of BPT offset analogs (blue solid line) are indicated. Right: Direct method oxygen abundance determined in Sect. \ref{sec:direct} versus the N2 oxygen abundance from \citet{pettini04}. The relations found by \citet{bian18} for typical SDSS star-forming galaxies (red dashed line) and their sample of BPT offset analogs (blue solid line) are indicated. In both figures, the direct-method abundances derived using the $T_e$(O\,{\sc ii}) derived from RO2 are shown as open circles, while those derived using $T_e$(O\,{\sc ii}) as inferred from $T_e$(O\,{\sc iii}) are shown as filled circles. Literature LBAs are indicated by the green pentagons.}
\end{figure*}

\subsection{Direct and strong line oxygen abundance}
\label{sec:direct}

Having determined electron densities and temperatures, we now use the formalism presented in \citet{izotov06} to estimate the `direct method' oxygen abundance, $Z_\mathrm{direct}=12 + \mathrm{log(O/H)}$ under the assumption that the total oxygen abundance can be approximated by $\mathrm{O/H}=(\mathrm{O}^++\mathrm{O}^{2+})/\mathrm{H}^+$ (see Eqns. 3 and 5 from \citet{izotov06}). This calculation depends on the electron temperatures $T_e$(O\,{\sc iii}) and $T_e$(O\,{\sc ii}), as well as the electron density for which we will use that determined from the \sii\ doublet in Sect. \ref{sec:Ne}. For the determination of $T_e$(O\,{\sc ii}) we have two choices. In the right-hand panel of Fig. \ref{fig:temperatures} we showed that the $T_e$(O\,{\sc ii}) estimated from RO2 and that inferred from $T_e$(O\,{\sc iii}) can show significant differences. 

These different estimates for the same temperature can lead to significant differences in the derived oxygen abundances. In order to evaluate this effect, we show in Fig. \ref{fig:ZOH_bothmethods} the difference one would obtain when choosing one or the other method. The figure plots O/H calculated using the RO2-based $T_e$(O\,{\sc ii}) and the RO3-based $T_e$(O\,{\sc iii}) versus the difference between O/H estimated using this method and that using $T_e$(O\,{\sc ii}) as inferred from $T_e$(O\,{\sc iii}) (using the electron density and the TO3-TO2 relation from \citet{perez-montero17}. The discrepancy in O/H can be up to $\sim$0.4 dex.  
In the analysis below, we will adopt the $T_e$(O\,{\sc ii}) values as inferred from $T_e$(O\,{\sc iii}). However, in all relevant plots, we also include a second set of symbols (open circles) that give the results for the case where $T_e$(O\,{\sc ii}) was estimated directly from RO2. While results on individual sources may vary, we show that our main conclusions are not affected by either choice. Further discussion of the possible sources of these discrepancies can be found in, e.g., \citet{yates19}. 

We find values of 12 + log(O/H) in the range $\sim$8.0--8.6 (with the exception of the extremely metal-poor object S01\_2 which has 12 + log(O/H)$\sim$7.4). These  direct method abundances can be compared to the strong line methods typically employed for objects at high redshift or objects with relatively shallow spectra at low redshift. We first compare two common strong line methods based on the $N2\equiv I(6584)/I(H\alpha)$ and $O3N2\equiv(I(5007)/I(H\beta))/(I(6584)/I(H\alpha))$ ratios. Using the \citet{pettini04} calibrations, which relate these line ratios to abundances determined using the direct method applied to a local calibration data set, the oxygen abundances for N2 and O3N2 are in general agreement (Fig. \ref{fig:ZN2_vs_ZO3N2}). We find slightly higher abundances based on N2 than on O3N2. A similar result was found for $z\sim2$ galaxies from the MOSDEF survey by \citet{sanders15} (red line in Fig. \ref{fig:ZN2_vs_ZO3N2}), which appears to be a good fit to the LBAs as well. This offset between N2 and O3N2 for LBAs and \citet{sanders15} is about half the size of that found for $z\sim2$ KBSS galaxies by \citet{steidel14}, especially at high metallicities. We also indicate the relation found for the local BPT offset analogs from \citet{bian18} (blue solid line). This relation was obtained by taking the best-fits between the direct method abundance and N2 and O3N2 given by \citet{bian18}, and converting these back to abundances in the \citet{pettini04} calibration. 

The comparison between the two strong line method abundance estimates shown in Fig. \ref{fig:ZN2_vs_ZO3N2} does not necessarily tell us anything about the accuracy of these strong line methods. Having determined the direct method abundances for LBAs, we can now compare these to either strong line method results. The results are shown in Fig. \ref{fig:Zdirect_vs_ZO3N2}, which compare $Z(T_e)$ with $Z(O3N2)$ (left panel) and $Z(N2)$ (right panel). The direct oxygen abundances do not appear to correlate very strongly with those based on O3N2 or N2, although the dynamic range in O/H of our sample is quite small making it difficult to look for such trends. Overall, however, the direct method abundances appear in general agreement with the strong line methods within an intrinsic scatter of 0.1--0.2 dex and a small systematic offset of at most 0.1 dex. These findings are very similar to those found at $z\sim2$ by \citet{patricio18}. The LBAs lie close to the relations for typical SDSS galaxies and BPT offset analogs from \citet{bian18}, which have a very similar range of abundances (7.8$\lesssim$12+log(O/H)$_{\mathrm{direct}}$$\lesssim$8.4) as our sources (blue and red lines in Fig. \ref{fig:Zdirect_vs_ZO3N2}). 

\begin{figure}
\includegraphics[width=\columnwidth]{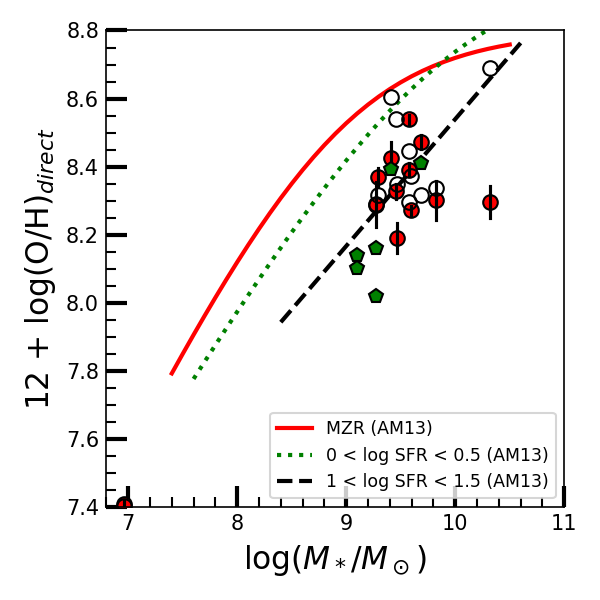}
\caption{\label{fig:mzr}The MZR relation based on stellar masses from SDSS and direct-method oxygen abundances determined in this paper. The red line shows the best-fit MZR for SDSS from \citet{andrews13}, while the green dotted and blue dashed lines show their best-fit MZRs for sub-samples of star-forming objects having $0<\mathrm{log(SFR)}<0.5$ and $1<\mathrm{log(SFR)}<1.5$, respectively. The direct-method abundances derived using the $T_e$(O\,{\sc ii}) derived from RO2 are shown as open circles, while those derived using $T_e$(O\,{\sc ii}) as inferred from $T_e$(O\,{\sc iii}) are shown as filled circles. Literature LBAs are indicated by the green pentagons.}
\end{figure}

In \citet{hoopes07} and \citet{overzier10} it was shown, based on strong line abundances, that LBAs lie below the local mass-metallicity relation of SDSS star-forming galaxies. We confirm this result using the direct method abundances in Fig. \ref{fig:mzr}.  At a given stellar mass, the LBAs for which direct method abundances could be determined are offset toward lower oxygen abundances compared to the local mass-metallicity relation (MZR). We compare the LBAs with the direct method abundances for stacked SDSS galaxies by \citet{andrews13}. The offset of the LBAs toward lower oxygen abundance is similar to the offsets observed for the SDSS stacks for high specific SFR objects (e.g., dashed black line in Fig. \ref{fig:mzr}). 

\subsection{Ionization parameter}

\begin{figure}
\includegraphics[width=\columnwidth]{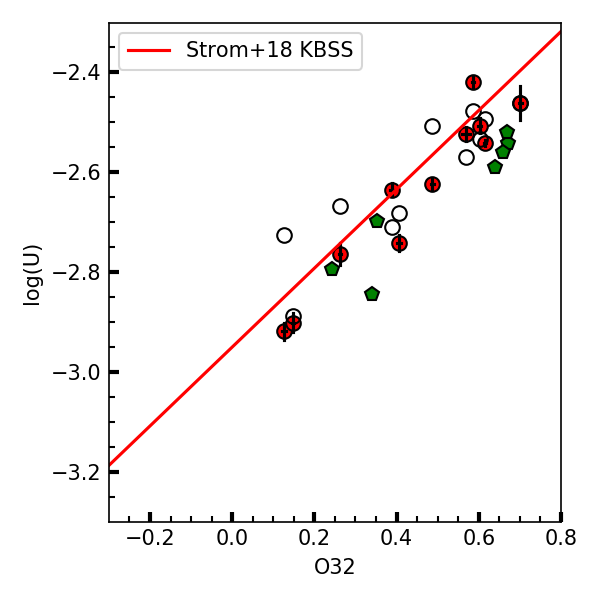}
\caption{\label{fig:O32_vs_logU}Dimensionless ionization parameter $U$ versus the $O_{32}$. The relation found for KBSS star-forming galaxies at $z\sim2$ from \citet{strom18} is indicated by the red solid line. Objects for which the direct-method abundances were derived using the $T_e$(O\,{\sc ii}) based on RO2 are shown as open circles, while those derived using $T_e$(O\,{\sc ii}) as inferred from $T_e$(O\,{\sc iii}) are shown as filled circles. Literature LBAs are indicated by the green pentagons.}
\end{figure}

The ionization state of the gas can be quantified by the ionization parameter $q\equiv Q(H^0)/n_H$, the ratio of the hydrogen ionizing photon rate at the surface of a plane-parallel slab and the density of hydrogen, also expressed as the dimensionless ionization parameter $U\equiv n_\gamma/n_H= q/c$, the ratio of the ionizing photon and hydrogen densities. We follow \citet{kojima17} in determining the ionization parameter using the ionization-sensitive $O_{32}\equiv(I(4959)+I(5007))/I(3727)$ ratio. Although $O_{32}$ depends both on $q$ and the oxygen abundance \citep{kewley02}, we can use our direct method estimate of the oxygen abundance to break this degeneracy. We calculate $q$ (and hence log(U)) using Eq. 13 from \citet{kobulnicky04} based on the \citet{kewley02} photoionization model grids. In Fig. \ref{fig:O32_vs_logU} we show the relation between $O_{32}$ and log(U), finding that LBAs lie a small distance (0--0.1 dex) below the relation determined by \citet{strom18} based on photoionization modeling of KBSS galaxies at $z\sim2$ (red line). \citet{strom18} showed that the log(U) or $q$ in KBSS galaxies correlate well with $O_{32}$, Ne3O2 and O3, but most strongly with $O_{32}$. 

\begin{figure}
\includegraphics[width=\columnwidth]{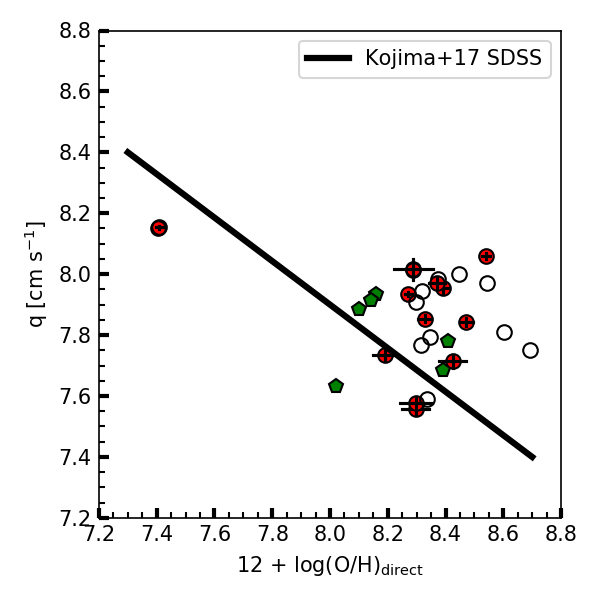}
\caption{\label{fig:Z_vs_q}Ionization parameter $q$ versus the direct method oxygen abundance. The solid black line indicates the \citet{kojima17} fit to local SDSS stacks from \citet{andrews13}. At a given 12 + log(O/H), LBAs tend to lie above the local relation. This is similar as that found for green pea galaxies and samples of $z\sim2$ star-forming galaxies studied by \citet{kojima17}. Objects for which the direct-method abundances were derived using the $T_e$(O\,{\sc ii}) based on RO2 are shown as open circles, while those derived using $T_e$(O\,{\sc ii}) as inferred from $T_e$(O\,{\sc iii}) are shown as filled circles. Literature LBAs are indicated by the green pentagons.}
\end{figure}

In typical local star-forming galaxies, the ionization parameter is strongly anti-correlated with oxygen abundance. \citet{kojima17} provide a best-fit relation based on the direct-method oxygen abundances measured in the SDSS stacks from \citet{andrews13}. This relation is indicated in Fig. \ref{fig:Z_vs_q} along with the LBA data. Although the dynamic range of our sample of LBAs is too small to see whether they follow the same general trend, the majority of the LBAs lie above the local relation at a given oxygen abundance. This result is very similar to that obtained by \citet{kojima17} for a sample of star-forming galaxies at $z=1.4-3.6$. They also noticed that this behavior is similar to that found for a sample of ``green pea'' galaxies \citep{amorin12,jaskot13}. As pointed out above, these include the 3 objects from \citet{amorin12} that are also in our LBA sample. The implications of this average offset in the ionization parameter $q$ at a fixed O/H will be discussed in \S3.6\ and \S3.7 below.  

\begin{figure*}
\begin{center}
\includegraphics[width=\textwidth]{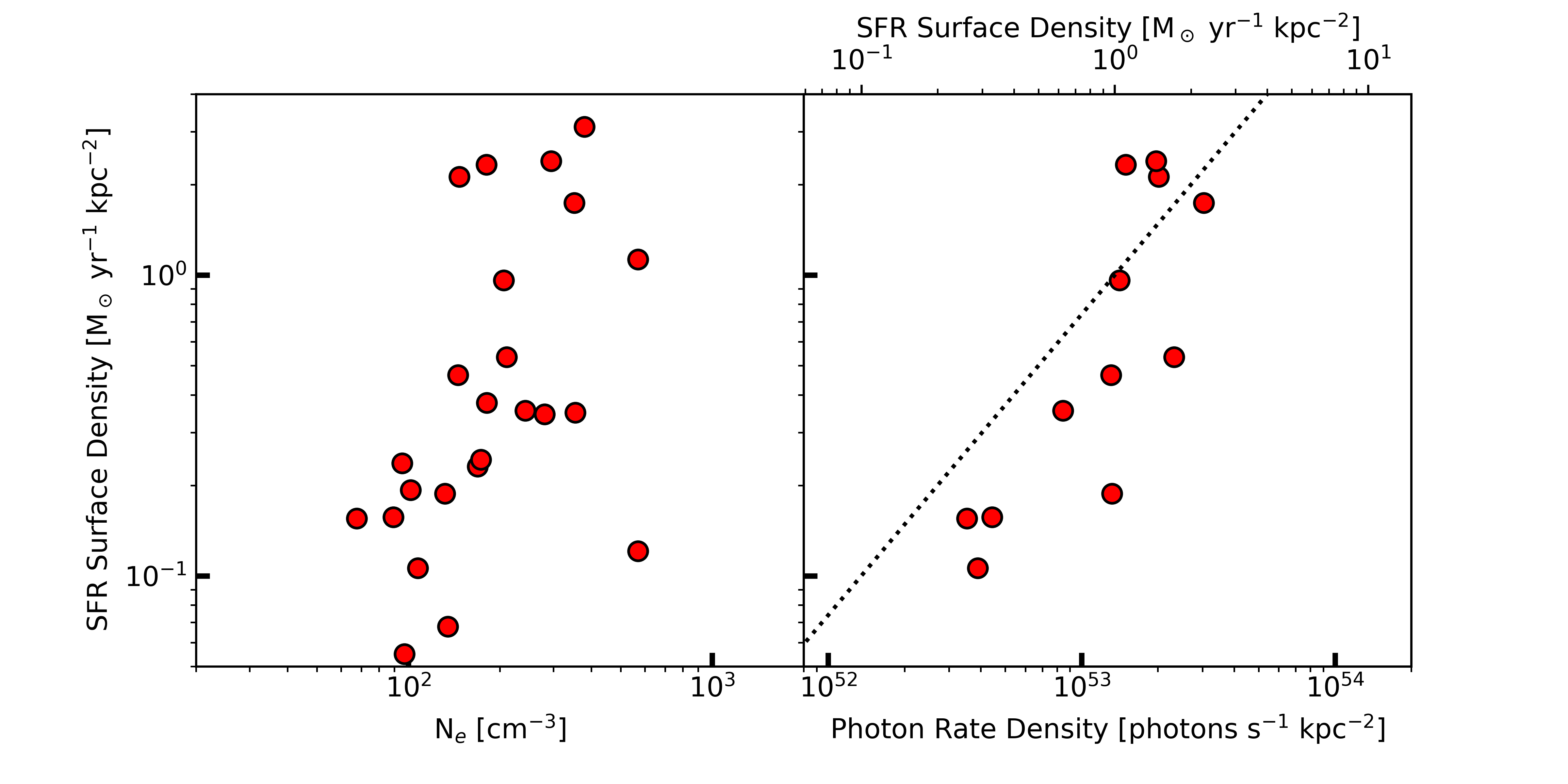}
\end{center}
\caption{\label{fig:SFRD}The SFR per unit area versus the electron density (left-hand panel) and the ionizing photon rate per unit area (right-hand panel). The SFR surface density on the vertical axis was measured from dust-corrected \ha-luminosity and the half-light radius measured from SDSS in the $r^\prime$-band. In the right-hand panel, the top axis gives another measure for the SFR surface density estimated directly from the ionizing photon rate density. The dotted line is the one-to-one relation between the two measures for the SFR surface density. See text of Sect. \ref{sec:SFRD} for details.}
\end{figure*}

\subsection{ISM pressure and feedback from massive stars}
\label{sec:SFRD}

It is expected that the pressure in the interstellar medium (ISM) of starburst galaxies is set by the feedback from massive stars \citep{chevalier85,heckman90}. The electron density is then the result of the compression of the shocked medium from the rate of momentum-injection from stellar winds and supernovae. The latter can be approximated by the SFR surface density. In the left-hand panel of Fig. \ref{fig:SFRD} we plot the electron density estimated from the \sii\ doublet against the SFR surface density. The SFR was estimated from \ha\ corrected for dust using the Balmer decrement, while the surface area was computed using the seeing-deconvolved half-light radius in the $r^\prime$-band from SDSS. The two quantities are indeed correlated, confirming our hypothesis that the star formation feedback sets the pressure in the ISM in these compact starburst galaxies. With knowledge of both the electron density and the ionization parameter we can furthermore estimate the total hydrogen-ionizing photon rate surface density, $N_ecU$. This is plotted in the right-hand panel of Fig. \ref{fig:SFRD}, again versus the SFR surface density. This ionizing photon rate surface density can be converted directly into a SFR surface density as well, assuming the standard relation between SFR and the ionizing photon rate of $\mathrm{SFR(Q(H}^0\mathrm{))}=7.4\times10^{-54}\mathrm{Q(H}^0\mathrm{)}$ \citep[e.g.][]{calzetti13}. These estimates for the SFR surface densities can be read from the top axis of Fig. \ref{fig:SFRD}, and they are of very similar order of magnitude as the more direct measure of SFR surface density shown on the vertical axis. The dotted line shows the one-to-one relation, indicating that most objects lie within a factor of a few or better from the expected values. This analysis confirms that the conditions in the ISM are directly related to the star formation activity in these compact starburst galaxies.  

\subsection{Nitrogen-to-oxygen abundance ratio}
\label{sec:NO}

Nitrogen and oxygen have similar ionization potentials, making $\mathrm{N}^+/\mathrm{O}^+$ a good proxy of the nitrogen-to-oxygen abundance ratio, N/O. We use the electron temperatures derived in Sect. \ref{sec:Ne} above together with the relations for N$^+$/H$^+$ and O$^+$/H$^+$ from \citet{izotov06} (Eqns. 3 and 6) to estimate N/O. Besides this direct method for N/O, there also exist convenient strong line methods for measuring N/O, such as N2O2$\equiv$\niisingle/\oiisingle, which can be easily measured in optical (low redshift) or near-infrared (high redshift) spectra of moderate signal-to-noise. Besides N2O2, N/O also correlates well with other strong line ratios involving nitrogen, such as N2S2$\equiv$\niisingle/\sii\ and N2$\equiv$\niisingle/\ha\ \citep[e.g.][]{perez-montero09,kojima17,strom17,strom18}. 

\begin{figure}
\begin{center}
\includegraphics[height=\columnwidth]{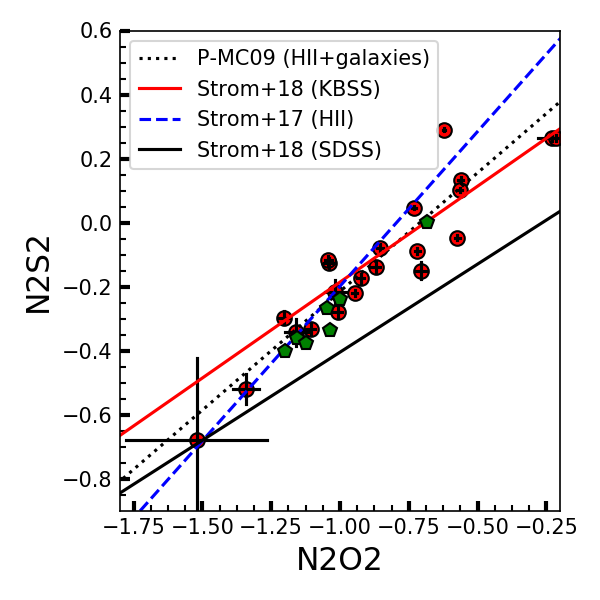}
\end{center}
\caption{\label{fig:N2O2_N2S2}The relation between strong line N/O abundance indicators N2O2 and N2S2. Fits from the literature are indicated. Black solid line: the median relation found for SDSS galaxies by \citet{strom18}. Black dotted line: the fit obtained for a sample of galactic and extragalactic \hii\ regions and galaxies from \citet{perez-montero09}. Blue dashed line: the fit to the sample of extra-galactic \hii\ regions by \citet{strom17}. Red solid line: the fit to KBSS galaxies at $z\sim2$ from \citet{strom18}. Literature LBAs are indicated by the green pentagons.}
\end{figure}

In Fig. \ref{fig:N2O2_N2S2} we first compare N2O2 with N2S2 with various relations from the literature. \citet{strom18} showed that KBSS galaxies at $z\sim2$ lie along a relation that is more similar to that found for local \hii\ regions \citep{perez-montero09,strom17} than typical star-forming galaxies from SDSS. LBAs occupy the region in between the relations for $z\sim2$ and \hii\ regions on one hand, and local SDSS galaxies on the other, with most LBAs being closer to the high redshift/local \hii\ lines. At a given N2O2, LBAs have higher N2S2 compared to typical SDSS galaxies. This behavior is very similar to that observed for local \hii\ regions and $z\sim2$ KBSS galaxies, and can be understood by the fact that LBAs and KBSS galaxies have star formation histories that are much simpler and more similar to \hii\ regions compared to typical SDSS star-forming galaxies \citep{sanders16a,kashino17,strom18}. Part of the [S\,{\sc ii}] emission may arise from diffuse emission regions in between the \hii\ regions, lowering N2S2 in the case of more complex galaxies \citep{sanders17}. 

\begin{figure*}
\begin{center}
\includegraphics[width=\textwidth]{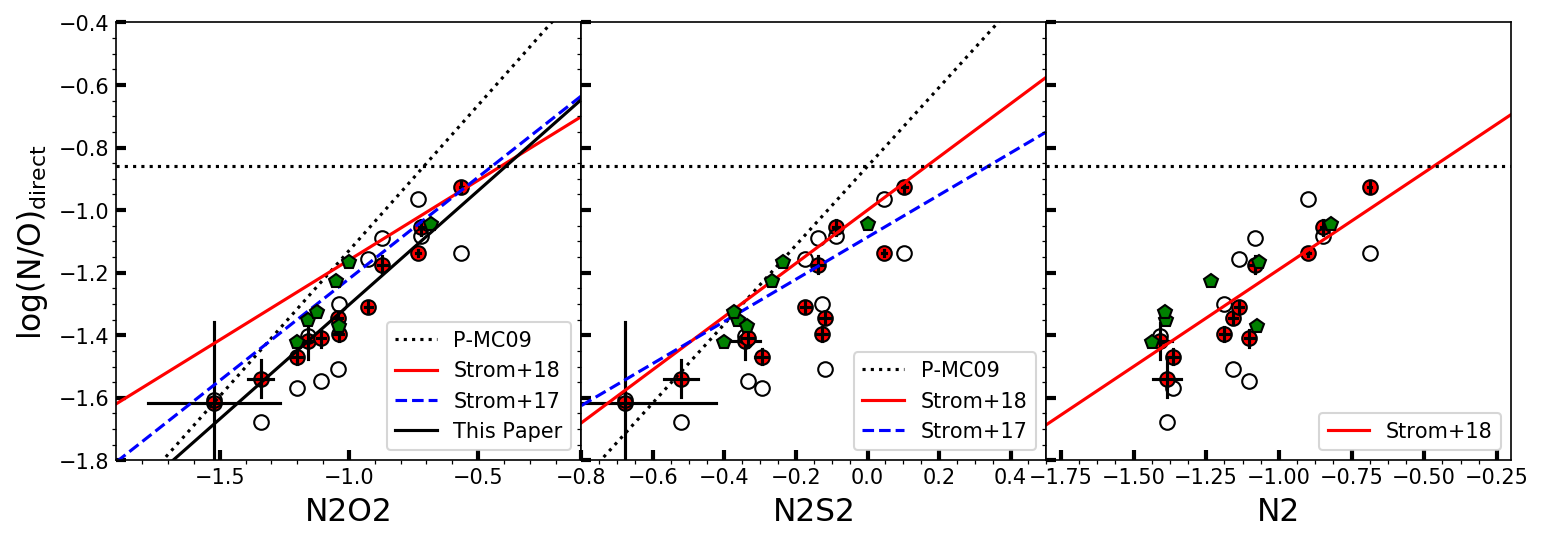}
\end{center}
\caption{\label{fig:NO_stronglines}Direct method N/O abundance versus strong line indicators N2O2 (left panel), N2S2 (middle panel) and N2 (right panel). Various fits from the literature are indicated. Black dotted line: the fit obtained for a sample of galactic and extragalactic \hii\ regions and galaxies from \citet{perez-montero09}. Blue dashed line: the fit to the sample of extra-galactic \hii\ regions by \citet{strom17}. Red solid line: the fit to KBSS galaxies at $z\sim2$ from \citet{strom18}. The solar N/O value is indicated by the black dotted lines. The solid black line in the left panel is the best-fit relation between N2O2 and log(N/O) determined for the LBAs presented in this paper ($\mathrm{log(N/O)_{direct}}=0.73\times\mathrm{N2O2} - 0.58$). Objects for which the direct-method abundances were derived using the $T_e$(O\,{\sc ii}) based on RO2 are shown as open circles, while those derived using $T_e$(O\,{\sc ii}) as inferred from $T_e$(O\,{\sc iii}) are shown as filled circles. Literature LBAs are indicated by the green pentagons.}
\end{figure*}

In Fig. \ref{fig:NO_stronglines} we plot the strong line indicators for N/O (N2O2, N2S2, and N2) versus the direct N/O ratio. The majority of LBAs are concentrated in the region between $\mathrm{(N/O)_{direct}}\gtrsim-1.5$ and the solar value (horizontal dotted line). The smallest scatter is found for N2O2, which is not surprising because it is the most direct measure of N$^+$/O$^+$ among these three indicators. The LBAs lie along the relations found for KBSS galaxies and local \hii\ regions. The N2O2--N/O relation for \hii\ regions and galaxies from \citet{perez-montero09} is known to be too steep, overestimating N/O for high N2O2 for both local \hii\ regions and high redshift galaxies \citep{strom18}. The relation between N/O and the strong line indicators found for the LBAs is more similar to the calibrations found for \hii\ regions and high redshift galaxies \citep{strom17,strom18}. Using only those objects in our sample for which the direct method N/O ratio is available, we fit the relation between N/O and N2O2, finding $\mathrm{log(N/O)_{direct}}=0.73 \times\mathrm{N2O2} - 0.58$. 

\begin{figure*}
\begin{center}
\includegraphics[width=\textwidth]{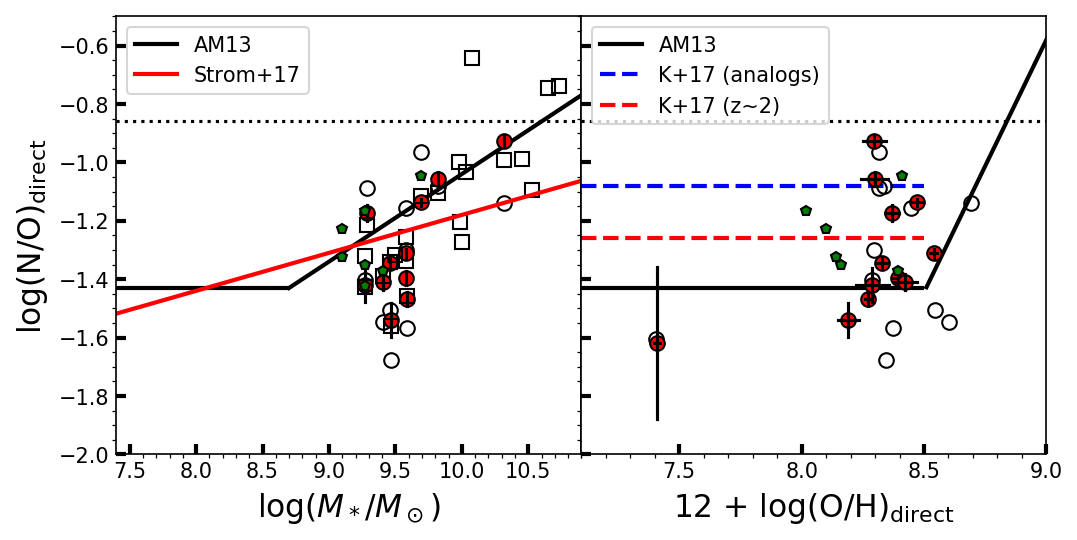}
\end{center}
\caption{\label{fig:NO_mass_OH}The direct method nitrogen-to-oxygen ratio log(N/O)$_{\mathrm{direct}}$ versus the stellar mass (left panel) and the direct method oxygen abundance. Solid black lines indicate the local relations based on the stacked SDSS galaxy spectra from \citet{andrews13}. In the panel on the left, the solid red line indicates the best-fit relation found for KBSS galaxies at $z\sim2$ from \citet{strom17}. In both panels, objects for which the direct-method abundances were derived using the $T_e$(O\,{\sc ii}) based on RO2 are shown as open circles, while those derived using $T_e$(O\,{\sc ii}) as inferred from $T_e$(O\,{\sc iii}) are shown as filled circles. In the left panel, open squares are LBAs with $\mathrm{log(N/O)_{direct}}$ determined from N2O2 and the best-fit correlation shown in Fig. \ref{fig:NO_stronglines}. In the right panel, the red dashed line marks the average value for $z\sim2$ galaxies ($\mathrm{log(N/O)_{direct}}=-1.26$, including upper limits) from \citet{kojima17}. The blue dashed line indicates the average value of $\mathrm{log(N/O)_{direct}}=-1.08$ determined for stacks of local galaxies that have comparable ($M_*$,SFR) as the $z\sim2$ galaxies, also from \citet{kojima17}. The solar N/O value is indicated by the black dotted lines. Literature LBAs are indicated by the green pentagons.}
\end{figure*}

\begin{figure*}
\begin{center}
\includegraphics[width=0.46\textwidth]{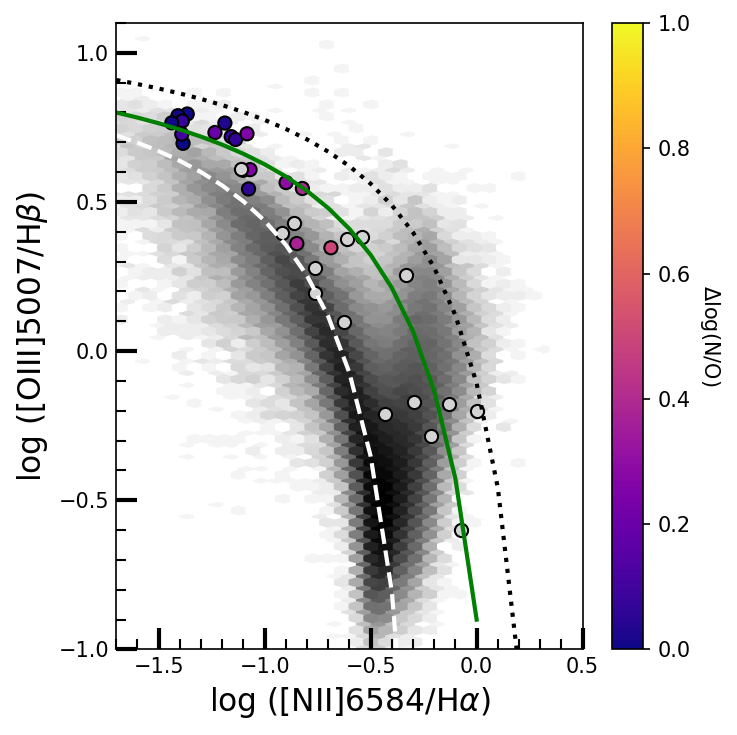}\hspace{1cm}
\includegraphics[width=0.46\textwidth]{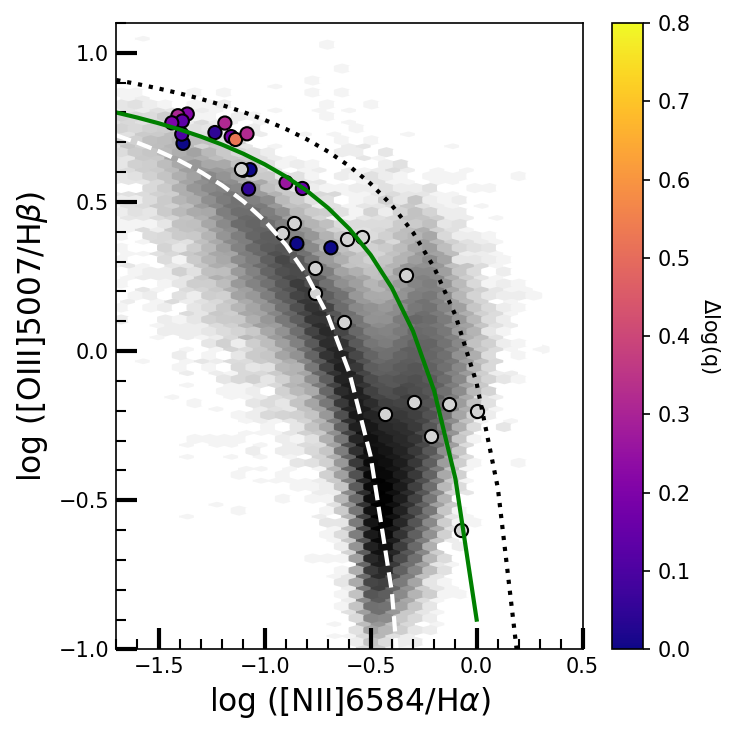}
\end{center}
\caption{\label{fig:BPT_offsets}BPT diagrams for objects with determinations of $q$ and N/O (based on $T_e$(O\,{\sc ii}) data inferred from $T_e$(O\,{\sc iii}) and electron density). A heavily binned density map of the SDSS local sample is shown in grayscale. The large points indicating LBAs are color-coded according to the magnitude of the offsets in log(N/O) (left panel) and log($q$) (right panel). Objects for which no values could be determined are marked in gray. In both panels, the white dashed line shows the locus of the star-forming sequence measured for local galaxies by \citet{steidel14}, the green solid line shows the locus of the $z\sim2$ KBSS-MOSFIRE galaxies from \citet{steidel14}, and the black dotted line shows the separation between composite objects and AGN from \citet{kewley01}.}
\end{figure*}

\begin{figure*}
\begin{center}
\includegraphics[width=0.7\textwidth]{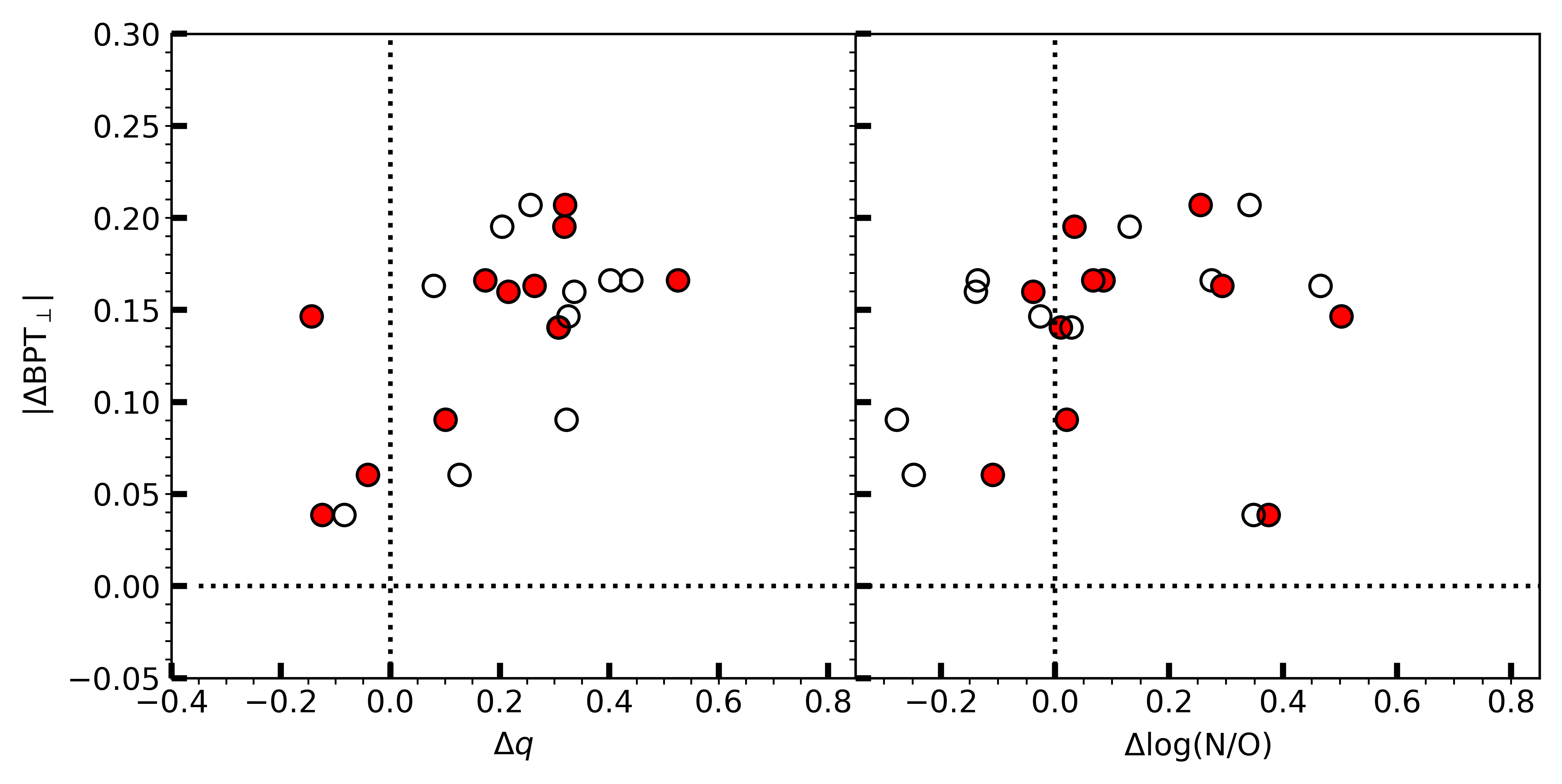}
\end{center}
\caption{\label{fig:DeltaBPT}BPT diagram offsets as a function of the offsets in ionization parameter and N/O. The BPT offsets, $\Delta$BPT$_\perp$, were measured perpendicularly with respect to the locus of local star-forming galaxies. In both panels, objects for which the direct-method abundances were derived using the $T_e$(O\,{\sc ii}) based on RO2 are shown as open circles, while those derived using $T_e$(O\,{\sc ii}) as inferred from $T_e$(O\,{\sc iii}) are shown as filled circles. Objects with the largest perpendicular offsets ($\Delta$BPT$_\perp$) indeed tend to have the largest offsets in $q$ (left panel) or log(N/O) (right panel) or both. }
\end{figure*}

We will now determine whether the nitrogen-to-oxygen ratios of LBAs are in any way markedly different compared to either typical local or high redshift sources. In Fig. \ref{fig:NO_mass_OH} we show the direct method nitrogen-to-oxygen ratio log(N/O)$_{\mathrm{direct}}$ versus the stellar mass (left panel), and the direct method oxygen abundances. In both panels, we again give two sets of points, one calculated using T(O\,{\sc ii}) as inferred from T(O\,{\sc iii}) (filled circles) and the other using T(O\,{\sc ii}) estimated from RO2 (open circles). In the left panel of Fig. \ref{fig:NO_mass_OH}, we are furthermore able to show a larger set of points by taking advantage of the best-fit relation between N/O and N2O2 determined for LBAs above (see left panel of Fig. \ref{fig:NO_stronglines}). This allows us to include LBAs for which no direct-method N/O abundances could be measured. These estimates are indicate by the open squares. In the left-hand panel, it can be seen that the LBAs scatter around the typical N/O found for stacks of SDSS galaxies of a similar stellar mass (solid black line), which was also determined through the direct method by \citet{andrews13}. \citet{kojima17} find that their small sample of $z\sim2$ galaxies have N/O values comparable or smaller than this local relation, while \citet{strom17} find a somewhat lower N/O on average by $\sim0.1-0.2$ dex for KBSS galaxies at $z\sim2$ compared to local SDSS galaxies with stellar masses in the range $\simeq10^{10}-10^{11}$ $M_\odot$ (red solid line in the left panel). However, in any case, these similar or somewhat lower values of N/O at fixed stellar mass are smaller than the typical downward offsets observed for LBAs in the O/H--$M_*$ relation (see Fig. \ref{fig:mzr}). This implies that LBAs, and the $z\sim2$ samples from \citet{kojima17} and \citet{strom17}, likely lie along a different N/O--O/H relation compared to typical local star-forming galaxies. We illustrate this in the right panel of Fig. \ref{fig:NO_mass_OH}, where we have indicated again the local sequence based on the SDSS stacks from \citet{andrews13} as determined by \citet{kojima17} (solid black lines). The LBAs are, on average, offset toward higher N/O at fixed O/H.  

The N/O excesses observed are of similar order of magnitude as those found for galaxies at $z\sim2$ by \citet{kojima17}. The red dashed line in the right panel of Fig. \ref{fig:NO_mass_OH} is the upper limit on the average N/O ($\mathrm{log(N/O)_{direct}}=-1.26$) for their $z\sim2$ sample, while the blue dashed line is the average N/O ($\mathrm{log(N/O)_{direct}}=-1.08$) they found for a sample of local galaxies selected to have similar (low) stellar masses and (high) star formation rates as their $z\sim2$ sample. \citet{kojima17} also noted that a stacked spectrum of the \citet{steidel16} sample of KBSS galaxies at $z\sim2$ falls close to their sample average (red dashed line), while green peas (which include several of the LBAs from our sample) span the range from no N/O excess to the average N/O for local, low mass, high star formation rate objects (blue dashed line). Our results for LBAs are in general agreement with those trends.    

\subsection{Origin of BPT offsets for Lyman Break Analogs}

It has been known for some time that intensely star-forming galaxies at high redshift such as Lyman break galaxies, star-forming BzK galaxies, and distant red galaxies are often offset toward higher line ratios with respect to the mean SDSS star-forming
population at low redshift in a standard BPT diagram \citep[e.g.][]{teplitz00,shapley05,erb06,liu08,lehnert09,hayashi09}. Such offsets were also found in some populations of low redshift galaxies, such as the ``warm'' infrared-luminous galaxies \citep{kewley01} and Wolf–Rayet galaxies \citep{brinchmann08a}. These BPT offsets have been observed for LBAs as well \citep[][and Fig. \ref{fig:bpt} in this paper]{hoopes07,overzier08,overzier09}, further strengthening the conclusion that they are good local analogs of the high redshift star-forming population. \citet{overzier09} showed that the BPT displacement is in the sense of enhancements in one or both of \niisingle/\ha\ and \oiiisingle/\hb, and that the size of the perpendicular offset from the local star-forming ridge increases with both the luminosity of the brightest starburst clump and the electron density, indicating that higher star formation rate densities, interstellar densities and pressures, and ionization parameters may be related to the BPT offsets \citep[e.g.][]{brinchmann08b,liu08,shirazi14}. 

In recent years, it has become feasible to address in more detail the physical origins of these BPT offsets, driven mainly by the wealth of near-infared spectroscopic data from new surveys of $z\sim2$ galaxies \citep[e.g.][]{masters14,steidel14,shapley15,kashino17}. Understanding how objects move through the BPT diagram, including the nature of the offsets, is extremely important, because the measurement of gas-phase abundances critically depends on the interpretation of ratios of the strong emission lines. This is particularly important at high redshift, where the temperature sensitive methods are often not available. There is therefore no guarantee that locally established  calibrations are directly applicable. Various authors have found that high redshift objects display BPT offsets that are often larger than that can be explained by their increased ionization parameters and densities, and have shown that the BPT offset objects often have an increased N/O abundance and/or ionization parameter at high redshift compared to local galaxies for the same oxygen abundance \citep[e.g.][]{masters14,shapley15,sanders16a,kojima17}. 
\citet{steidel16}, \citet{strom17} and \citet{strom18} showed that $z\sim2$ galaxies indeed have a higher N/O (at fixed excitation), and that N/O varies with O/H in a similar manner as observed for local \hii\ regions. Besides a higher N/O, they furthermore show the need for a harder ionizing spectrum at high redshift to fully explain the BPT offsets. These harder spectra could be a result of the differences in the star formation (and thus chemical enrichment) histories between typical local and high redshift galaxies. 

In the previous sections, we have shown that at fixed O/H, LBAs have higher ionization parameters (Fig. \ref{fig:Z_vs_q}) and higher N/O (Fig. \ref{fig:NO_mass_OH}, right panel) compared to typical local star-forming galaxies. \citet{kojima17} showed that galaxies at $z\sim2$, as well as a small sample of local analogs consisting of LBAs, green peas and low mass/high SFR objects, display BPT offsets when either one of N/O or $q$ are increased, or both are increased. Positions along the BPT diagram are further modulated by the different O/H of the galaxies. The change in N/O (at fixed O/H) only affects the \niisingle/\ha\ ratio by an amount that directly corresponds to that change (a 0.3 dex increase in N/O gives a 0.3 dex change in \niisingle/\ha). The increase in $q$ is in the opposite direction as N/O, with some component along \oiiisingle/\hb\ that is roughly aligned along the  star-forming ridge. At 12 + log(O/H)=8.10, a positive change of 0.3 dex in $q$ gives an increase in  \oiiisingle/\hb\ of $\sim0.1$ dex but a decrease in \niisingle/\ha\ of $\sim0.2$ dex. Combined with the 0.3 dex increase in N/O, the total offset in the BPT diagram is toward higer values of both \oiiisingle/\hb\ and \niisingle/\ha, and in the direction observed for the $z\sim2$ galaxies studied.

In Fig. \ref{fig:BPT_offsets} we show again the location of LBAs in the BPT diagram, but with a color-coding determined by the offsets measured in N/O and $q$ relative to the local relations expectation given the O/H determined for each object through the direct method. The objects with the largest offsets in either parameter (or both) are expected to have the largest BPT offsets, which is indeed the case. The objects in our sample with the largest offsets lie around the average relation found by \citet{steidel14} for galaxies at $z\sim2$. In the bottom right corner of the diagram, we find a number of massive and (likely) more metal-rich LBAs for which the \oiiiweak\ flux could not be determined, and thus were not included in the analysis. We show the offsets measured perpendicularly from the local star-forming ridge as a function of the offsets in ionization parameter and N/O in Fig. \ref{fig:DeltaBPT}. Objects with the largest perpendicular offsets ($\Delta$BPT$_\perp$) indeed tend to have the largest offsets in $q$ (left panel) or log(N/O) (right panel) or both. From the analysis presented here, we thus conclude that the physical origin of the BPT offsets in LBAs are likely to be the same as those that explain the  galaxies at $z\sim2$ and other types of local analogs thereof. 

\begin{figure}
\begin{center}
\includegraphics[height=\columnwidth]{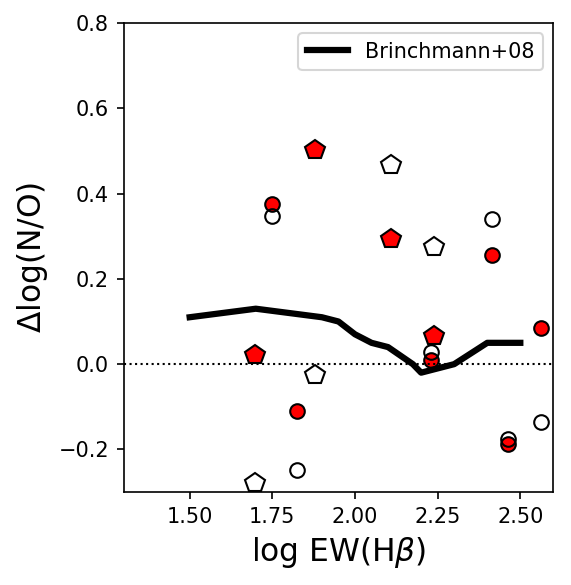}
\end{center}
\caption{\label{fig:ewhb}Offset in log(N/O) versus the equivalent width in \hb. The black solid line shows the average N/O excess measured for WR galaxies compared to non-WR galaxies at the same equivalent width of \hb\ from \citet{brinchmann08a}, which reaches a maximum value of $\Delta \mathrm{log(N/O)}\sim0.13\pm0.1$ dex at EW(\hb) $\lesssim$ 100 \AA. Objects for which the direct-method abundances were derived using the $T_e$(O\,{\sc ii}) based on RO2 are shown as open symbols, while those derived using $T_e$(O\,{\sc ii}) as inferred from $T_e$(O\,{\sc iii}) are shown as filled symbols. Objects marked with large pentagons are LBAs identified with WR features of class 2 or class 3 according to the classification scheme of \citet{brinchmann08a}.}
\end{figure}

\subsection{Possible origin of the excess N/O}

As shown in Fig. \ref{fig:NO_mass_OH}, LBAs span a wide range of almost 1 dex in N/O values at a given O/H ratio. This wide scatter is similar to that observed for local \hii\ regions and the sample of KBSS galaxies at $z\sim2$ by \citet{strom18}. These N/O values furthermore lie, on average, several tenths of dex above the typical N/O expected for local galaxies. This is similar to the trends observed for $z\sim2$ galaxies and other local analogs \citep[e.g.][]{kojima17,strom18}. What could be responsible for this N/O excess? The two most straightforward explanations are that N/O is increased either due to the presence of WR stars that enhance nitrogen on a time-scale that is shorter than the SNe enrichment \citep[e.g.][]{hawley12}, or due to a decrease in O/H from the infall of a large quantity of low metallicity gas with the consequence that N/O appears spuriously high for the resulting O/H \citep[e.g.][]{amorin10,masters14}. 

\subsubsection{Evidence for WR stars}

\begin{figure}
\begin{center}
\includegraphics[width=0.9\columnwidth]{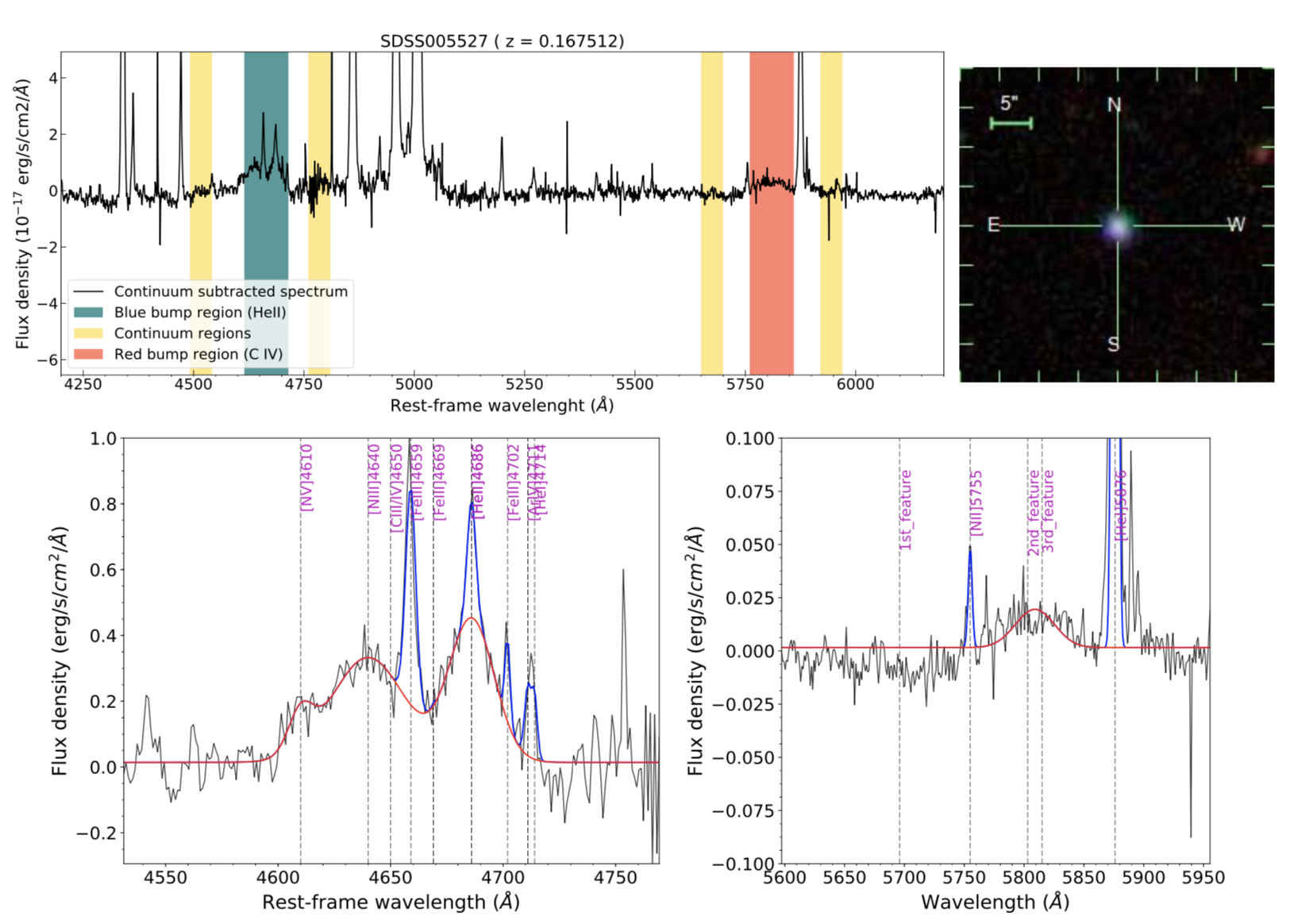}
\includegraphics[width=0.9\columnwidth]{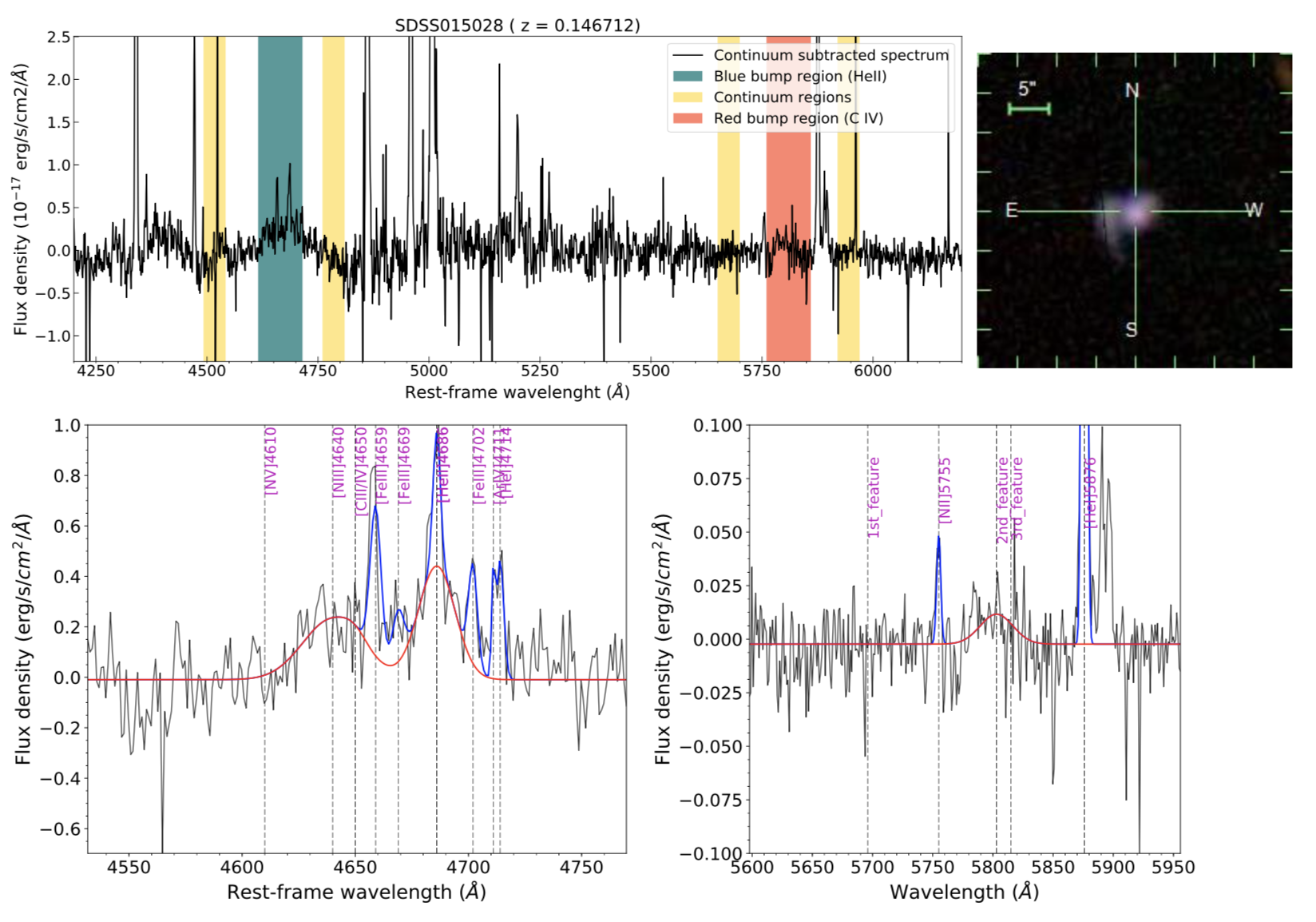}
\end{center}
\caption{\label{fig:wr}Strong Wolf-Rayet features detected in the LBAs SDSS005527 (left panel) and SDSS015028 (right panel). On the left hand side of the top row of panels we show the spectral regions used to measure the flux of the WR spectral features known as the blue (blue shaded region) and red bump (red shaded region). The spectral ranges marked in yellow on either side of the two bumps are used to subtract the continuum flux. On the right hand side of the top row of panels we show an SDSS image stamp of the sources. The bottom row of panels show the best-fit obtained for the blue (left) and red (right) bumps with the red lines showing the characteristic WR bump emission and the blue lines showing the nebular emission lines.}
\end{figure}

Galaxies with WR features often have enhanced N/O. \citet{brinchmann08a} compared the average N/O between WR-enhanced galaxies and galaxies without WR features as a function of the equivalent width of \hb\ in order to control for difrerences that could arise due to secondary nitrogen production. In Fig. \ref{fig:ewhb} we show the excess N/O as a function of the EW(\hb) for our sample. \citet{brinchmann08a} found that the WR phase can be responsible for about $0.15\pm0.1$ dex excess N/O for objects with EW(\hb) below $\sim$100 \AA, while above that the starbursts are typically younger than the phase where enrichment by WR winds occurs (thick solid line in Fig. \ref{fig:ewhb}). Several LBAs are known to have WR features. We have measured the strengths of the WR blue and red bumps in our sample, with two examples of objects with strong bumps shown in Fig. \ref{fig:wr}. In total we find 6 objects with clearly identifiable WR features (005527, 015028, 020356, 040208, BPT10 and HST03). These are marked in Fig. \ref{fig:ewhb} with the large pentagons. \citet{brown14} also showed evidence for WR features in two additional LBAs (004054 and 092600). While some of the objects with the largest N/O excesses include objects with strong WR features in their spectra, the N/O excess is much larger than expected for WR galaxies. Furthermore, there appears to be no trend in Fig. \ref{fig:ewhb} that relates the N/O excess with WR features or EW(\hb). 

\subsubsection{Evidence for infall of metal-poor gas}

\begin{figure}
\begin{center}
\includegraphics[height=\columnwidth]{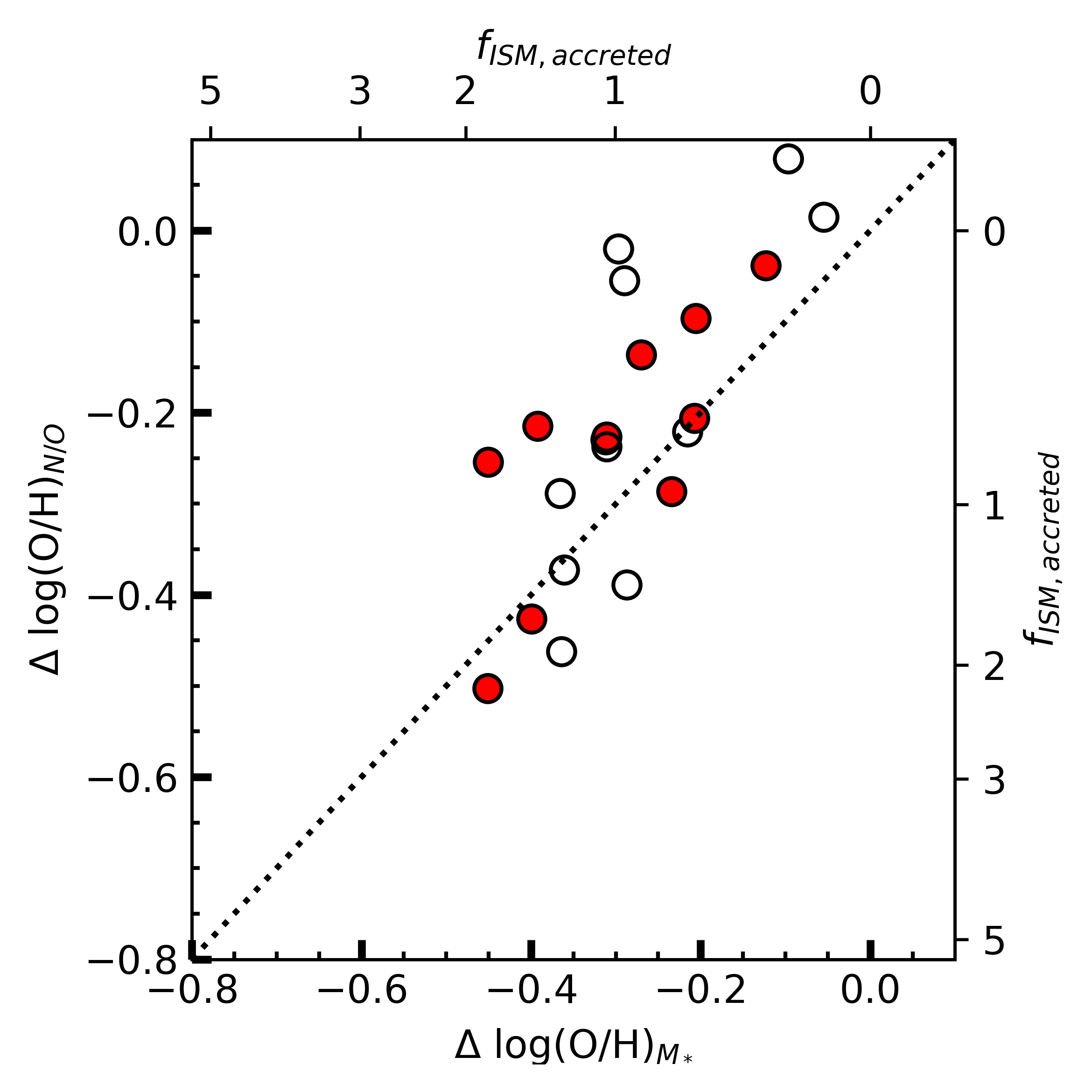}
\end{center}
\caption{\label{fig:doh}The offset in O/H with respect to the local MZR (from Fig. \ref{fig:mzr}) versus the offset in O/H with respect to the local average at a fixed N/O (from the middle panel of Fig. \ref{fig:NO_mass_OH}). Objects for which the direct-method abundances were derived using the $T_e$(O\,{\sc ii}) based on RO2 are shown as open symbols, while those derived using $T_e$(O\,{\sc ii}) as inferred from $T_e$(O\,{\sc iii}) are shown as filled symbols. The good one-to-one correspondence between them implies that these offsets are one and the same, and related to the infall of a large quantity of metal-poor gas. On the secondary axes we have translated the offsets into the quantities of accreted metal-poor gas needed (in units of the pre-infall gas mass).}
\end{figure}

Besides nitrogen enhancement from WR stars, the rapid inflow of relatively metal-poor gas is an attractive alternative (or additional) scenario \citep[e.g.][]{hwang19}, especially given that most LBAs appear to be starbursts triggered by a recent interaction event \citep{overzier08,overzier09}. We showed that at a given stellar mass, LBAs have relatively low O/H with respect to the local MZR (Fig. \ref{fig:mzr}), while they have normal N/O for their stellar mass (left panel of Fig. \ref{fig:NO_mass_OH}). They also have a lower O/H than expected based on their N/O (from the right panel of Fig. \ref{fig:NO_mass_OH}). This is exactly what would be expected for the accretion scenario. We can show this more clearly in Fig. \ref{fig:doh}, where we plot the offset in O/H with respect to the local MZR versus the offset in O/H with respect to the local average at a fixed N/O (from the right panel of Fig. \ref{fig:NO_mass_OH}). The good one-to-one correspondence between them implies that these offsets are one and the same, and related to the infall of a large quantity of metal-poor gas. In order to decrease O/H by 0.3 dex, the galaxy would need to accrete a quantity of metal poor gas equal to the mass of the ISM prior to the infall. On the secondary axes of Fig. \ref{fig:doh} we have translated the offsets in O/H observed to the quantities of accreted metal-poor gas in units of the pre-infall gas mass. We have shown that the effect of the presence of WR stars on N/O is expected to be relatively limited, and furthermore is at odds with the normal N/O versus stellar mass relation (left panel of Fig. \ref{fig:NO_mass_OH}). We conclude that the recent gas accretion scenario can largely, or completely, explain the apparent N/O excesses observed. 

\section{Summary}

In this paper, we analyzed VLT/XShooter spectra of a sample of LBAs to study the physical parameters of the interstellar medium in luminous UV-selected sources that share many properties with star-forming galaxies at high redshift. Our main findings are summarized as follows.

$\bullet$ We estimated the electron densities and measured the electron temperatures T([OIII]) and T([OII]) for a subset of our sample, which allowed us to estimate direct oxygen abundances. The oxygen abundances of LBAs are in the range $12+\mathrm{log(O/H)}\simeq8.0-8.6$, confirming previous results based on strong-line methods that LBAs are, on average, offset from the local MZR (Fig. \ref{fig:mzr}). 

$\bullet$ Comparing the direct method abundance estimates with those based on the O3N2 and N2 strong line ratios, we find general agreement within a scatter of 0.1--0.2 dex. This suggests that the strong line methods can be used for LBA-like galaxies at higher redshifts, as shown by other authors. 

$\bullet$ We determined the ionization parameter based on the $O_{32}$ index and the oxygen abundance, finding that LBAs have ionization parameters that are typically higher by up to 0.5 dex than typical star-forming galaxies of the same O/H (Fig. \ref{fig:Z_vs_q}). We show that the SFR surface densities are correlated with the electron densities as expected for an ISM in which the pressure is regulated by the feedback from massive stars and supernovae. From the electron densities and the ionization parameter, we furthermore estimate the hydrogen-ionizing photon flux and show that it agrees remarkably well with the observed SFR surface density (Fig. \ref{fig:SFRD}).  

$\bullet$ We analyzed nitrogen abundance sensitive strong line ratios (N2O2, N2S2 and N2) and compare them with the N/O ratio determined through the temperature-sensitive method (Figs. \ref{fig:N2O2_N2S2} and \ref{fig:NO_stronglines}). The LBAs tend to follow the same relations between these parameters found for both typical star-forming galaxies at $z\sim2$ as well as local \hii-regions \citep[e.g.][]{strom18}.

$\bullet$ On avearge, LBAs lie on or close to the relation between N/O and stellar mass found for typical star-forming galaxies (left panel of Fig. \ref{fig:NO_mass_OH}), but above the typical N/O expected for galaxies at fixed O/H (right panel of Fig. \ref{fig:NO_mass_OH}). 

$\bullet$ We show that the offsets observed for LBAs in the BPT diagram are linked to the excesses in $q$ and/or N/O (Figs. \ref{fig:BPT_offsets} and \ref{fig:DeltaBPT}). The relatively high ionization parameters, relatively low oxygen abundances, excess N/O and BPT offsets are of a similar order of magnitude as those observed by other authors for star-forming galaxies at $z\sim2-3$ as well as previously studied local analogs \citep[e.g.][]{amorin10,brown14,masters16,sanders16a,steidel14,steidel16,strom17,kojima17,bian18,strom18}.  

$\bullet$ Finally, we explore the origin of the N/O excess considering the two main scenarios proposed in the literature. We show that WR features observed in the spectra of some LBAs can explain at most a small fraction of the nitrogen enhancement (Figs. \ref{fig:ewhb} and \ref{fig:wr}). The majority of the excess N/O, however, appears to be related to the recent inflow of large quantities of relatively metal-poor gas, which lowers O/H while leaving N/O unchanged. The relative decrease in O/H at fixed N/O is similar to the decrease in O/H at fixed stellar mass, and suggests that LBAs have experienced accretion of quantities of gas up to several times their original gas mass (Fig. \ref{fig:doh}). This is consistent with the conclusions of \citet{amorin10} based on strong line method abundances determinations for Green Pea galaxies and Lyman Break Analogs of \citet{overzier09}.\\

The analysis performed in this paper highlights some of the difficulties of determining fundamental parameters such as ionization parameter, O/H, and N/O even in relatively nearby galaxies observed with 8-m telescopes. This illustrates some of the challenges we are faced with in determining these parameters and, more importantly, the physical processes that caused them, at much higher redshifts. This is important for upcoming observations with the {\it James Webb Space Telescope} which will give access to similar rest-frame optical emission line diagnostics for large samples of high redshift galaxies for the first time.

\begin{acknowledgements}
We are grateful to Andrew Humphrey, Fuyan Bian, Irene Shivaei, Masami Ouchi, Ricardo Amorin, Roberto Cid Fernandes, Tomoko Suzuki, and the anonymous referee for helpful comments. This study was financed in part by the Coordena\c{c}\~ao de Aperfei\c{c}oamento de Pessoal de N\'ivel Superior - Brasil (CAPES) - Finance Code 001. RAO is grateful for financial support from FAPERJ (202.876/2015), CNPq (400738/2014-7, 309456/2016-9), CAPES (88881.156185/2017-01) and FAPESP (2018/02444-7). The STARLIGHT project is supported by the Brazilian agencies CNPq, CAPES and FAPESP and by the France-Brazil CAPES/Cofecub program. Based on observations made with ESO Telescopes at the La Silla Paranal Observatory under programme IDs 085.B-0784 and 096.B-0192. 
\end{acknowledgements}

\clearpage

\begin{table*}
\begin{small}
\footnotesize
\centering
\caption{Details on the sample selected for the VLT/XShooter observations.\label{tab:sample}} 
\begin{tabular}{lllllrllll} 
\hline
\hline
SDSS ID & ID$^a$ & Selection$^b$ & COS$^c$ & R.A.    & Dec.    & $z^d$ & Date & Seeing$^e$ & Mode$^f$ \\ 
        &    &           &         & (J2000) & (J2000) &              &      &  FWHM (\arcsec) &        \\                            
\hline  
J001009.97--004603.6 & 001009   &  LBA           & n & 00:10:09.97 & --00:46:03.66 & 0.243094 & 2010-11-07    & 1.7 & N \\ 
J001054.85+001451.3$^\dagger$   & 001054   &  LBA           & n & 00:10:54.85 &   00:14:51.35 & 0.243141 & 2010-11-07    & 1.3 & O \\ 
J004054.32+153409.6  & 004054   &  LBA           & n & 00:40:54.33 &   15:34:09.66 & 0.283241 & 2010-11-08    & 1.4 & N \\ 
J005439.78+155446.9$^\dagger$   & 005439   &  LBA           & n & 00:54:39.80 &   15:54:46.93 & 0.236400 & 2010-11-06    & 2.0 & N \\ 
J005527.45--002148.7 & 005527   &  LBA           & y & 00:55:27.46 & --00:21:48.71 & 0.167449 & 2010-11-07    & 1.1 & N \\ 
J015028.39+130858.4  & 015028   &  LBA           & y & 01:50:28.41 &   13:08:58.40 & 0.146712 & 2010-11-06    & 1.3 & N \\ 
J020356.91--080758.5 & 020356   &  LBA           & n & 02:03:56.91 & --08:07:58.51 & 0.188335 & 2010-11-07    & 1.1 & N \\ 
J021348.53+125951.4  & 021348   &  LBA           & y & 02:13:48.54 &   12:59:51.46 & 0.218962 & 2010-09-07    & 1.7 & N \\ 
J032845.99+011150.8  & 032845   &  LBA           & n & 03:28:45.99 &   01:11:50.85 & 0.142181 & 2010-10-05    & 1.8 & N \\ 
J035733.99--053719.6 & 035733   &  LBA           & n & 03:57:34.00 & --05:37:19.70 & 0.203746 & 2010-10-05    & 1.8 & N \\ 
J040208.86--050642.0 & 040208   &  LBA           & n & 04:02:08.87 & --05:06:42.06 & 0.139291 & 2010-10-05    & 1.9 & N \\ 
J143417.15+020742.5  & 143417   &  LBA           & n & 14:34:17.16 &   02:07:42.58 & 0.180325 & 2010-04-27    & 1.1 & O \\
J210358.74--072802.4 & 210358   &  LBA           & y & 21:03:58.75 & --07:28:02.45 & 0.136840 & 2010-08-10    & 1.6 & O \\
J214500.25+011157.5  & 214500   &  LBA           & n & 21:45:00.26 &   01:11:57.58 & 0.204321 & 2010-11-07    & 1.4 & N \\
J231812.99--004126.1 & 231812   &  LBA           & n & 23:18:13.00 & --00:41:26.10 & 0.251682 & 2010-11-06    & 1.7 & N \\
J232539.22+004507.2  & 232539   &  LBA           & n & 23:25:39.23 &   00:45:07.25 & 0.277000 & 2010-11-06    & 2.4 & N \\
J235347.69+005402.0  & 235347   &  LBA           & n & 23:53:47.69 &   00:54:02.08 & 0.223431 & 2010-10-08    & 1.4 & N \\
J124423.37+021540.4  & BPT03    &  LBA2, BPT      & n & 12:44:23.28 &   02:15:40.40 & 0.238964 & 2016-03-13    & 1.1 & N \\
J082247.66+224144.0  & BPT08    &  LBA2, BPT      & n & 08:22:47.75 &   22:41:44.10 & 0.216226 & 2016-03-12    & 2.1 & N \\  
J101629.88+073404.9  & BPT09    &  LBA2, BPT      & n & 10:16:30.00 &   07:34:04.90 & 0.182710 & 2016-03-13    & 0.8 & N \\ 
J124509.05+104340.1  & BPT10    &  LBA2, BPT      & n & 12:45:09.12 &   10:43:40.00 & 0.165569 & 2016-03-12    & 1.4 & N \\  
J084034.10+134451.3  & BPT11    &  LBA2, BPT      & n & 08:40:34.07 &   13:44:51.30 & 0.226961 & 2016-03-13    & 1.3 & N \\  
J120735.77+082215.5  & BPT15    &  LBA2, BPT      & n & 12:07:35.76 &   08:22:15.50 & 0.204993 & 2016-03-13    & 1.3 & N \\ 
J102355.73+232338.6  & BPT20    &  LBA2, BPT      & n & 10:23:55.67 &   23:23:38.70 & 0.254211 & 2016-03-12    & 2.1 & N \\ 
J101009.90+205035.2  & BPT23    &  LBA2, BPT      & n & 10:10:09.83 &   20:50:35.50 & 0.209547 & 2016-03-12    & 2.1 & N \\ 
J120721.44+021657.7  & BPT26    &  LBA2, BPT      & n & 12:07:21.35 &   02:16:57.70 & 0.221747 & 2016-03-15    & 0.9 & N \\ 
J141612.87+122340.4  & HST03    &  LBA2, COS      & y & 14:16:12.96 &   12:23:40.50 & 0.123122 & 2016-03-12    & 2.1 & N \\ 
J104457.79+035313.1  & S01$\_$2 &  S2-Deficit  & y & 10:44:57.84 &   03:53:13.10 & 0.012879 & 2016-03-13    & 1.3 & O \\ 
J095343.89--000524.7 & S04$\_$1 &  S2-Deficit  & y & 09:53:43.91 & --00:05:24.60 & 0.083360 & 2016-03-15    & 1.4 & O \\ 
J122627.93+094456.6  & S09$\_$1 &  S2-Deficit  & y & 12:26:27.84 &   09:44:56.70 & 0.090481 & 2016-03-15    & 0.9 & O \\ 
\hline
\hline
\multicolumn{10}{l}{$\dagger$ Originally part of the LBA sample, objects 001054 and 005439 were found to be AGN based on broad Mg II lines detected with XShooter.}\\
\multicolumn{10}{l}{a. The number between parentheses is used to identify objects in certain figures.}\\
\multicolumn{10}{l}{b. Sample selection (see text for details).}\\
\multicolumn{10}{l}{c. Observed with the HST/COS spectrograph: 'y' for yes, 'n' for no.}\\
\multicolumn{10}{l}{d. The systemic redshift was obtained from the SDSS spectra.}\\
\multicolumn{10}{l}{e. Average seeing (FWHM) measured from the telluric standard star spectrum in the UVB arm.}\\
\multicolumn{10}{l}{f. Observing mode: 'N' for nodding, 'O' for offset.}
\end{tabular}
\end{small}
\end{table*}

\begin{sidewaystable}
\begin{center}
\begin{scriptsize}
\caption{Emission line fluxes.\label{tab:fluxes}} 
\begin{tabular}{lrrrrrrrrrrrrr} 
\hline
\hline
ID       &  \multicolumn{13}{c}{Emission line fluxes$^{a,b,c}$}\\  
         &  \multicolumn{1}{c}{[O\,{\sc ii}]$\lambda\lambda$3727} & \multicolumn{1}{c}{[O\,{\sc iii}]$\lambda$4363}  & \multicolumn{1}{c}{\hb} & \multicolumn{1}{c}{[O\,{\sc iii}]$\lambda$4959} & \multicolumn{1}{c}{[O\,{\sc iii}]$\lambda$5007} & \multicolumn{1}{c}{[N\,{\sc ii}]$\lambda$5755} & \multicolumn{1}{c}{[N\,{\sc ii}]$\lambda$6548} & \multicolumn{1}{c}{\ha} & \multicolumn{1}{c}{[N\,{\sc ii}]$\lambda$6584} & \multicolumn{1}{c}{[S\,{\sc ii}]$\lambda$6717} & \multicolumn{1}{c}{[S\,{\sc ii}]$\lambda$6731} & \multicolumn{1}{c}{[O\,{\sc ii}]$\lambda$7319} & \multicolumn{1}{c}{[O\,{\sc ii}]$\lambda$7330}\\
\hline   
BPT03   &   1762.1$\pm$ 8.6 &  75.8$\pm$1.4 &  867.9$\pm$ 6.7 & 1857.4$\pm$6.3 & 5412.4$\pm$14.8  &  --          &  38.6$\pm$1.0 & 2562.2$\pm$3.0 &  110.5$\pm$ 4.7 & 120.3$\pm$1.4 &  98.4$\pm$1.2 & 20.6$\pm$0.5 & 16.9$\pm$0.5\\
BPT08   &   1879.5$\pm$40.5 &  60.9$\pm$1.6 &  899.4$\pm$ 9.6 & 1730.0$\pm$6.8 & 5229.3$\pm$13.7  &  --          &  85.5$\pm$1.6 & 2645.3$\pm$4.8 &  171.8$\pm$ 5.8 & 119.5$\pm$1.7 & 110.2$\pm$1.6 & 33.7$\pm$1.9 & 24.8$\pm$1.7\\
BPT09   &   1655.3$\pm$11.4 &  42.5$\pm$1.7 &  728.8$\pm$10.8 & 1260.1$\pm$6.4 & 3816.3$\pm$13.8  &  --          &  51.7$\pm$1.1 & 2160.2$\pm$3.3 &  150.2$\pm$ 4.5 & 107.5$\pm$1.5 &  89.7$\pm$1.3 & 16.5$\pm$0.7 & 13.5$\pm$0.8\\
BPT10   &   1534.9$\pm$ 6.3 &  24.6$\pm$0.7 &  864.8$\pm$ 5.5 & 1485.5$\pm$5.8 & 4429.9$\pm$10.6  &  2.7$\pm$0.5 &  63.1$\pm$1.0 & 2502.9$\pm$3.8 &  182.4$\pm$ 5.2 & 148.2$\pm$1.3 & 124.1$\pm$1.1 & 24.7$\pm$0.5 & 19.4$\pm$0.5\\
BPT11   &    464.3$\pm$ 4.1 &  11.8$\pm$0.6 &  260.1$\pm$ 4.5 &  463.9$\pm$2.6 & 1392.3$\pm$ 5.3  &  --          &  20.3$\pm$0.6 &  758.4$\pm$1.8 &   62.6$\pm$ 2.3 &  47.9$\pm$0.9 &  38.1$\pm$0.8 &  7.2$\pm$0.3 &  5.7$\pm$0.5\\
BPT15   &    557.8$\pm$ 4.8 &   --          &  213.8$\pm$ 4.1 &  171.0$\pm$2.7 &  507.9$\pm$ 4.6  &  --          &  49.3$\pm$1.1 &  626.8$\pm$2.0 &  153.3$\pm$ 2.6 &  59.9$\pm$1.1 &  52.8$\pm$1.0 &  6.1$\pm$0.6 &  5.1$\pm$0.8\\
BPT20   &    341.2$\pm$ 7.3 &   --          &  206.2$\pm$ 4.9 &  124.9$\pm$3.0 &  369.9$\pm$ 4.7  &  --          & 104.6$\pm$1.6 &  606.0$\pm$2.2 &  280.9$\pm$ 2.5 &  51.6$\pm$1.5 &  47.7$\pm$1.5 &  7.3$\pm$1.2 &  4.8$\pm$0.9\\
BPT23   &    605.6$\pm$70.6 &   --          &  237.4$\pm$ 8.9 &   53.9$\pm$4.8 &  159.7$\pm$ 7.4  &  --          & 122.3$\pm$2.5 &  703.6$\pm$3.5 &  357.8$\pm$ 3.4 & 107.4$\pm$2.4 &  87.1$\pm$2.2 &  1.2$\pm$1.8 & 13.2$\pm$1.6\\
BPT26   &    380.8$\pm$29.3 &   --          &  304.7$\pm$17.7 &   43.3$\pm$8.0 &  158.6$\pm$14.8  &  --          & 204.2$\pm$4.0 &  934.6$\pm$4.9 &  569.2$\pm$ 5.7 &  98.8$\pm$3.0 & 102.3$\pm$3.1 &  --          &  --         \\
HST03   &   7803.5$\pm$46.1 &   --          & 2149.8$\pm$13.7 & 1718.9$\pm$38.5& 5172.3$\pm$35.4  & 28.4$\pm$5.0 & 667.6$\pm$5.2 & 6465.2$\pm$8.0 & 1864.6$\pm$11.2 & 493.0$\pm$4.8 & 462.6$\pm$4.3 & 59.2$\pm$2.0 & 44.0$\pm$2.1\\
S01$\_2$   &    876.5$\pm$10.1 & 522.0$\pm$4.0 & 2907.3$\pm$ 9.9 & 3820.4$\pm$8.9 &10006.9$\pm$ 18.0 &  --          &   9.7$\pm$1.2 & 8509.2$\pm$7.2 &    --           &  70.9$\pm$1.6 &  56.4$\pm$1.4 & 19.2$\pm$0.8 & 13.7$\pm$0.8\\
S04$\_1$   &      --           &   --          &   48.7$\pm$ 3.0 &   13.5$\pm$1.9 &   30.6$\pm$ 3.0  &  --          &  35.0$\pm$1.6 &  106.7$\pm$1.7 &  106.7$\pm$ 1.9 &   --          &   --          &  --          &  --         \\
S09$\_1$   &      --           &   --          &  681.0$\pm$11.0 &   65.4$\pm$5.5 &  170.9$\pm$ 9.2  &  --          & 716.5$\pm$8.1 & 2041.0$\pm$7.2 & 1718.7$\pm$ 9.0 & 110.3$\pm$4.2 & 134.5$\pm$4.4 &  8.6$\pm$2.4 & 12.4$\pm$2.5\\
001009  &    122.3$\pm$ 2.7 &   --          &   47.6$\pm$ 2.3 &   33.6$\pm$1.4 &   90.7$\pm$ 2.2  &  --          &   7.7$\pm$0.5 &  138.5$\pm$1.1 &   24.0$\pm$ 1.0 &  19.2$\pm$1.1 &  14.8$\pm$0.9 &  --          &  --         \\
004054  &    277.4$\pm$ 2.0 &  12.0$\pm$1.8 &  169.7$\pm$ 2.2 &  349.0$\pm$2.2 & 1044.6$\pm$ 4.7  &  --          &   5.8$\pm$0.5 &  493.5$\pm$1.3 &   19.2$\pm$ 1.7 &  23.7$\pm$0.7 &  18.6$\pm$0.6 &  3.9$\pm$0.8 &  3.3$\pm$1.0\\
005527  &   3234.8$\pm$30.0 &  35.9$\pm$1.8 & 1611.3$\pm$10.1 & 1991.9$\pm$8.7 & 5918.5$\pm$16.2  &  9.9$\pm$1.8 & 230.0$\pm$2.1 & 4751.8$\pm$5.4 &  599.6$\pm$ 7.2 & 284.8$\pm$2.3 & 253.4$\pm$2.1 & 56.6$\pm$0.9 & 43.8$\pm$0.9\\
015028  &   1344.9$\pm$10.3 &  10.0$\pm$1.2 &  605.9$\pm$16.1 &  449.8$\pm$4.4 & 1345.5$\pm$11.0  &  --          & 119.8$\pm$1.8 & 1786.2$\pm$3.4 &  366.9$\pm$ 4.1 & 155.7$\pm$1.6 & 133.6$\pm$1.5 & 11.4$\pm$0.5 &  9.5$\pm$0.8\\
020356  &    732.9$\pm$ 4.1 &   6.7$\pm$0.6 &  248.3$\pm$ 4.3 &  338.2$\pm$2.8 & 1002.4$\pm$ 6.3  &  2.0$\pm$0.3 &  19.7$\pm$0.6 &  722.9$\pm$1.6 &   57.3$\pm$ 2.0 &  70.2$\pm$1.1 &  53.0$\pm$1.0 &  6.5$\pm$0.3 &  5.5$\pm$0.4\\
021348  &      --           &   --          &  147.9$\pm$13.6 &    --          &   98.1$\pm$10.2  &  --          & 136.1$\pm$3.4 &  444.8$\pm$4.2 &  330.3$\pm$ 4.3 &  42.5$\pm$2.6 &  44.1$\pm$2.8 &  --          &  --         \\
032845  &    605.7$\pm$ 3.3 &   3.6$\pm$0.5 &  281.2$\pm$ 2.6 &  206.2$\pm$1.9 &  645.0$\pm$ 4.0  &  --          &  35.7$\pm$0.9 &  815.4$\pm$1.6 &  115.6$\pm$ 2.1 &  80.1$\pm$1.0 &  61.6$\pm$0.8 &  6.5$\pm$0.3 &  5.4$\pm$0.4\\
035733  &    847.2$\pm$ 8.1 &   --          &  233.1$\pm$ 5.4 &  118.4$\pm$3.9 &  363.9$\pm$ 5.1  &  --          &  40.5$\pm$1.2 &  690.9$\pm$2.6 &  118.8$\pm$ 2.8 &  78.4$\pm$1.4 &  64.2$\pm$1.3 &  5.3$\pm$0.8 &  5.0$\pm$1.2\\
040208  &    373.9$\pm$ 2.8 &   --          &  104.0$\pm$ 2.1 &   89.3$\pm$1.7 &  259.1$\pm$ 3.4  &  --          &  11.6$\pm$0.6 &  303.0$\pm$1.2 &   36.6$\pm$ 1.4 &  41.3$\pm$1.1 &  28.3$\pm$0.8 &  3.0$\pm$0.4 &  1.9$\pm$0.6\\
143417  &    533.8$\pm$ 8.6 &   --          &  297.8$\pm$10.2 &   63.0$\pm$3.2 &  183.7$\pm$ 6.5  &  --          & 109.3$\pm$2.1 &  880.5$\pm$2.9 &  324.5$\pm$ 3.9 &  98.7$\pm$2.0 &  77.8$\pm$1.6 &  --          &  --         \\
214500  &    554.3$\pm$ 4.1 &   --          &  213.3$\pm$ 3.4 &   88.2$\pm$2.3 &  266.4$\pm$ 3.3  &  --          &  48.3$\pm$0.9 &  623.2$\pm$1.9 &  148.2$\pm$ 2.3 &  93.8$\pm$1.2 &  71.3$\pm$1.0 &  4.8$\pm$0.6 &  3.6$\pm$1.1\\
231812  &    953.9$\pm$ 5.9 &   --          &  267.2$\pm$ 3.6 &  239.2$\pm$2.6 &  720.7$\pm$ 4.5  &  --          &  35.6$\pm$0.9 &  784.1$\pm$1.7 &  107.8$\pm$ 2.3 & 101.3$\pm$1.5 &  77.4$\pm$1.2 &  8.6$\pm$0.5 &  6.4$\pm$0.7\\
232539  &    237.6$\pm$ 2.7 &   --          &  101.4$\pm$ 2.9 &  135.0$\pm$1.7 &  413.3$\pm$ 3.5  &  --          &   6.8$\pm$0.7 &  295.8$\pm$1.6 &   22.9$\pm$ 1.9 &  22.8$\pm$0.9 &  15.0$\pm$0.7 &  --          &  --         \\
235347  &    352.6$\pm$ 3.9 &   9.1$\pm$0.6 &  133.6$\pm$ 4.4 &  233.6$\pm$2.0 &  664.5$\pm$ 4.3  &  --          &   5.8$\pm$0.5 &  390.6$\pm$1.4 &   16.1$\pm$ 1.7 &  30.6$\pm$0.8 &  22.6$\pm$0.7 &  4.2$\pm$0.7 &  2.9$\pm$0.4\\
\hline
\hline
\multicolumn{14}{l}{$^a$ All line fluxes are given in units of $10^{-17}$ erg s$^{-1}$ cm$^{-2}$.}\\ 
\multicolumn{14}{l}{$^b$ Flux errors were determined by propagating the error spectra during the flux measurements, and do not include the uncertainties in the absolute flux calibration.}\\
\multicolumn{14}{l}{$^c$ Line fluxes were corrected for Galactic and internal reddening, as well as stellar continuum absorption.}\\
\end{tabular}
\end{scriptsize}
\end{center}
\end{sidewaystable}

\begin{table*}
\begin{small}
\footnotesize
\centering
\caption{Spectroscopic LBA sample taken from the literature.\label{tab:literature_sample}} 
\begin{tabular}{lllllrll} 
\hline
\hline
SDSS ID & ID$^a$ & Selection$^b$ & COS$^c$ & R.A.    & Dec.    & $z^d$ & Reference$^e$ \\ 
        &        &               &         & (J2000) & (J2000) &       &           \\                            
\hline   
\hline
J004054.32+153409.6  & 004054 (L1)  &  LBA           & n & 00:40:54.33 & +15:34:09.66  & 0.283241 & \citet{amorin12}\\ 
J113303.80+651341.3  & 113303 (L2)  &  LBA           & n & 11:33:03.80 & +65:13:41.31  & 0.241    & \citet{amorin12}\\ 
J232539.22+004507.2  & 232539 (L3)  &  LBA           & n & 23:25:39.23 & +00:45:07.25  & 0.277000 & \citet{amorin12}\\
\hline
J004054.32+153409.6  & 004054 (L4)  &  LBA           & n & 00:40:54.33 & +15:34:09.66  & 0.283241 & \citet{brown14}\\ 
J005527.45--002148.7 & 005527 (L5)  &  LBA           & y & 00:55:27.46 & --00:21:48.71 & 0.167449 & \citet{brown14}\\ 
J020356.91--080758.5 & 020356 (L6)  &  LBA           & n & 02:03:56.91 & --08:07:58.51 & 0.188335 & \citet{brown14}\\ 
J092600.41+442636.1  & 092600 (L7)  &  LBA           & y & 09:26:00.41 & +44:27:36.13  & 0.18072  & \citet{brown14}\\
\hline
\hline
\multicolumn{8}{l}{a. The number between parentheses is used to identify objects in certain figures.}\\
\multicolumn{8}{l}{b. Selection of the target: 'LBA' for objects from the sample of \citet{heckman05}.}\\
\multicolumn{8}{l}{c. Observed with the HST/COS spectrograph: 'y' for yes, 'n' for no.}\\
\multicolumn{8}{l}{d. The systemic redshift was obtained from the SDSS spectra.}\\
\multicolumn{8}{l}{e. Reference from which the extinction-free, continuum absorption-corrected line fluxes were taken.}
\end{tabular}
\end{small}
\end{table*}

\begin{table*}
\begin{small}
\footnotesize
\centering
\caption{Basic parameters of LBAs used in this paper.\label{tab:basic_pars}} 
\begin{tabular}{lccccccc} 
\hline
\hline
ID       & log($M_*/M_\odot$) & log($L_{FUV}/L_\odot$) & log($I_{FUV}/L_\odot$) & log([OIII]/H$\beta$)  & log([NII]/H$\alpha$) & log($N_e$[SII]) & log($N_e$[OII])\\  
         &                    &                        &       (kpc$^{-2}$)     &                       &                      &  (cm$^{-3}$)    & (cm$^{-3}$)    \\
\hline  
BPT03     &      9.59  &     10.7  &     9.66  &    0.79   &   -1.37   &    2.26    &   2.39\\
BPT08     &      9.58  &     10.5  &     9.15  &    0.77   &   -1.19   &    2.55    &    --\\
BPT09     &      9.46  &     10.5  &     9.28  &    0.72   &   -1.16   &    2.31    &   2.34\\
BPT10     &      9.58  &     10.7  &     9.42  &    0.71   &   -1.14   &    2.32    &    2.5\\
BPT11     &      9.29  &     10.4  &     9.20  &    0.73   &   -1.08   &    2.16    &   2.23\\
BPT15     &      10.4  &     10.5  &     9.81  &    0.38   &   -0.61   &    2.45    &   2.72\\
BPT20     &      10.1  &     10.6  &     9.66  &    0.25   &   -0.33   &    2.55    &    2.9\\
BPT23     &      10.7  &     10.6  &     9.74  &   -0.17   &   -0.30   &    2.23    &    --\\
BPT26     &      10.8  &     10.6  &     9.84  &   -0.28   &   -0.22   &    2.76    &    2.8\\
HST03     &      10.0  &     10.8  &     10.3  &    0.38   &   -0.54   &    2.58    &   2.86\\
S01$\_2$  &      6.96  &     8.46  &     10.1  &    0.54   &   -2.50   &    2.17    &   2.64\\
S09$\_1$  &      10.1  &     10.1  &     10.1  &   -0.60   &   -0.07   &    3.01    &    --\\
001009    &      10.5  &     10.5  &     9.47  &    0.28   &   -0.76   &    2.05    &   1.58\\
004054    &      9.27  &     10.3  &     9.07  &    0.79   &   -1.41   &    2.12    &   2.43\\
005527    &      9.69  &     10.6  &     10.1  &    0.57   &   -0.90   &    2.47    &   1.47\\
015028    &      10.3  &     10.6  &     9.44  &    0.35   &   -0.69   &    2.39    &   2.27\\
020356    &      9.41  &     10.6  &     9.30  &    0.61   &   -1.10   &    1.95    &    --\\
021348    &      10.4  &     10.7  &     10.1  &   -0.18   &   -0.13   &    2.76    &    --\\
032845    &      9.82  &     10.4  &     9.08  &    0.36   &   -0.85   &    2.03    &   2.19\\
035733    &      9.99  &     10.6  &     9.57  &    0.19   &   -0.77   &    2.26    &   2.19\\
040208    &      9.50  &     10.4  &     9.51  &    0.40   &   -0.92   &    1.99    &   1.99\\
143417    &      10.7  &     10.6  &     9.59  &   -0.21   &   -0.43   &    2.13    &   2.27\\
214500    &      9.98  &     10.5  &     9.93  &    0.10   &   -0.62   &    1.98    &   2.26\\
231812    &      10.0  &     11.0  &     9.46  &    0.43   &   -0.86   &    2.01    &    2.1\\
232539    &      9.27  &     10.6  &     9.72  &    0.61   &   -1.11   &    2.24    &   2.24\\
235347    &      9.47  &     10.5  &     9.04  &    0.70   &   -1.38   &    1.83    &   1.95\\
\hline
\hline
\end{tabular}
\end{small}
\end{table*}

\begin{table*}
\begin{small}
\footnotesize
\centering
\caption{Temperatures, abundances and ionization parameters determined in this paper.\label{tab:Z_pars}} 
\begin{tabular}{lcccccccc} 
\hline
\hline
ID       &   $T_e$(O\,{\sc iii})        & $T_e$(O\,{\sc ii})        &     $T_e$(N\,{\sc ii})      & 12 + log(O/H) [N2]   &    12 + log(O/H) [O3N2]     &  12 + log(O/H) [$T_e$]$^a$    &  log(N/O) [$T_e$]$^a$ & $q^b$ \\
         &   ($\times10^4$ K) & ($\times10^4$ K)&  ($\times10^4$ K) & & & & \\
\hline  
BPT03     &       1.20$\pm$ 0.01  &     0.94$\pm$ 0.02  &  --                    &    8.12$\pm$0.01    &   8.04$\pm$0.01   &    8.27$\pm$0.01     &   -1.47$\pm$0.02  &   7.94$\pm$0.01 \\
BPT08     &       1.14$\pm$ 0.01  &     1.09$\pm$ 0.05  &  --                    &    8.22$\pm$0.01    &   8.11$\pm$0.01   &    8.39$\pm$0.02     &   -1.40$\pm$0.02  &   7.95$\pm$0.01 \\
BPT09     &       1.13$\pm$ 0.01  &     0.83$\pm$ 0.03  &  --                    &    8.24$\pm$0.01    &   8.13$\pm$0.01   &    8.33$\pm$0.02     &   -1.34$\pm$0.02  &   7.85$\pm$0.01 \\
BPT10     &       0.94$\pm$ 0.01  &     1.14$\pm$ 0.02  &    0.99$\pm$ 0.09      &    8.25$\pm$0.01    &   8.14$\pm$0.01   &    8.54$\pm$0.01     &   -1.31$\pm$0.03  &   8.06$\pm$0.01 \\
BPT11     &       1.05$\pm$ 0.02  &     1.16$\pm$ 0.05  &  --                    &    8.28$\pm$0.01    &   8.15$\pm$0.01   &    8.37$\pm$0.03     &   -1.17$\pm$0.03  &   7.97$\pm$0.02 \\
BPT15     &    --                 &     0.84$\pm$ 0.06  &  --                    &    8.55$\pm$0.01    &   8.41$\pm$0.01   &   --                 &   --              &         --      \\
BPT20     &    --                 &     1.21$\pm$ 0.16  &  --                    &    8.71$\pm$0.01    &   8.54$\pm$0.01   &   --                 &   --              &         --      \\
BPT23     &    --                 &     1.06$\pm$ 0.16  &  --                    &    8.73$\pm$0.01    &   8.69$\pm$0.01   &   --                 &   --              &         --      \\
BPT26     &    --                 &  --                   &  --                  &    8.78$\pm$0.01    &   8.75$\pm$0.02   &   --                 &   --              &         --      \\
HST03     &    --                 &     0.60$\pm$ 0.02  &    0.99$\pm$ 0.08      &    8.59$\pm$0.01    &   8.44$\pm$0.01   &   --                 &   --              &         --      \\
S01$\_2$  &       1.91$\pm$ 0.01  &     1.47$\pm$ 0.06  &  --                    &    7.47$\pm$0.15    &   7.76$\pm$0.08   &    7.41$\pm$0.01     &   -1.62$\pm$0.25  &   8.15$\pm$0.01 \\
S04$\_1$  &    --                 &  --                   &  --                  &    8.90$\pm$0.01    &   8.79$\pm$0.02   &   --                 &   --              &         --      \\
S09$\_1$  &    --                 &     0.86$\pm$ 0.39  &  --                    &    8.86$\pm$0.01    &   8.90$\pm$0.01   &   --                 &   --              &         --      \\
001009    &    --                 &  --                   &  --                  &    8.47$\pm$0.01    &   8.40$\pm$0.01   &   --                 &   --              &         --      \\
004054    &       1.14$\pm$ 0.05  &     1.12$\pm$ 0.17  &  --                    &    8.10$\pm$0.02    &   8.03$\pm$0.01   &    8.29$\pm$0.06     &   -1.42$\pm$0.06  &   8.02$\pm$0.03 \\
005527    &       0.96$\pm$ 0.01  &     1.14$\pm$ 0.02  &     1.03$\pm$ 0.09     &    8.39$\pm$0.01    &   8.26$\pm$0.01   &    8.47$\pm$0.02     &   -1.14$\pm$0.01  &   7.84$\pm$0.01 \\
015028    &       1.01$\pm$ 0.03  &     0.74$\pm$ 0.02  &  --                    &    8.51$\pm$0.01    &   8.40$\pm$0.01   &    8.30$\pm$0.05     &   -0.93$\pm$0.02  &   7.56$\pm$0.02 \\
020356    &       0.98$\pm$ 0.03  &     0.86$\pm$ 0.03  &     1.51$\pm$  0.15    &    8.27$\pm$0.01    &   8.18$\pm$0.01   &    8.43$\pm$0.05     &   -1.41$\pm$0.03  &   7.71$\pm$0.02 \\
021348    &    --                 &  --                   &  --                  &    8.83$\pm$0.01    &   8.75$\pm$0.02   &   --                 &   --              &         --      \\
032845    &       0.94$\pm$ 0.03  &     0.94$\pm$ 0.03  &  --                    &    8.42$\pm$0.01    &   8.34$\pm$0.01   &    8.30$\pm$0.06     &   -1.06$\pm$0.02  &   7.58$\pm$0.02 \\
035733    &    --                 &     0.69$\pm$ 0.05  &  --                    &    8.46$\pm$0.01    &   8.42$\pm$0.01   &   --                 &   --              &         --      \\
040208    &    --                 &     0.76$\pm$ 0.06  &  --                    &    8.38$\pm$0.01    &   8.31$\pm$0.01   &   --                 &   --              &         --      \\
143417    &    --                 &  --                   &  --                  &    8.65$\pm$0.01    &   8.66$\pm$0.01   &   --                 &   --              &         --      \\
214500    &    --                 &     0.81$\pm$ 0.07  &  --                    &    8.54$\pm$0.01    &   8.50$\pm$0.01   &   --                 &   --              &         --      \\
231812    &    --                 &     0.83$\pm$ 0.03  &  --                    &    8.41$\pm$0.01    &   8.32$\pm$0.01   &   --                 &   --              &         --      \\
232539    &    --                 &  --                   &  --                  &    8.27$\pm$0.02    &   8.18$\pm$0.01   &   --                 &   --              &         --      \\
235347    &       1.19$\pm$ 0.03  &     0.98$\pm$ 0.08  &  --                    &    8.11$\pm$0.03    &   8.06$\pm$0.02   &    8.19$\pm$0.04     &   -1.54$\pm$0.06  &   7.73$\pm$0.02 \\
\hline
\hline
\multicolumn{9}{l}{a. Values were calculated by inferring $T_e$(O\,{\sc ii}) from $T_e$(O\,{\sc iii}) and the electron density.}\\
\multicolumn{9}{l}{b. Ionization parameter $q$ as shown in Fig. \ref{fig:Z_vs_q}.}\\
\end{tabular}
\end{small}
\end{table*}

\end{document}